\documentclass[11pt,a4paper]{article} 
 \pdfoutput=1
 \usepackage[utf8x]{inputenc}  
 \usepackage{jheppub}
 \usepackage{enumerate}
 \usepackage{epsfig} 
 \usepackage{float} 
 \usepackage[caption = false]{subfig}
 \usepackage{wasysym} 
 \usepackage{mathrsfs} 
 \usepackage{amsfonts} 
 \usepackage{amsbsy} 
 \usepackage{amscd} 
 \usepackage{pstricks} 
 \usepackage{multirow} 
 \usepackage{tikz}
 \usepackage{color}
 \usepackage{slashed}
 \usepackage{array}
 \usepackage{tablefootnote}
 
 \usetikzlibrary{arrows,positioning,shapes.geometric} 
 \usepackage[compat=1.1.0]{tikz-feynman}       
 \usepackage[font=small,labelfont=bf]{caption}
 \usepackage{slashed}
 \usepackage{multirow}
 \usepackage{tabularx}
 \usepackage{soul} 
 \usepackage{listings} 
 \usepackage{color}
  
 \title{Invisible Higgs search through Vector Boson Fusion: \\ A deep learning approach}
 
 \author[a,b]{Vishal S. Ngairangbam,}  
 \author[a]{Akanksha Bhardwaj,} 
 \author[a]{Partha Konar,}
 \author[c]{Aruna Kumar Nayak }
 
 \affiliation[a]{Physical Research Laboratory, Ahmedabad - 380009, Gujarat, India}
 \affiliation[b]{Indian Institute of Technology, Gandhinagar - 382424, Gujarat, India}
 \affiliation[c]{Institute of Physics, Bhubaneswar - 751005, Odisha, India}
 
 \emailAdd{vishalng@prl.res.in}
 \emailAdd{akanksha@prl.res.in}
 \emailAdd{konar@prl.res.in}
 \emailAdd{nayak@iopb.res.in}
\abstract{ 
	Vector boson fusion proposed initially as
	an alternative channel for finding heavy Higgs has now
	established itself as a crucial search scheme to probe different properties of the Higgs boson or for new physics.
	We explore the merit of deep-learning entirely
	from the low-level calorimeter data in the search
	for invisibly decaying Higgs. Such an effort supersedes decades-old faith in the remarkable event
	kinematics and radiation pattern as a signature
	to the absence of any color exchange between incoming partons in the vector boson fusion mechanism. We investigate among different neural network architectures, considering both low-level and high-level
	input variables as a detailed comparative analysis. To have a consistent comparison with existing techniques, we closely follow a recent experimental study of CMS
	search on invisible Higgs with 36 fb$^{−1}$ data. We find
	that sophisticated deep-learning techniques have the
	impressive capability to improve the bound on invisible branching ratio by a factor of three, utilizing the same amount of data. Without relying on any exclusive event reconstruction, this novel technique can provide the most stringent bounds on the invisible branching ratio of the SM-like Higgs boson. Such an outcome has the ability to constraint many different BSM models severely.
}

\allowdisplaybreaks

\begin{document}
\maketitle
\flushbottom

\newpage
\section{Introduction}
\label{sec:intro}
With the emergence of deep learning frameworks, a plethora of machine learning applications have gained immense importance in high-energy physics (HEP) recently,  in collider and neutrino physics \cite{Albertsson:2018maf,Aurisano:2016jvx,Yates:2017lxa}. Supported by substantial multilateral developments in this field, efforts are being poured in to explore different aspects of HEP phenomenology, especially in the context of the Large Hadron Collider (LHC) \cite{Radovic2018,Guest:2018yhq,Bourilkov:2019yoi,Kim:2019wns,Amacker:2020bmn,Baldi:2014pta}. In recent years, deep learning applications have been widely explored to understand hadronic jets' formation and properties, the most common structured object found in any event at LHC, created from QCD fragmentation and hadronization of fundamental quarks and gluons. 
More interestingly, boosted heavy particles like Higgs, top or massive gauge bosons can also produce similar jet objects after the hadronization of their decay products. Prior to the advent of deep-learning approaches, the realization that the internal dynamics of different jet objects are dissimilar received intense scrutiny \cite{Asquith:2018igt,Ellis:2009su,Butterworth:2008iy,Salam:2009jx,Shelton:2013an,Marzani:2019hun}  looking into the underlying structures as probes for new physics \cite{Sirunyan:2016ipo,Das:2017gke,Bhardwaj:2018lma,Bhardwaj:2019mts,Patrick:2016rtw,Kang:2015nga,Bhardwaj:2020llc,Banerjee:2018bio}.
For jet substructure studies, the primary deep-learning approach is to employ calorimeter energy deposits of a jet in  $\eta - \phi$ pixel tower converted into the pictorial description of such `jet-images' \cite{Cogan:2014oua} as input to Convolutional Neural Network (CNN)
\cite{deOliveira:2015xxd,Barnard:2016qma,Komiske:2016rsd}. Very successful n-prong taggers are developed for  Z /W bosons \cite{Baldi:2016fql} and the top tagging \cite{Kasieczka:2017nvn,Macaluso:2018tck,Roy:2019jae} by utilizing this idea, which is further extended to discriminate between quark and gluons \cite{Komiske:2016rsd}. Contrary to jet-images, various other approaches have also been explored for the input space. These include looking for the optimal basis of substructure variables in N-body phase space \cite{Datta:2017rhs}, forming the jet-spectra with two-point correlations  at different angular ranges \cite{Lim:2018toa,Chakraborty:2019imr}, and making an analogy of collider events with natural language thereby using recursive neural networks for feature extraction \cite{Louppe:2017ipp}.  Deep Neural Networks (DNN) have established their importance for classification of signal and background using low/high-level variables \cite{Metodiev:2017vrx,Guo:2018hbv,DeSimone:2018efk,Hajer:2018kqm,Bhattacherjee:2019fpt,Blance:2019ibf,Jung:2019iii,Windischhofer:2019ltt,Ren:2017ymm,Abdughani:2020xfo}. 
Although there are some studies \cite{Diefenbacher:2019ezd,Bhimji:2017qvb,Andrews:2019wng,Lin:2018cin}  of utilizing the inclusive event information at hadron colliders as input for deep-learning neural networks, their full potential are yet to be explored extensively. For the benefit of the readers, many more such exciting approaches in the machine learning framework can be followed in the recent review \cite{Larkoski:2017jix,Carrazza:2017qro,Abdughani:2019wuv} and references therein. 

\begin{figure}[t]
	\centering		
	\includegraphics[scale=0.25]{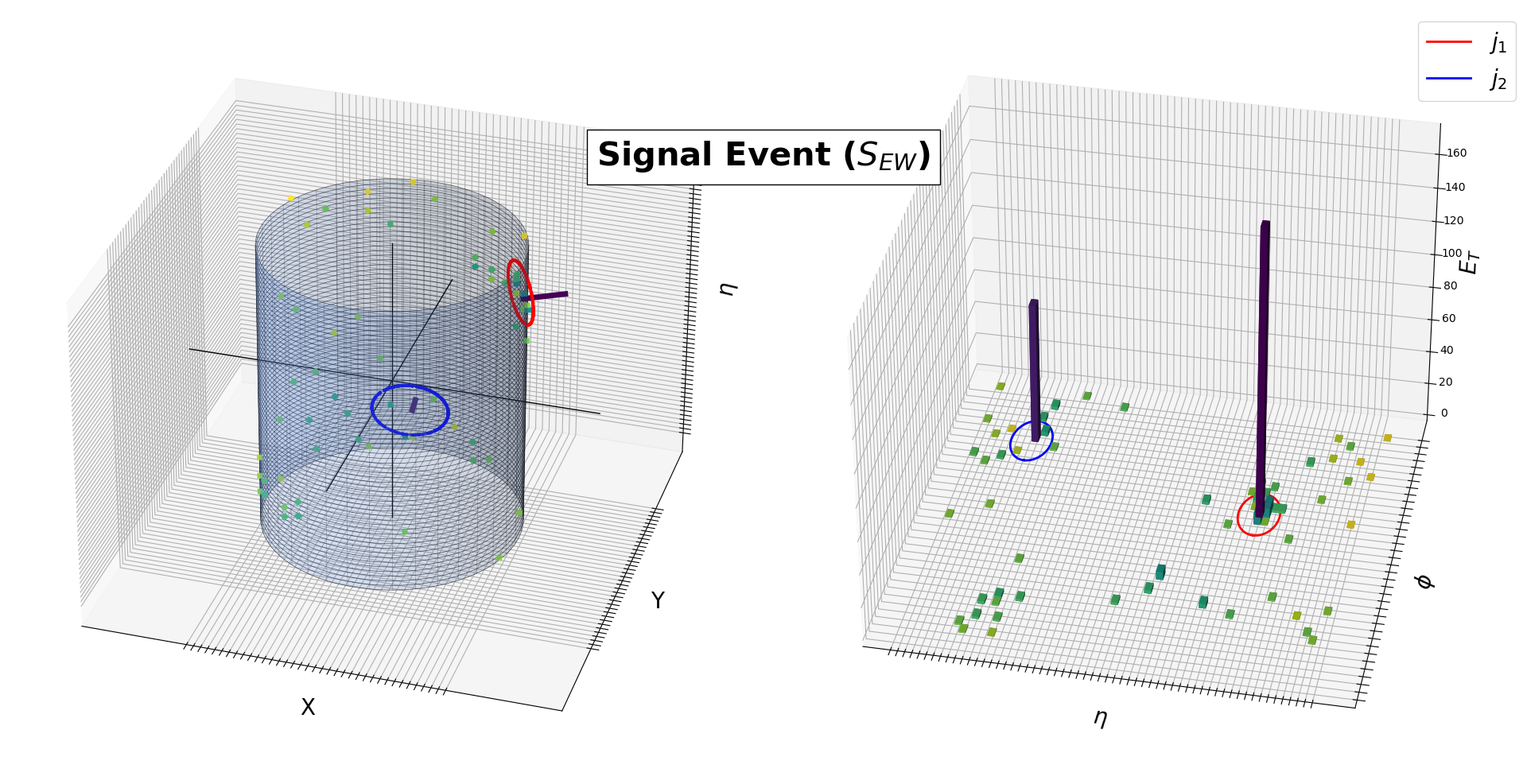}
	\caption{The figure shows a 3D depiction of a prototype signal event originated from an electroweak VBF Higgs production in a naive detector geometry in left plot. The same event is flattened in a convenient $\eta-\phi$ plane in right plot, where the transverse projection of calorimeter energy deposits in different pixels are drawn. Two reconstructed primary jets are shown with color circles, and corresponding transverse energy deposits are visible from height of the bars.
	}
	\label{fig:sig_3D}				
\end{figure}

\begin{figure}[t]
	\centering		
	\includegraphics[scale=0.25]{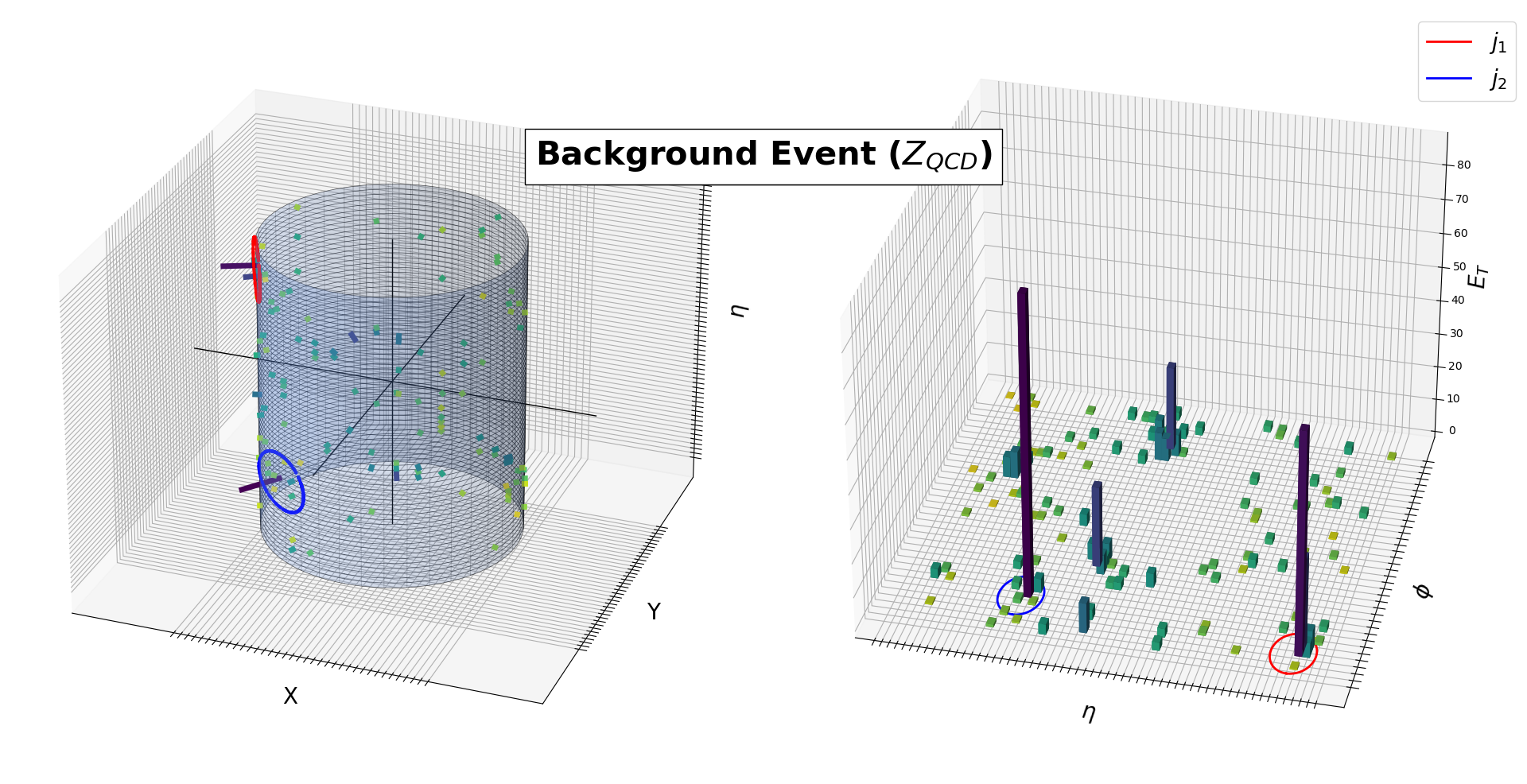}
	\caption{ Same as figure \ref{fig:sig_3D}, but for a prototype background event originated from a  $Z(\nu\bar{\nu})+jets$ production, where the jets originate from QCD vertices.
	}
	\label{fig:bckg_3D}				
\end{figure}

Taking an analogy from jet-image classification, we use the full calorimeter image to study the invisible Higgs production in association with a pair of jets. Vector-boson fusion (VBF) production of color singlet particles provide a unique signature in hadron colliders. First studied in reference  \cite{first_vbf,PhysRevD.47.101,PhysRevD.48.5162}, they are characterized by the presence of two hard jets in the forward regions with a large rapidity gap, and a relative absence of hadronic activity in the central regions, when the singlet particle decays non-hadronically. For illustration, the left panel of figure \ref{fig:sig_3D}  shows an event of a Higgs produced in VBF channel decaying invisibly in a simplistic tower geometry, while the same event is mapped in a flattened ($\eta,\phi)$ plane by rolling out the $\phi$-axis, with the height of the bars corresponding to the magnitude of the transverse projection of calorimeter energy deposits in each pixel. In order to highlight the differences with non-VBF processes, it is instructive to show one such example in figure \ref{fig:bckg_3D}. This is a representative event from $Z(\nu\bar{\nu})+jets$ background, where the jets arise from QCD vertices, which inherently has a much higher hadronic activity in the central regions between the two leading jets. 
Even though the rapidity gap vanishes when the singlet particle decays hadronically, the absence of color connection between the two forward jets and the central region persists and has been used in the experimental analysis \cite{Khachatryan:2015bnx}, in searches of the Higgs boson decaying to bottom quarks. The VBF process was proposed as the most important mechanism for heavy Higgs searches \cite{CAHN1984196} thanks to a much slower fall in cross-section compared to the s-channel mediated process. Usefulness for intermediate to light mass scalar was also subsequently realized  \cite{Rainwater:1997dg} due to its unique signature at the collider. VBF process holds great importance to measure Higgs coupling with gauge bosons and fermions as it allows independent observations of Higgs decay like $h^0\rightarrow WW$ \cite{Rainwater:1999sd}, $h^0 \rightarrow \tau\tau$ \cite{Rainwater:1998kj}. Therefore, it also plays a vital role in determining anomalous coupling to vector boson \cite{Plehn:2001nj,Hankele:2006ma} or the CP properties of the Higgs \cite{Han:2016bvf,Zanji:CPhigss}. Its clean features make it the most sensitive channel for searching invisible decay of the Higgs boson \cite{Eboli:2000ze} and in search for physics beyond the standard model \cite{Datta:2001cy,Choudhury:2003hq,Konar:2006qx}.  
As the Higgs can decay invisibly only through a pair of $Z$ bosons producing neutrinos with minuscule branching ratio in the Standard Model (SM), observation of any significant deviation can provide a strong indication towards a theory beyond the Standard Model (BSM) \cite{Shrock:1982kd}. Hence, this search plays a crucial role to constrain many BSM scenarios, like dark-matter \cite{Arcadi:2019lka,Djouadi:2011aa,Djouadi:2012zc,Han:2016gyy,Baek:2012se}, massive neutrinos \cite{Belotsky:2002ym,Bambhaniya:2014kga}, supersymmetric \cite{B_langer_2001,Datta:2001hv}, and extra-dimensional models \cite{Giudice_2001,Hagiwara:2008iv}.

Although being one of the most promising channels, the production of invisible Higgs is challenging to probe as only a few observables can be constructed over the unique features of VBF. 
Ensuring a color quiet central region by so-called `central jet veto,' and rather specific choices related to the jets,  the separation in pseudorapidity $|\Delta\eta_{jj}|$ and the dijet invariant mass $m_{jj}$ are the significant ones. A central jet veto essentially discards events with additional jets in the region between the two forward tagging jets. Electroweak VBF production of Higgs can satisfy such criteria naturally with excellent efficiency. These same criteria can also ensure the elimination of vast QCD backgrounds up to a large extent, where jets are produced with a massive $W$ or $Z$ boson decaying (semi)invisibly. Finally,  the much weaker electroweak backgrounds coming in the form of VBF production of $W$ or $Z$, become the dominant factor for such study.
However, we must note at the same time that there is a significant drop in signal contribution from other dominant non-VBF Higgs production modes, such as higher-order in $\alpha_s$ correction to gluon fusion initiated processes for Higgs productions \cite{Klamke:2007cu}.  

A natural order of inquiry, therefore, calls for the investigation of whether deep machine learning vision in the form of CNN, together with other neural networks, can have the ability to recognize the characteristics of VBF by learning from the data itself. Networks would map the probability distribution functions to characterize each process by utilizing low-level or high-level variables. Moreover, we would like to understand how useful these learned features are or how they correlate with our traditional characteristics of VBF. Finally, there can be enough scope to engage this very sophisticated tool to get some hybrid output in terms of maximizing the efficiency in selecting signal events, rather than classifying in separate clusters as VBF and non-VBF types.

While the present study can easily be extended for other decay modes of Higgs, we choose the invisible channel for our study to showcase the importance of deep learning quantitatively using different neural network architectures. We propose to study the full event topology of VBF by examining the calorimeter tower-image using CNN, which utilizes the low-level variables. We also consider the performance of event classification using dense Artificial Neural Networks(ANN), which employ high-level variables. In total, we investigate seven different neural network architectures and provide a  comparative study of the performance of networks. The performance of networks is quantified in terms of expected constraints on the invisible branching ratio (BR ($h^0\to$ inv)) of the Higgs boson. 

The latest report from ATLAS collaboration \cite{ATLAS-CONF-2020-008} puts an upper limit on BR($h^0\to$inv) at 95\% confidence level (CL) to 0.13, from an integrated luminosity of 139 fb$^{-1}$ at the LHC. The CMS analysis also puts an upper limit at 95\% CL to 0.19 for combined data set of 7, 8 and 13 TeV for 4.9 fb$^{-1}$, 19.7 fb$^{-1}$ and 38.2 fb$^{-1}$ integrated luminosities respectively \cite{Sirunyan:2018owy}. These bounds still allow the significant presence of BSM physics. Our principal aim, therefore, is to study the viability of CNNs to improve these results using low-level variables in the form of the entire calorimeter image, as well as to compare its performance to DNN/ANN architecture with high-level variables as input. We find that the bounds on the BR($h^0\to$inv) can indeed be significantly improved using these networks.

The rest of this paper is organized as follows. In section \ref{sec:vbf}, we discuss the Higgs production mechanism via the VBF channel and different SM backgrounds contributing to this process. We also discuss the generation of simulated data consistent with the VBF search strategy. In section \ref{sec:data_rep}, we describe the details of the data representation used in the present study. Here, different classes of high-level variables are also defined.  Preprocessing methods of feature spaces are addressed in section \ref{sec:preprocess}. We discuss the seven different neural network architecture and its performance in section  \ref{sec:net_run}. The results, interpreted in terms of expected bounds on the invisible branching ratio, for all the architectures are presented in section \ref{sec:results}. There, we also discuss the impact of pileup on the result of our analysis. 
Finally, we close our discussion with the summary and conclusion in the last section.

\section{Vector Boson Fusion production of Higgs and analysis set-up}
\label{sec:vbf}

VBF production of the SM Higgs has the second-highest production cross-section after gluon-fusion at the LHC. Loop induced Higgs production and decay depend on the presence of contributing particles and different modifiers in fermions and gauge boson coupling with the scalar. Hence, both production cross-section and decay branching ratios are modified in the presence of new physics. 
In this present work, we consider the production of SM like Higgs boson and constrain its invisible decay width. Such constraint is essential in many new physics scenarios, such as Higgs portal dark matter  \cite{Arcadi:2019lka,Djouadi:2011aa,Djouadi:2012zc,Han:2016gyy,Baek:2012se}, where new particles do not modify their couplings with SM particles.

\begin{figure} [t]
	\begin{center} 
		\begin{tikzpicture}[scale=0.8, transform shape]
		\begin{feynman}
		\vertex (i1) {$q$};
		\vertex [ right= of i1] (a);
		\vertex[below = of a] (b);
		\vertex [ right= 1.5 cm  of a] (f1) {$q$};
		\vertex[below right= 1.0 cm of a] (c);
		\vertex[right=of c] (f) {$h^0$};
		\vertex [left = of b](i2) {$q'$};
		\vertex[right of=b](f2){$\bar{q}$};
		
		\diagram*{(i1)--[fermion](a)--[fermion](f1),
			(a)--[photon](c)--[scalar](f),
			(c)--[photon](b),(i2)--[fermion](b)--[fermion](f2),};
		\end{feynman}
		\end{tikzpicture}
		\hspace{0.3cm}
		\begin{tikzpicture}[scale=0.8, transform shape]
		\begin{feynman}
		\vertex (i1) {g};
		\vertex [right= of i1] (a);
		\vertex [right=0.8 cm of a](x);
		\vertex[ right= of x](y);
		\vertex[below = 1.5 cm of a] (b);
		\vertex [ below of=i1] (i2) {g};
		\vertex[right = 0.8 cm of b] (d);
		\vertex[above right= 1 cm of d] (c);
		\vertex[ right = of d](e);
		\vertex[right=of c] (f) {$h^0$};
		\diagram*{(i1)--[gluon](a)--[fermion](x)--[fermion](c),(i2)--[gluon](b)--[fermion](a),
			(c)--[fermion](d)--[fermion](b),(d)--[gluon](e),(c)--[scalar](f),(x)--[gluon](y)};
		\end{feynman}
		\end{tikzpicture}
	\end{center} 
	\caption{Representative diagrams for production of Higgs signal through (left) electroweak VBF channel and (right) a higher-order QCD process in gluon fusion where two QCD jets can be detected along with a sizable missing transverse-energy from invisible Higgs decay.}
	\label{fig:sg_diag}
\end{figure}
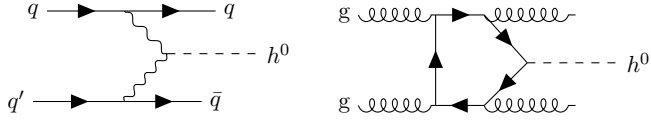

The electroweak production of Higgs is dominated by the fusion of two massive vector bosons, which are radiated off two initial (anti-)quarks, as represented in figure \ref{fig:sg_diag} (left plot). This exchange of color singlet state between two quarks ensures no color connection between two final jets, typically produced in a forward (backward) region of the opposite hemisphere. The central region - between these two jets remain color quiet, lacking any jet activity even after radiation and fragmentation of the two scattered quarks while looking at the hadronic final states. 
As we have already discussed, an agnostic viewpoint requires a serious re-examination after the inclusion of all other processes, such as non-VBF Higgs signal from gluon fusion. One such sample diagram is shown in figure \ref{fig:sg_diag} (right plot). Additional radiation from gluons can provide a typical VBF type signal, once again, in the absence of the key attributes like color-quiet central region, etc.

Another interesting feature of VBF Higgs production is that the corresponding cross-section has very modest correction under higher-order QCD, which has been known for a long time \cite{Han:1992hr,Figy:2003nv}. Integrated and differential cross sections for VBF Higgs production have now been calculated up to very high levels of accuracy. QCD corrections are known up to N3LO \cite{Dreyer:2016oyx}, reducing the scale-uncertainty up to $2\%$, while Electroweak corrections are known up to NLO \cite{Ciccolini:2007jr}. Moreover, non-factorizable contributions have also been calculated for the first time \cite{Liu:2019tuy}, and show up to percent level corrections compared to the leading order (LO) distributions.

At hadron colliders, traditional searches  \cite{Datta:2000ja,Konar:2003pn,Delannoy:2013ata,Berlin:2015aba} of non-hadronically decaying color-singlet particles in the VBF production channel focus on rejecting the large QCD backgrounds from $Z+jets$, and $W+jets$ background via a \emph{central jet-veto},  after a hard cut on the separation of the two forward  jets in pseudorapidity $|\Delta \eta_{jj}|$, and the dijet invariant mass $m_{jj}$.  This opens up the possibility of using inclusive event-shape variables like N-jettiness \cite{Stewart:2010tn}, to improve the selection efficiency \cite{Braren:2015epq}. In this study, we explore the feasibility of using deep-learning techniques instead of event-shape variables.  We study the invisible decay of the Higgs boson as a prototype channel for gauging the power of deep-learning methods in VBF since there is no contamination on the radiation patterns between the two forward jets from the decay products. 
We closely follow the shape-based analysis performed by the CMS experiment at CERN-LHC  \cite{Sirunyan:2018owy}\footnote{For ATLAS analysis with similar data, see reference~\cite{Aaboud:2018sfi}.}.
As already commented, the central jet veto played a critical role in the usual searches of VBF to control the vast QCD background. The role of additional information from QCD radiation between the tagging jets and within the jet itself was explored in references \cite{Bernaciak:2014pna,Biekotter:2017gyu}. It was found that relaxation of the minimum $p_T$ requirement of the central jet improved the sensitivity, and the inclusion of subjet level information resulted in further suppression of backgrounds. However, the present analysis does not rely on a central jet veto, as the main aim is to study the VBF topology with the low-level data, made possible with modern deep-learning algorithms.  Therefore, with the relaxed selection requirements on $|\Delta\eta_{jj}|$ and $m_{jj}$,
the selected signal gets a significant contribution from the gluon-fusion production of Higgs on top of VBF processes. Due to the relaxed selection criteria, we also get a substantial contribution from QCD backgrounds.

\subsection{Signal topology}
The present study relies on all dominant contributions to Higgs coming both from electroweak VBF processes and also higher-order in QCD gluon fusion processes. Here at least two jets should be reconstructed along with sizable missing transverse-energy from invisible decay of Higgs. Hence, we classify the full signal contribution in two channels:
\begin{itemize}
	\item $S_{QCD}$: Gluon-fusion production of Higgs with two hard jets, where the Higgs decays invisibly.
	\item $S_{EW}$: Vector-Boson fusion production of Higgs decaying invisibly.
\end{itemize}
The subscript $EW (QCD)$ denotes the absence (presence) of strong coupling $\alpha_S$, at leading order(LO) for the interested topology. This also segregates the channels with absence or presence of color exchange between the two incoming partons at LO. Figure \ref{fig:sg_diag} shows a representative Feynman diagram of the signal channels in each class.

\begin{figure}[t]
	\begin{center} 
		\begin{tikzpicture}[scale=0.87, transform shape]
		\begin{feynman}
		\vertex (i1) {$q$};
		\vertex [right= of i1] (a);
		\vertex[below = of a] (b);
		\vertex [ right of=a] (f1) {$q$};
		\vertex[below right= 1.0 cm of a] (c);
		\vertex[right=1.2 cm of c] (f) {$V$};
		\vertex [left of =b](i2) {$\bar{q}$};
		\vertex[right of=b](f2){$\bar{q}$};
		
		\diagram*{(i1)--[fermion](a)--[fermion](f1),
			(a)--[photon](c)--[photon](f),
			(c)--[photon](b),(i2)--[anti fermion](b)--[anti fermion](f2),};
		\end{feynman}
		\end{tikzpicture}
		\hspace{1.0cm}
		\begin{tikzpicture}[scale=0.87, transform shape]
		\begin{feynman}
		\vertex (i1) {$g$};
		\vertex [ right= of i1] (a);
		\vertex[below = of a] (b);
		\vertex [ right=  of a] (f1) {$q$};
		\vertex[below right= 1.0 cm of a] (c);
		\vertex[right= 1.2 cm of c] (f) {$V$};
		\vertex [left = of b](i2) {$g$};
		\vertex[right of=b](f2){$\bar{q}$};
		\diagram*{(i1)--[gluon](a)--[fermion](f1),
			(a)--[fermion](c)--[photon](f),
			(c)--[fermion](b),(i2)--[gluon](b)--[fermion](f2),};
		\end{feynman}
		\end{tikzpicture}
	\end{center} 
	\caption{Representative diagrams for dominant background processes through (left) VBF type weak production and (right) QCD production of massive vector-bosons $V$, such as $W$ or $Z$ which decay invisibly by producing undetected lepton or neutrinos.}
	\label{fig:bg_diag}
\end{figure}
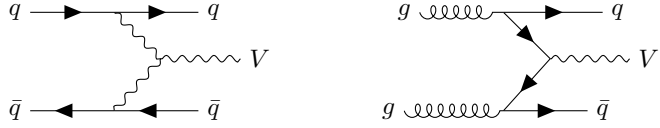  

\subsection{Backgrounds}
The major backgrounds contributing to the invisible Higgs VBF signature can come from the different standard model processes. Among them, VBF type electroweak, and QCD production of massive vector-bosons ($W$ or $Z$) contribute copiously. All these processes ensure a pair of reconstructed jets along with considerable missing transverse energy from invisible decay of these gauge bosons. A substantial fraction of $W$ and $Z$ can produce neutrinos or a lepton which remain undetected at the detector. We consider the following backgrounds in all our analyses: 

\begin{itemize}		
	\item $Z_{QCD}$: $Z(\nu\bar{\nu})+jets$ process contributes as the major SM background due to high cross section. 
	\item $W_{QCD}$: $W^\pm(l^\pm\nu)+jets$ process also contribute to the SM background when the lepton is not identified.
	\item $Z_{EW}$ : Electroweak production of Z decaying invisibly along with two hard jets is topologically identical with the electroweak signal and contributes significantly to the background.
	\item $W_{EW}$ : Electroweak production of $W^\pm$ with two hard jets can also produce an identical signal when the lepton does not satisfy the identification criteria.
	
\end{itemize} 
Similar to the signal processes, the subscript $EW(QCD)$ denotes the absence (presence) of strong coupling $\alpha_S$, at LO for the interested topology having at least two reconstructed jets in the final state.
Figure \ref{fig:bg_diag} shows representative Feynman diagrams of the background channels divided into four different classes. 

There are also other background processes like top-quark production, diboson processes, and QCD multijet backgrounds whose contribution would be highly suppressed compared to these four backgrounds. The top and diboson backgrounds would contribute to leptonic decay channels where charged leptons, if present, are not identified, while the QCD multijet background would contribute when there is severe mismeasurement of the jet energies.

\subsection{Simulation details}
We used MadGraph5\_aMC@NLO(v2.6.5)  \cite{Alwall:2014hca} to generate parton-level events for all processes at 13 TeV LHC. These events are then showered and hadronized with Pythia(v8.243)  \cite{Sjostrand:2014zea}. Delphes (v3.4.1)  \cite{deFavereau:2013fsa} is used for fast-detector simulation of the CMS working conditions. Jets are clustered using the FastJet(v3.2.1) \cite{Cacciari:2011ma} package. The signal processes are generated using a modified version of the Higgs Effective Field Theory (HEFT) model \cite{Alloul:2013bka,PhysRevD.22.178,PhysRevD.59.057504}, where the Higgs boson  can decay to a pair of scalar dark matter particle at tree level.  We are interested in probing high transverse momentum of Higgs, where the finite mass of top quark in gluon fusion becomes essential. Hence, we have taken into account such effect by reweighting the missing transverse energy (\textsc{met}) distribution of the events with recommendations from reference \cite{deFlorian:2016spz}. The parton level cross-sections of $Z_{QCD}$ and $W_{QCD}$ were also matched up to four and two jets, respectively, via the MLM procedure  \cite{Hoche:2006ph}. Since the $W^\pm$ backgrounds contribute when the leptons are missed within the range of tracker or when they are not reconstructed at the detector, the parton level cuts on the generated leptons are removed to cover the whole range in pseudorapidity ($\eta$).

For a consistent comparison with current experimental results, we repeat the shape-analysis of reference \cite{Sirunyan:2018owy} with our simulated dataset. The \textsc{met} cut for the deep-learning study is relaxed from 250 GeV to 200 GeV.\\

\begin{figure*}[t]
	\centering		
	\includegraphics[height=0.31\textwidth, width=0.49\textwidth]{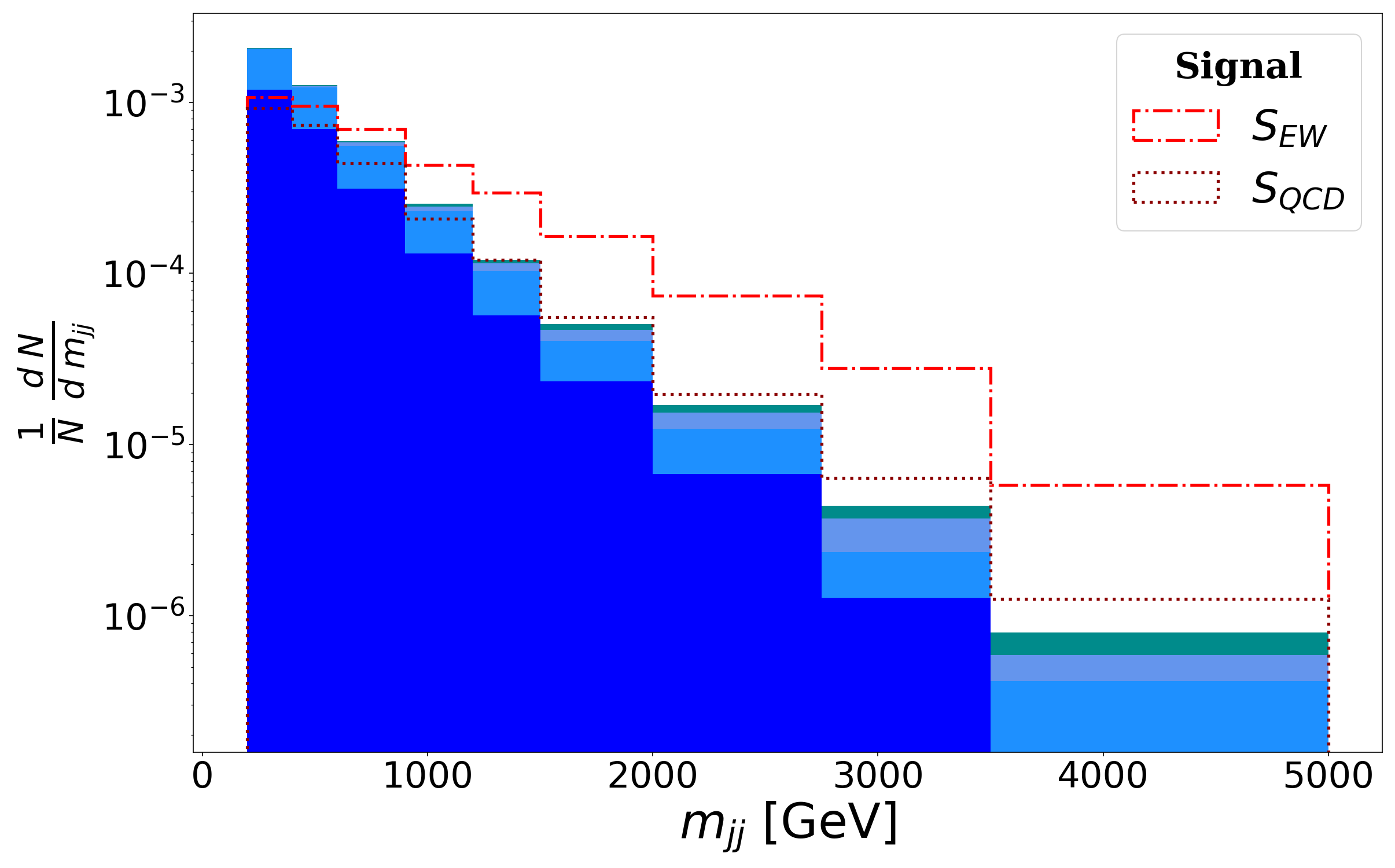}~
	\includegraphics[height=0.31\textwidth, width=0.49\textwidth]{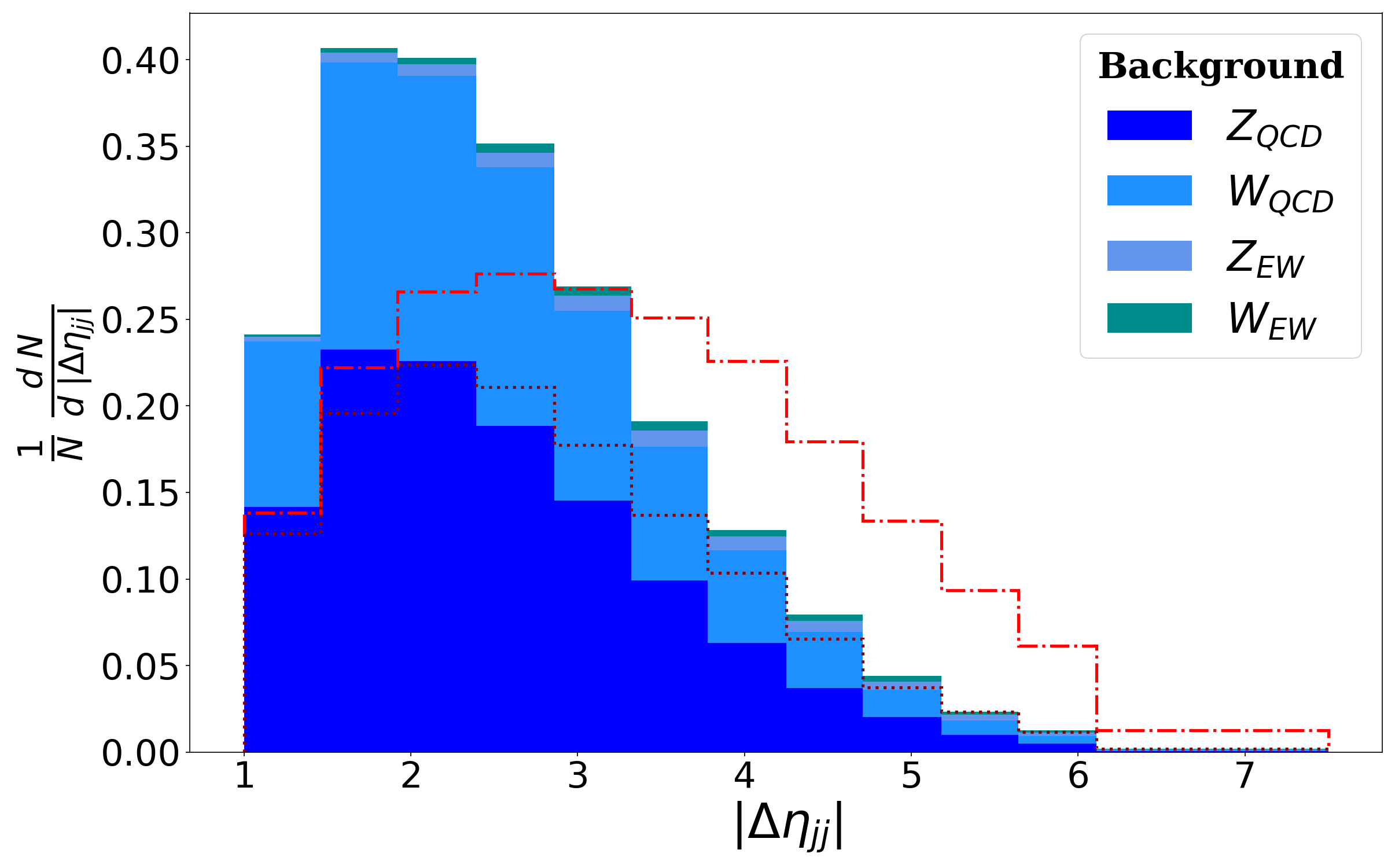}
	\caption{Distribution of (left) $m_{jj}$  and  (right) $\Delta \eta_{jj}$ of events passed after the passing the tighter selection requirement ($\textsc{met}>250 $ GeV). The contribution of each channel to its parent class has been weighted by their cross-sections and the baseline efficiency at 13 TeV. The signal and backgrounds are then individually normalized, and the lines/color show the contribution of each channel to its parent class.}
	\label{fig:cms_shape}				
\end{figure*}

\noindent
{\bf Baseline selection criteria: } We apply the following pre-selections: 
\begin{itemize} 
	\item \textbf{Jet $p_T$}: At least two jets with leading (sub-leading) jet having minimum transverse momentum $p_T>80 \;(40)$ GeV.
	\item \textbf{Lepton-veto:} We veto events with the reconstructed electron (muon) with minimum transverse momentum $p_T > 10$ GeV in the central region, {\sl i.e.} $|\eta|<2.5\,(2.4)$. This rejects leptonic decay of single $W^\pm$, and semi-leptonic $t\bar{t}$ backgrounds.
	\item \textbf{Photon-veto:} Events having any photon with $p_T > 15$ GeV in the central region, $|\eta|< 2.5$ are discarded.	
	\item \textbf{$\tau$ and b-veto}: No tau-tagged jets in $|\eta|<2.3$ with $p_T> 18$ GeV, and no b-tagged jets in $|\eta|<2.5$ with $p_T>20$ GeV are allowed. This rejects leptonic decay of single $W^\pm$, semi-leptonic $t\bar{t}$ and single top backgrounds.
	\item \textbf{MET}: Total missing transverse momentum for the event must satisfy $\textsc{met} >200 $ GeV for all our deep-learning study, whereas we compared CMS shape-analysis consistent with $\textsc{met} >250 $ GEV.
	\item \textbf{Alignment of MET with respect to jet directions}: Azimuthal angle separation between the reconstructed jet with the missing transverse momentum to satisfy $\min(\Delta \phi(\vec{p}_T^{\textsc{met}},\vec{p}_T^{j})) > 0.5$ for up to four leading jets with $p_T > 30$ GeV and $|\eta| < 4.7 $. QCD multi-jet background that arises due to severe mismeasurement is reduced significantly via this requirement.
	\item \textbf{Jet rapidity}: We require both jets to have produced with $|\eta_j| < 4.7$, and at least one of the jets to have $|\eta_{j_i}| < 3$, since the L1 triggers at CMS do not use the information from the forward regions.
	\item \textbf{Jets in opposite hemisphere}: Those events which have the two leading jets reside in the opposite hemisphere in $\eta$ are selected. This is done by imposing the condition $\eta_{j_1}\, \times \;\eta_{j_2}<0$. 
	\item \textbf{Azimuthal angle separation between jets}: Events with $|\Delta \phi_{jj}|<1.5$ are selected. This helps in reducing all non-VBF backgrounds.
	\item \textbf{Jet rapidity gap}: Events having minimum rapidity gap between two leading jets $|\Delta\eta_{jj}|>1$ are selected. 
	\item \textbf{Di-jet invariant mass}: We required a minimum invariant mass of two leading jets, $m_{jj}>200$ GeV. Note that, this along with the previous selection requirements are relatively loose compared to traditional selection criteria of VBF topologies, which result in significant enhancement of the signal from $S_{QCD}$, although at the cost of increased QCD backgrounds ($Z_{QCD}$ and $W_{QCD}$).
\end{itemize}
Interestingly, one can notice that a relaxed selection requirement may give rise to additional contamination from Higgs-strahlung type topologies to the $S_{EW}$ channel, which is included in our EW generation of events. However, these events are not expected to survive a selection of di-jet invariant masses of more than 200 GeV. 
After extracting the events passing the above selection requirements and the respective selection efficiency (calculated from the weights) for $S_{QCD}$, the pre-selected events are unweighted again so that we get equal weights for individual events.\footnote{See \ref{app:gf_plots}, for distribution of the important kinematic-variables and details of the re-weighting and unweighting of events.} The background and signal classes are formed by mixing the channels with the expected proportions using appropriate k-factors, cross-sections, and the baseline selection efficiencies. We use cross-sections quoted in reference \cite{deFlorian:2016spz} for both signal processes. For instance, the $S_{QCD}$ is calculated up to NNLL +NNLO accuracy \cite{Stewart:2013faa}, while for $S_{EW}$ it is calculated up to NNLO \cite{Cacciari:2015jma} in QCD and NLO in electroweak. We use the LO distributions with their overall normalizations increased to accommodate the total cross-section at higher perturbative accuracies without accounting for the possible change in shape. Similarly, all background cross sections are calculated by scaling the LO result with global NLO k-factors \cite{Lindert:2017olm,Oleari:2003tc}. We generated 200,000 training and 50,000 validation balanced dataset of events for the deep-learning classifier.  The signal class consists of 44.8\%  $S_{EW}$ and the 55.2\% $S_{QCD}$ channels; while the background class consists of 51.221\% $Z_{QCD}$, 44.896\% $W_{QCD}$, 2.295\% $Z_{EW}$ and 1.587\% $W_{EW}$ channels.

We also extract event sample for all channels with the harder selection requirement on missing transverse momentum ($\textsc{met}>250 $ GeV), the value used in reference  \cite{Sirunyan:2018owy}, from the same set of generated events used for the deep-learning analysis. The extracted dataset contains:  39\%  $S_{EW}$ and the 61\% $S_{QCD}$ channels for the signal class; and  54.43\% $Z_{QCD}$, 40.92\% $W_{QCD}$, 3.05\% $Z_{EW}$ and 1.58\% $W_{EW}$ channels for the background class. The bin-wise stacked histogram of all channels for $m_{jj}$ and $|\Delta\eta_{jj}|$ are shown in figure \ref{fig:cms_shape}. The properties of the $EW$ and the $QCD$ subsets are evident from these distributions: $EW$ contribute more at higher $m_{jj}$ and $|\Delta \eta_{jj}|$, while the opposite is true for $QCD$.

\section{Data Representation for the Network}
\label{sec:data_rep}

\begin{figure*}[t]
	\centering		
	\includegraphics[height=0.31\textwidth, 
	width=0.49\textwidth]{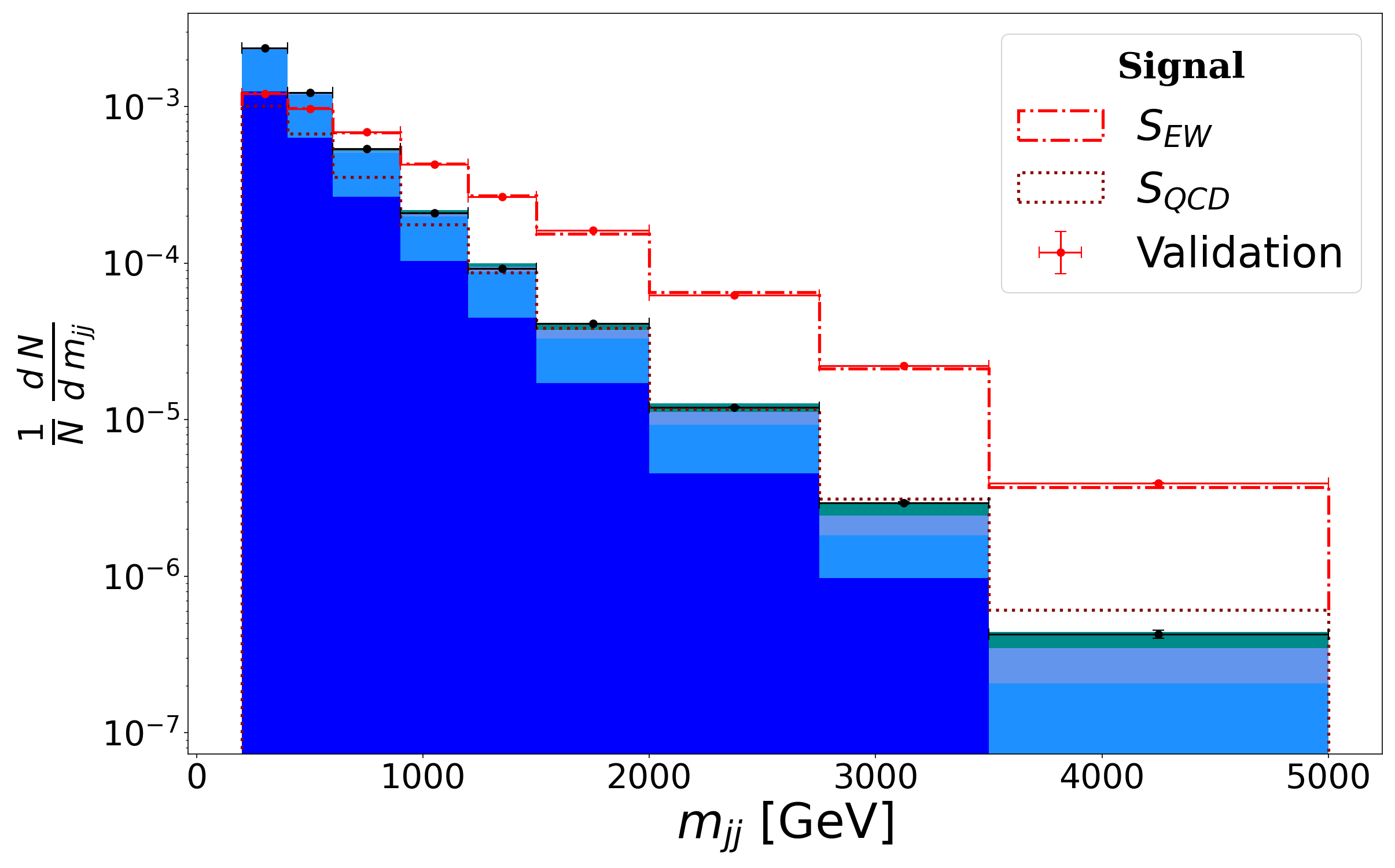}
	\includegraphics[height=0.31\textwidth, width=0.49\textwidth]{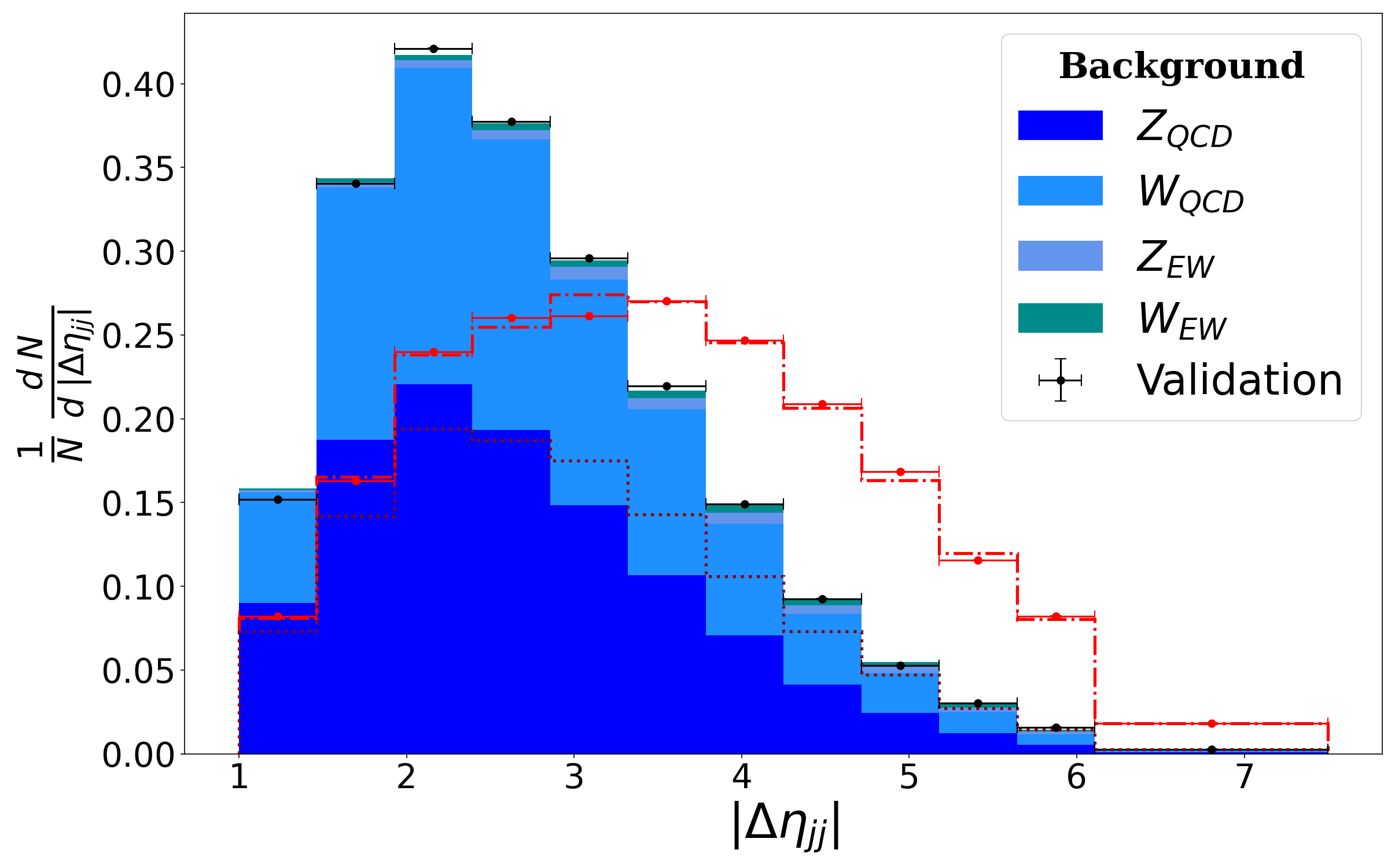}\\  \vspace{0.1 cm}
	\includegraphics[height=0.31\textwidth, width=0.49\textwidth]{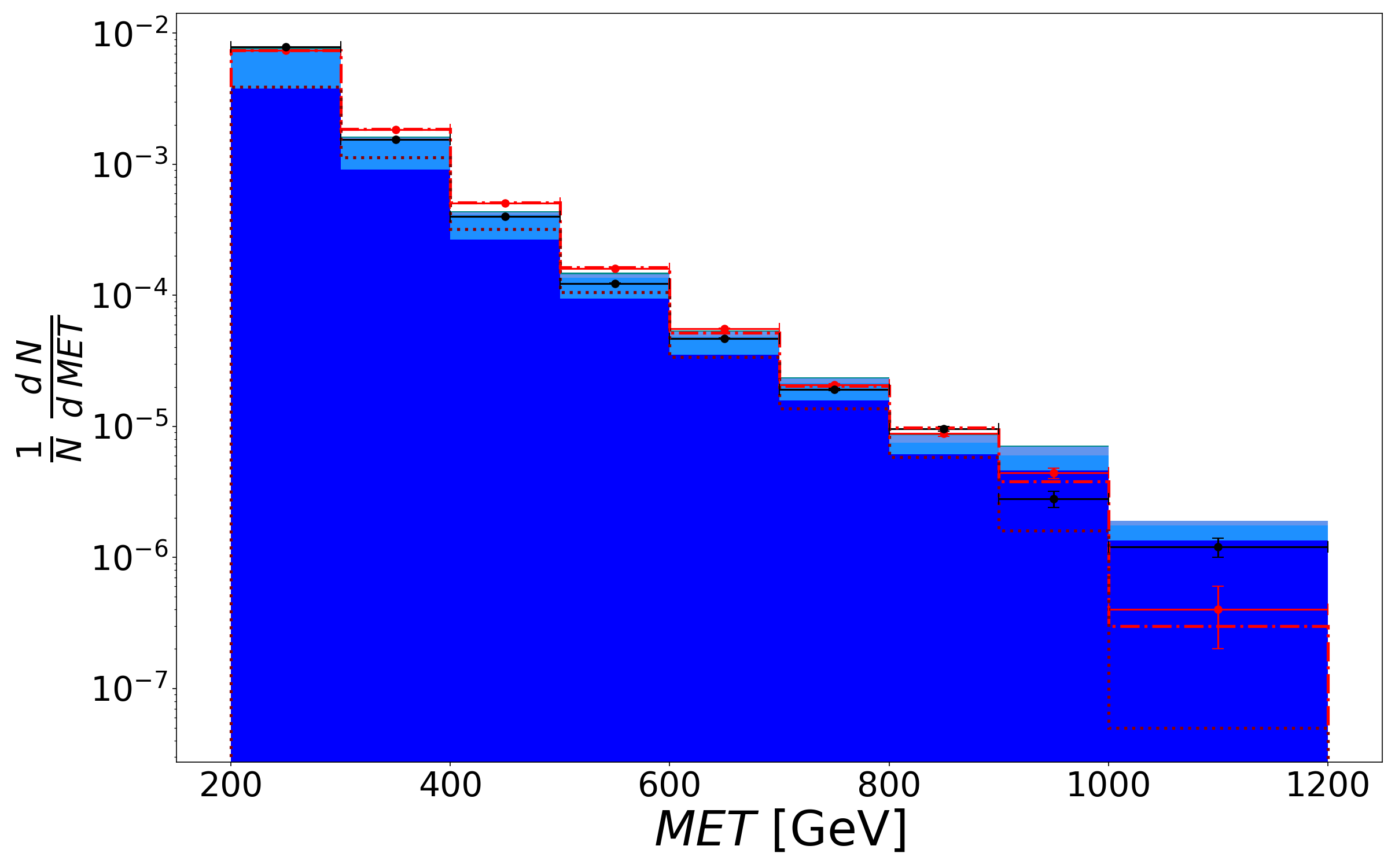}
	\includegraphics[height=0.31\textwidth,width=0.49\textwidth]{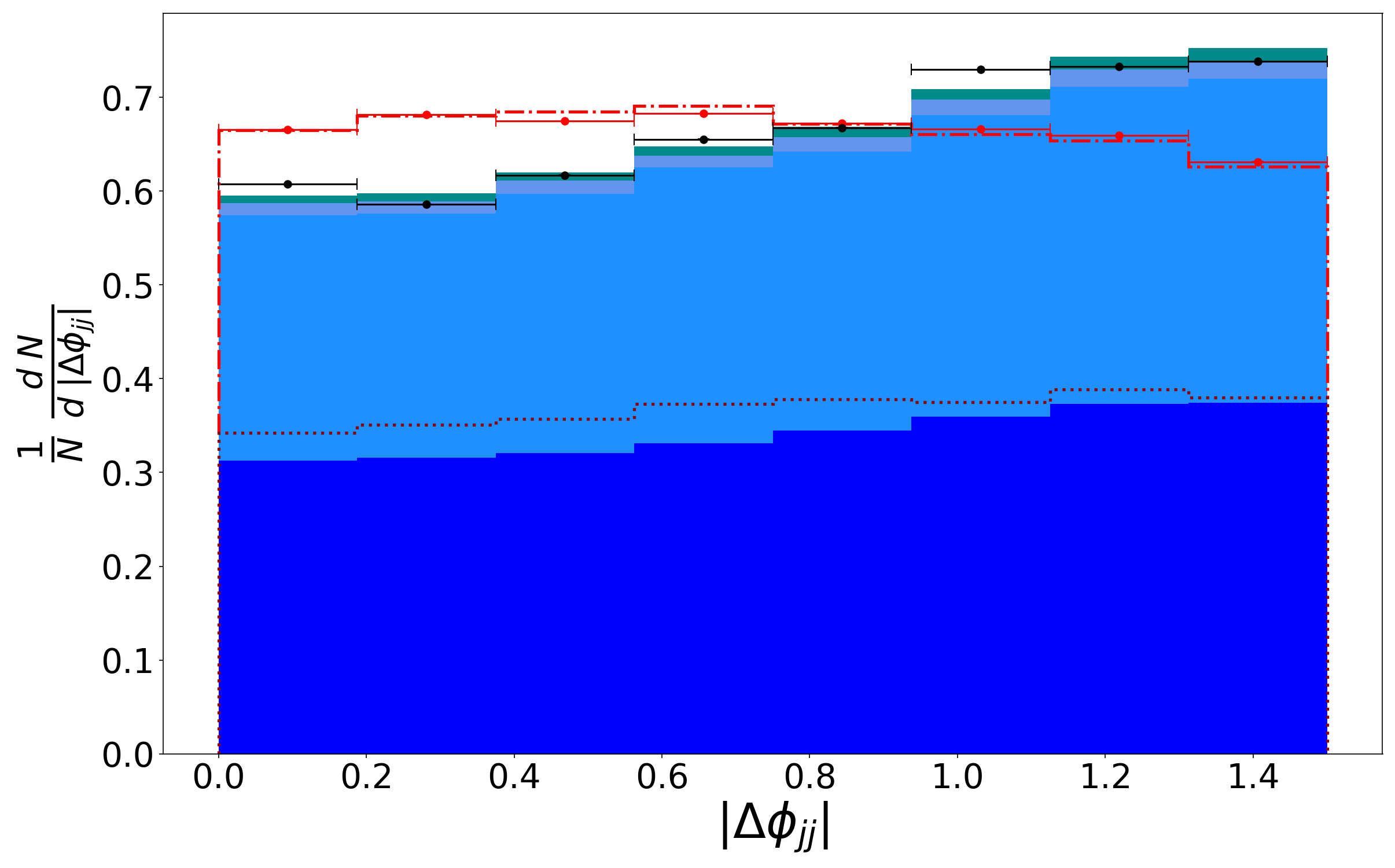}
	\caption{Similar to figure \ref{fig:cms_shape}, some of the basic input high-level kinematic variables used for our analysis ($\textsc{met}>200 $ GeV) are shown for signal and background.}
	\label{fig:kinematic}				
\end{figure*}  

Neural network architectures for deep-learning are mostly designed with two blocks. The first stage generally consists of locally-connected layers (with or without weight sharing) with some particular domain level specifications which extract the features. The second stage consists typically of densely connected layers, whose function is to find a direction in the learned feature-space, which optimally satisfies the particular target of the network locally by learning its projections in different representations at each subsequent layer. For instance, in classification problems, it finds the decision boundary between different classes. At the same time, in an unsupervised clustering, it compresses the feature-space so that the modes become localized in a smaller volume.
A synergy between the representation of data and the network architecture is a must for efficient feature extraction. This is evident from the fact that convolutional neural networks perform best with data structures that have an underlying Euclidean structure, while recurrent networks work best with sequential data structures. In the context of classifying boosted heavy particles like $W$, Higgs, top quark or  heavy scalars  decaying to large-radius jets from QCD background, a lot of efforts \cite{Cogan:2014oua,Baldi:2016fql,Almeida:2015jua,Kasieczka:2017nvn,Komiske:2016rsd,deOliveira:2015xxd} went into representing the data like an image in the $(\eta,\phi)$ plane to use convolutional layers for feature extraction, while some others \cite{Erdmann:2018shi,Butter:2017cot}, use physics-motivated architectures. Convolutional architectures work in these cases because the differences between the signal jet and the background (QCD) follows a Euclidean structure.\footnote{Most high-level variables designed from QCD knowledge are functions of $\Delta R=\sqrt{\Delta \eta^2+\Delta\phi^2}$.} The Minkowski structure of space-time prohibits a direct use of convolutional architectures. Although geometric approaches \cite{geom_cnn_rev} exist to counter the non-Euclidean nature, the number of dimensions makes it computationally expensive. Graph neural networks \cite{Ren:2019xhp,Abdughani:2018wrw,Mullin:2019mmh,Shlomi:2020gdn} provide a possible workaround which is computationally less intensive, for feature learning in non-Euclidean domains.

In the present work, we want to study the difference in radiation patterns between the two forward jets for signal and background events; hence, we primarily choose a convolutional architecture for automatic feature extraction. Therefore, the low-level feature space we prefer is the \emph{tower-image}, in the  $(\eta,\phi)$-plane, with the transverse energy $E_T$, as the pixel values. One can take into account the different resolutions in the central and forward regions of calorimeter towers in LHC detectors. For simplicity, and also to demonstrate the resolution dependence, we construct two images - a high-resolution image with bin size $0.08\times0.08$, and a low-resolution  image with bin size $0.17\times0.17$, in the full range of the tower, $[-5,5]$ for $\eta$ and while $[-\pi,\pi)$ for $\phi$. Convolutional neural-networks, in general, look at global differences, and increasing the resolution does not play as important a role. We examine CNNs in these two different resolutions to inspect this for our particular case. 
The procedure of forming a tower-image does not naturally take the periodicity of the $\phi$ axis into account. In order to let the network know this inherently, we expand the image obtained after binning, in the $\phi$ axis such that the connectivity between the two edges is not broken. This is done by taking a predetermined number of $\phi$-rows from each edge of the original image and forming a new image where these rows are padded \cite{Diefenbacher:2019ezd,Lin:2018cin} in their corresponding opposite sides, thereby mimicking the periodicity. This is similar to cutting the cylinder at two different points in $\phi$ for each edge, such that there is an overlapping region in the final image.  Taking the jet radius $R=0.5$, which have a regular geometry since they are clustered with anti-$k_t$ algorithm  \cite{Cacciari:2008gp}, we choose the number of rows to be 4 (8) for the low (high)-resolution images, with one bin as a buffer. This gives a low-resolution (LR) image of $59\times 45$ and a high-resolution (HR) one of $125\times 95$. 

\begin{figure}[t]
	\centering		
	\includegraphics[width=0.5\textwidth]{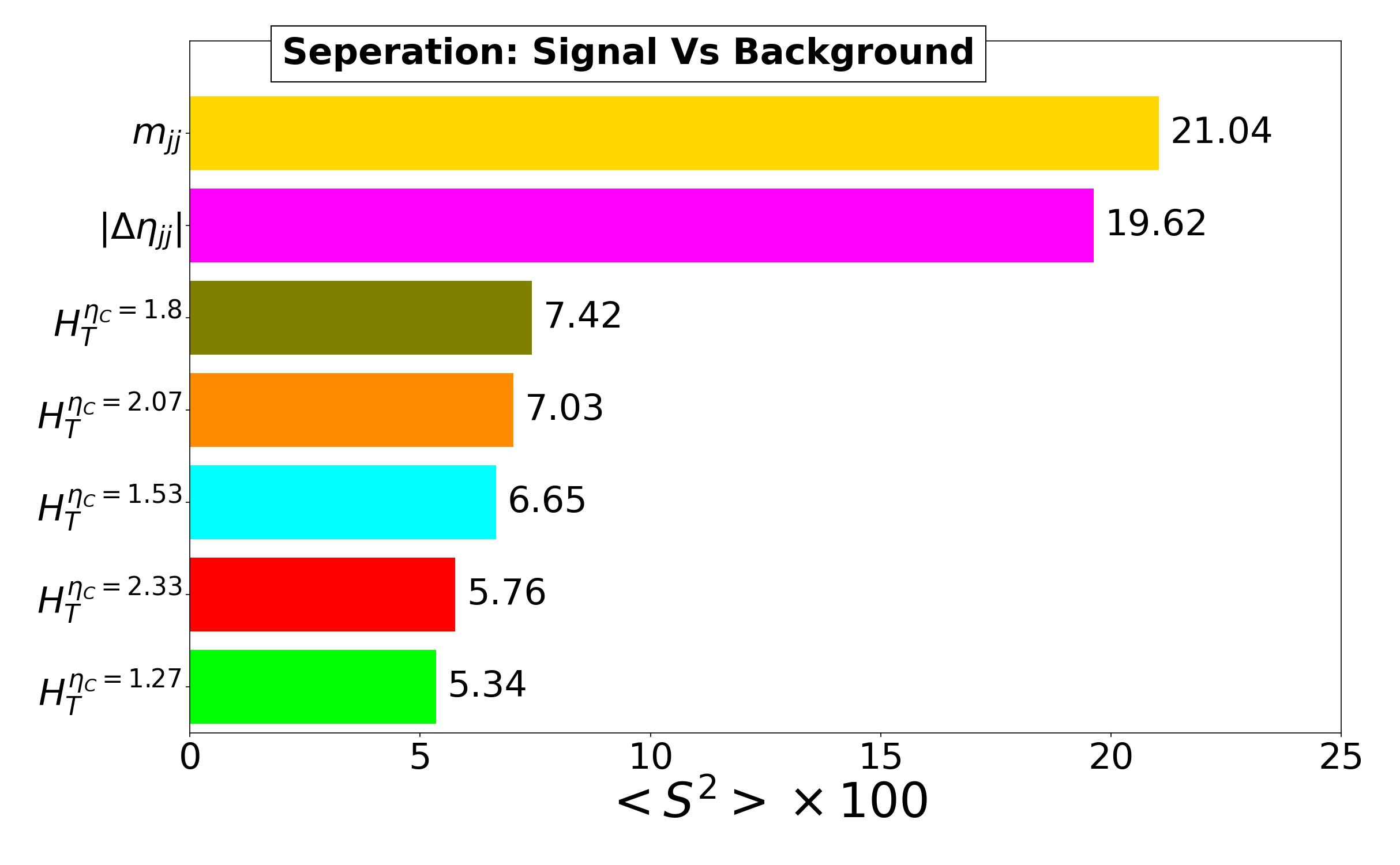}
	\caption{The separation of the 7 highest performing variables (given in percentage).}
	\label{fig:var_imp}				
\end{figure} 
A significant difference between low-level and high-level feature spaces is that the modes of the data in low-level representations are not distinct. Although this is marginally enhanced by preprocessing, high-level features derived from the said low-level features have distinctly localized modes in their distribution. An exemplary ability of deep-learning algorithms is to by-pass this step and learn their own representations which perform better than the high-level variables developed by domain-specific methods. To analyze the relative performance of physics-motivated variables derived from the calorimeter deposits, we consider two classes of high-level variables. The first one consists of the following kinematic variables: 

\begin{eqnarray}
\label{eq:high_level} \nonumber
\mathcal{K}\equiv(\;|\Delta\eta_{jj}|,\; |\Delta\phi_{jj}|\;,\;m_{jj}\;,\;\textsc{met}\;,\;\phi_{\textsc{met}}\;, \;\Delta \phi^{j_1}_{\textsc{met}}\;,\quad  \;\Delta\phi^{j_2}_{\textsc{met}}\;,\;\Delta\phi^{j_1+j_2}_{\textsc{met}}\;) \quad 
\end{eqnarray}
$\phi_{\textsc{met}}$ is the azimuthal direction of $\textsc{met}$ in the lab-frame. $\Delta \phi^{j_1}_{\textsc{met}}$, $\Delta \phi^{j_2}_{\textsc{met}}$ and $\Delta \phi^{j_1+j_2}_{\textsc{met}}$ are the azimuthal separation of  $\textsc{met}$ with the direction of the leading, sub-leading and the vector sum of these two jets, respectively.	Clearly, these do not contain any information about the radiation pattern between the tagging jets. The second class of variables: the sum of $E_T$ of the tower constituents in the interval $[-\eta_C,\eta_C]$, incorporates this information:
\begin{equation} 
\label{eq:ht}	
\mathcal{R}\equiv(H^{\eta_C}_T | \eta_C \in \mathcal{E}) \quad,\quad H^{\eta_C}_T = \sum_{\eta<|\eta_C|} E_T\quad.
\end{equation}
$\mathcal{E}$ denotes the set of chosen $\eta_C$'s. We vary $\eta_C$ uniformly in the interval [1,5]: 
\begin{eqnarray} \nonumber
\label{eq:eta_set}
\mathcal{E}=\{1, 1.27, 1.53, 1.8, 2.07,2.33, 2.6, 2.87, 3.13, 3.4, 3.67, 3.93, 4.2, 4.47, 4.73,5\},  \quad 
\end{eqnarray}
to get 16 such variables. 
Their inclusion helps us to provide a thorough comparison of the high-level and low-level feature spaces. Figure \ref{fig:kinematic} shows the signal vs background distribution of some important kinematic-variables. The channel-wise contributions to the parent class are also stacked with different colors/lines.  We see that the characteristics of the $m_{jj}$ and $|\Delta\eta_{jj}|$ are the same with figure \ref{fig:cms_shape}, with the electroweak processes contributing more at higher values. A feature seen for $|\Delta\phi_{jj}|$ is the shape of the signal and background distributions. Clearly, the difference is due to the $S_{EW}$ contribution since $S_{QCD}$ has a very similar shape as that of the background. This is another characteristic of VBF processes that the leading jets, originating from electroweak vertices, have lower separation in $\phi$ compared to those originating from QCD. Similar plots for the remaining four kinematic variables and the $\mathcal{R}$ set of variables are shown in figure \ref{fig:rest_kin} and figure \ref{fig:ht} in \ref{app:hl_plots}.  A brief discussion of the two feature spaces  (mainly $\mathcal{R}$) is also presented. We denote the combined high-level feature-space as $\mathcal{H}$, which is 24-dimensional. 

In order to gauge the discriminating power of each feature $x$, we determine the separation \cite{Harrison:1998yr} defined as,
\begin{equation}
\label{eq:separation}
<S^2>=\frac{1}{2}\int \frac{\left(p_S(x)-p_B(x)\right)^2}{p_S(x)+p_B(x)}\;dx\quad.
\end{equation}
$p_S(x)$ and $p_B(x)$ denote the normalized probability distribution of the signal and background classes. It gives a classifier-independent discrimination power of the feature $x$. A  value of zero (one) denotes identical (non-overlapping) distributions. We plot the separation (in percentage) of the seven highest important variables out of the 24 features in figure \ref{fig:var_imp}.   It is interesting to note that out of these, there are five variables from $\mathcal{R}$, even though the first and the second are from $\mathcal{K}$, and they are much greater in magnitude.

\section{Preprocessing of feature space}
\label{sec:preprocess}

\begin{figure*}[t]
	\centering	
	\includegraphics[height=0.4\textwidth, width=\textwidth]{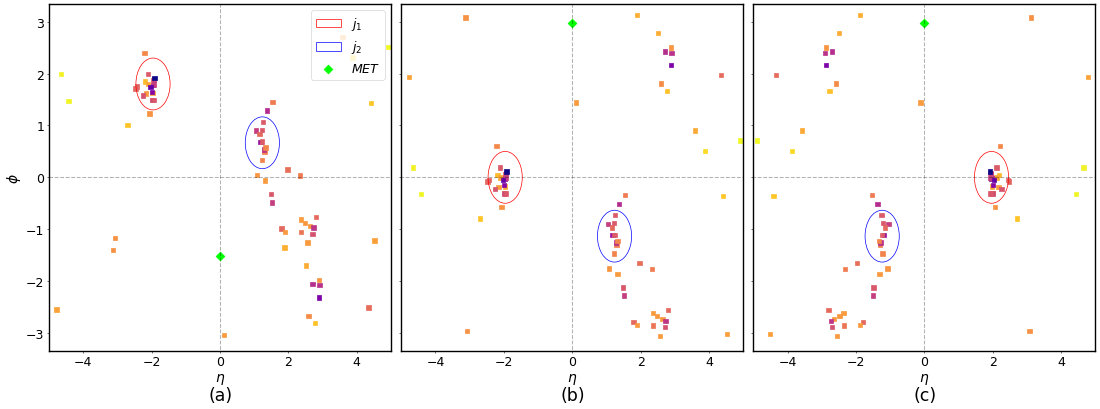}
	\caption{Scatter plot of tower constituents of an event in the $(\eta,\phi)$-plane showing:  (a) the raw event; and the effects of (b) rotation ($\phi_{j_1}=0$), and (c) reflection ($\eta_{j_1}>0$) operations. The pseudorapidity of $\textsc{met}$ has been set to zero for illustration. It is important to note that the points here are not binned into pixels and the values are the ones extracted from the Delphes Tower constituents.}
	\label{fig:preproc_steps}				
\end{figure*} 					

Preprocessing of features is indispensable for shallow machine learning as it helps maximize the statistical output from smaller data sizes. In deep-learning applications, it helps in faster convergence of the training and in approaching optimal accuracy with a lesser amount of data using simpler architectures. Even though the primary aim of our model is to learn the differing QCD radiation patterns, we can only devise preprocessing operations that preserve the Lorentz symmetries of the event.	The spatial orientation of the events, in general, can be regularized by the following procedure:
\begin{enumerate}
	\item\textbf{Identify principal directions:} Choose three final-state directions $\{\hat{n}_1 ,\hat{n}_2 , \hat{n}_3\}$. These can be any three final state objects, which are the interest of our studies like photons, leptons, and jets, or they can be chosen to be generic directions in the lab frame. 
	\item \textbf{First Rotation:} Rotate the event such that: $$\hat{n}_1\to \hat{n}'_1=(0,0,1)\equiv\hat{n}^a\quad,\quad \hat{n}_2\to \hat{n}'_2\quad,\quad \hat{n}_3\to \hat{n}'_3\quad\quad.$$ 
	After this operation, the orientation of $\hat{n}_1$ is the same for all events. 
	\item \textbf{Second Rotation:} Rotate the event along $\hat{n}^a$ such that: 
	$$\hat{n}'_2\to \hat{n}''_2=(0,n^b_y,n^b_z)\equiv\hat{n}^b\quad,\quad \hat{n}'_3\to \hat{n}''_3\quad\quad.$$ The plane formed by $\hat{n}_1$ and $\hat{n}_2$ has the same orientation for all events after this operation. 
	\item \textbf{Reflection:} Reflect along yz-plane such that:
	$$n''_{3}\to (|n^c_x|,n^c_y,n^c_z)\equiv\hat{n}^c\quad.$$ The half-space containing $\hat{n}_3$ becomes the same for all events after this step.
\end{enumerate}

These are passive operations which affect the orientation of the reference frame without changing the physics. For most event topologies, we can see that there will be better feature regularisation when $\hat{n}_2$ and $\hat{n}_3$ are equal. In hadron colliders, due to the unknown partonic center-of-mass energy $\sqrt{\hat{s}}$, we set the z-axis as $\hat{n}_1$,  preserving the transverse momentum of all final state particles. We choose two different instances of $\hat{n}_2\in\{\hat{n}_{\textsc{met}},\hat{n}_{j_1}\}$. For our choice of $\hat{n}_1$, the z-direction of $\hat{n}_2$ does not matter and we can take its value for $\hat{n}_{\textsc{met}}$ to be zero. However, the z-direction becomes important for the third operation and we choose $\hat{n}_3=\hat{n}_{j_1}$. This translates to applying the following operations to the four-momenta of each events:
\begin{enumerate}
	\item Rotate along z-axis such that $\phi_0=0$. We choose two instances of $\phi_0\in\{\phi_{\textsc{met}},\phi_{j_1}\}$. 
	\item Reflect along the xy-plane, such that the leading jet's $\eta$ is always positive. 
\end{enumerate}

\begin{figure*}[t]
	\centering		
	\includegraphics[height=0.6\textwidth, width=\textwidth]{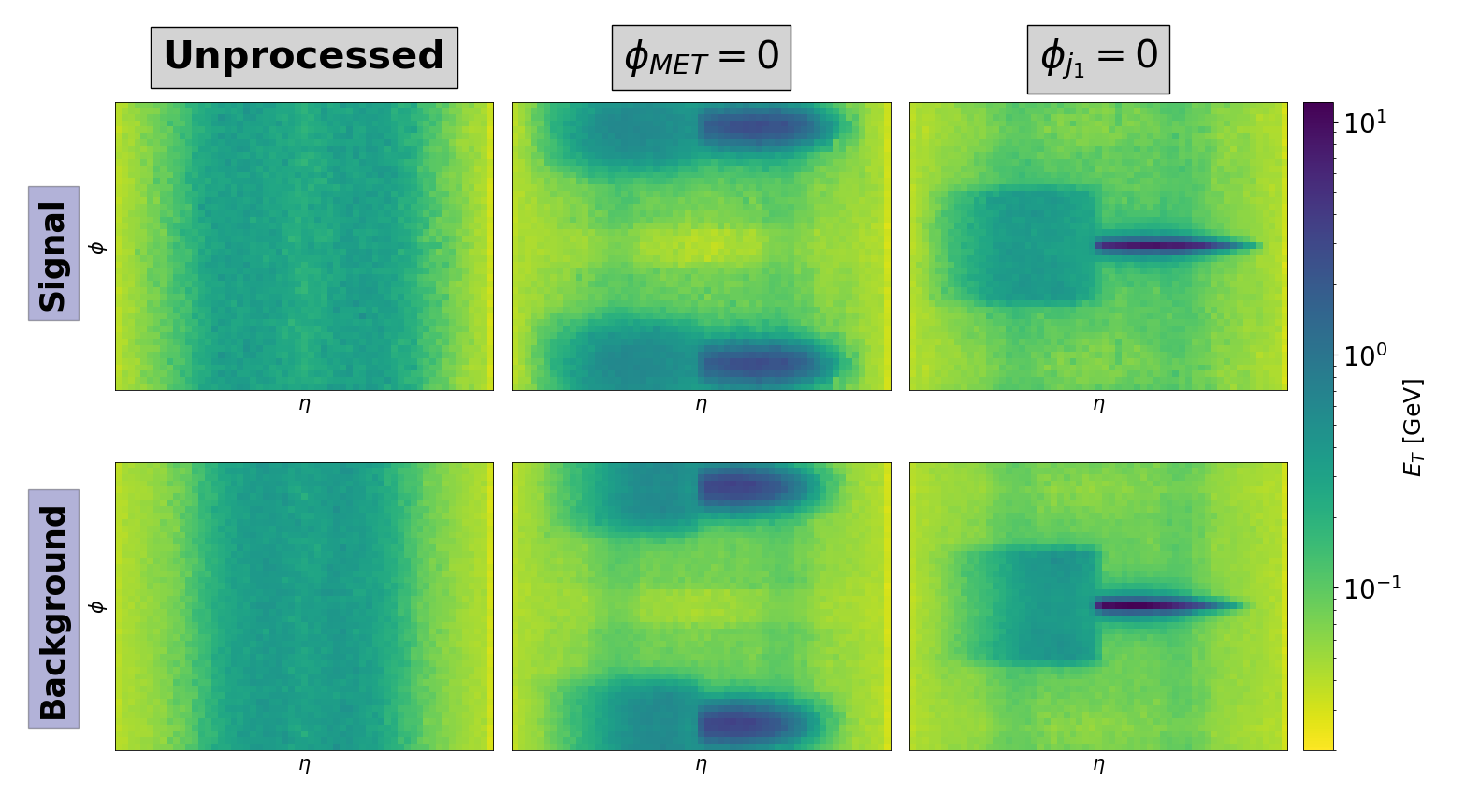}
	\caption{Average of 25,000 low-resolution tower-images of (left) unprocessed, (center) processed image with $\phi_{\textsc{met}}=0$ and (right) $\phi_{J}=0$ for (top panel) signal and (bottom panel) background classes. The images are binned in the full range of the tower: $\eta\in[-5,5]$ and $\phi\in[-\pi,\pi)$. We can see that as we go from left to right, there is a discernible improvement in regularization of the features. There are no distinctly localized hard regions for the unprocessed case, while there are some for the $\phi_{\textsc{met}}=0$ instance, which becomes harder for $\phi_{j_1}=0$ case with the hardest region around the leading jet.}
	\label{fig:avg_img}				
\end{figure*}

After these two steps, the tower-constituents are binned in the resolutions as mentioned earlier and then padded on the $\phi$-axis. We denote the feature-spaces obtained after preprocessing with the two instances of $\phi_0$ as $\mathcal{P}_{\textsc{met}}$ and $\mathcal{P}_{J}$. 
Figure \ref{fig:preproc_steps} shows the different steps of preprocessing steps for an event taking $\phi_0=\phi_{j_1}$. 
Averaged low-resolution image of the validation dataset of each class without preprocessing, and for both instances of $\phi_0$ are shown in figure \ref{fig:avg_img}. As emphasized earlier, it is seen that there is a better regularization when $\hat{n}_2=\hat{n}_3$ ($\phi_{j_1}=0,\;\eta_{j_1}>0$).  Clearly, the dominant features are the jets, and while for $\mathcal{P}_{J}$, these lie in the center; for $\mathcal{P}_{\textsc{met}}$ they lie at the $\phi$-boundary. Thus, the effect of padding is much more pronounced in $\mathcal{P}_{\textsc{met}}$. In analogy, it becomes crucial when the Higgs boson decays in a hadronic channel (say $h^0\to b\bar{b}$ or even $h^0\to \tau^+\tau^-$), where we would desire the jets arising from Higgs -- be it normal or large-radius, to be at the center of the image. Combining the instances of preprocessing and resolutions, there are four low-level feature spaces, namely: $\mathcal{P}^{LR}_{\textsc{met}}$, $\mathcal{P}^{HR}_{\textsc{met}}$, $\mathcal{P}^{LR}_{J}$ and $\mathcal{P}^{HR}_{J}$. The superscripts $LR$ and $HR$ denote the low and high-resolutions. We notice that all the high-level variables except $\phi_{\textsc{met}}$, are invariant under the two preprocessing operations, although, for our purpose, we extract them prior to their application. This follows from the usual physical intuition that absolute positions in the lab-frame are of no particular importance, and the useful information comes from the relative position of the different final-states.

We regularize the high-level features by mapping the distribution of each variable to their z-scores. Calculating the mean $\bar{x}^j$, and the standard deviation $\sigma^j$ for each feature of the whole dataset (training and validation data of both classes together), we perform the following operation on each variable of all events,
\begin{equation}
\label{eq:high_level_z_score}
z^j_i=\frac{x^j_i-\bar{x}^j}{\sigma^j}\quad.
\end{equation}
The superscript $j$ denotes the feature index, and the subscript $i$ denotes the per-event index. It is particularly useful since the features have very different ranges (for instance, $m_{jj}$ and $|\Delta\eta_{jj}|$), and the operation minimizes this disparity. Furthermore, the features of $z^j$ are now dimensionless. A caveat here is that the values of mean and standard deviations used are calculated from a balanced dataset. In experimental data, the presence of both classes, if at all, there is a positive signal, is never balanced. We justify our choice by their class independence, by virtue of which the relative differences in the shape of the signal and background distributions are conserved, and the same set of values can be used when applying to unknown data with no labels.

\section{Neural Network architecture and performance}
\label{sec:net_run}

\begin{figure*}[t]
	\centering	
	\includegraphics[height=0.78\textwidth, width=\textwidth] {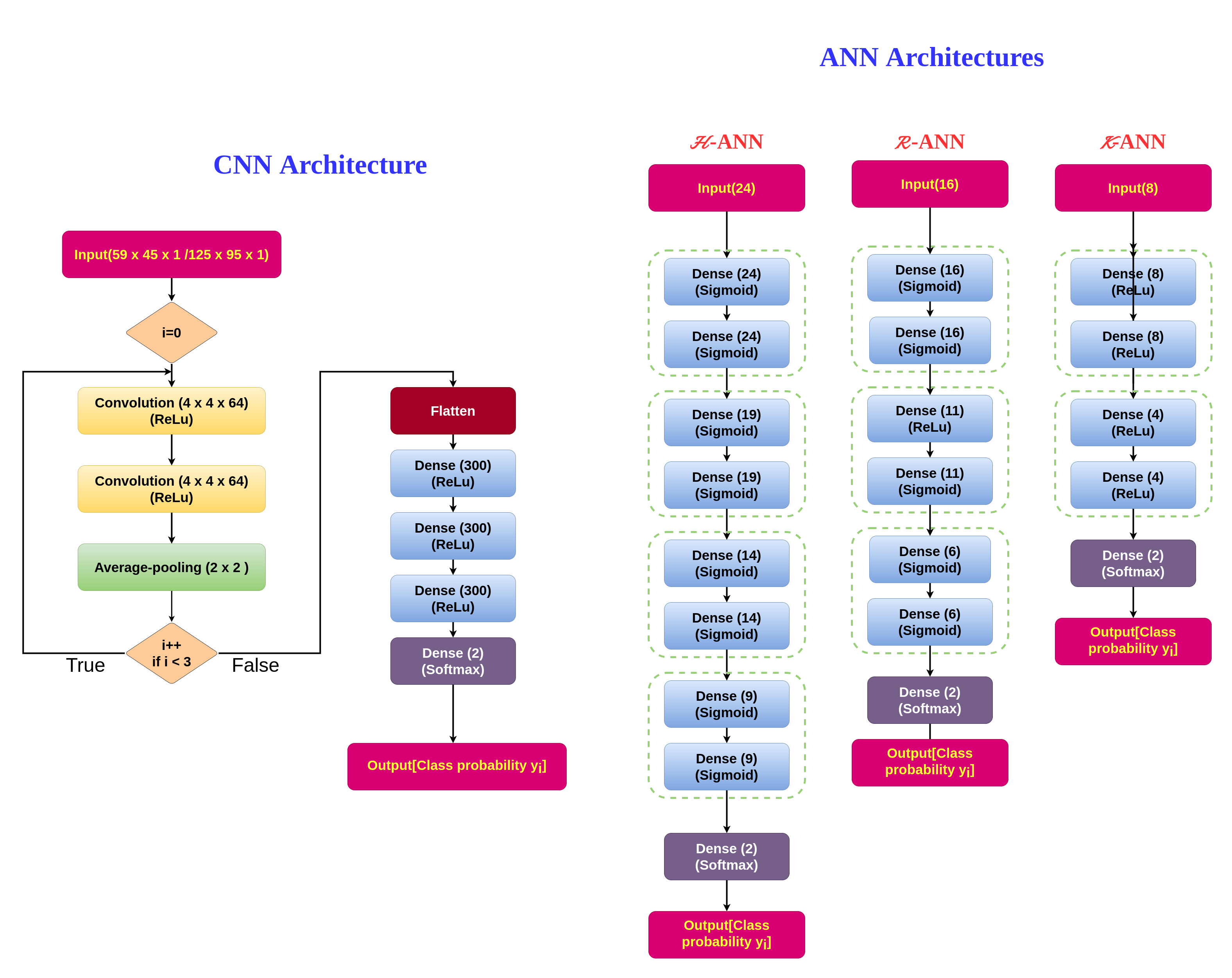} 
	\caption{Simplified architecture of (left) CNNs and (right) ANNs.}
	\label{fig:NN_archi}
\end{figure*}

In the previous sections, we have defined seven feature spaces, which are broadly grouped into high-level classes comprising of  $\mathcal{K}$ (kinematic), $\mathcal{R}$ (QCD-radiative) and $\mathcal{H}$ (a combination of the two previous spaces); while low-level spaces are: $\mathcal{P}^{LR}_{\textsc{met}}$, $\mathcal{P}^{HR}_{\textsc{met}}$, $\mathcal{P}^{LR}_{J}$ and $\mathcal{P}^{HR}_{J}$. With these as inputs, we train neural-networks for classification. The generic architecture chosen for the high-level feature spaces are dense Artificial Neural Networks (ANNs) while for low-level ones are Convolutional Neural Networks. Hence, we name the 7 networks as: $\mathcal{K}$-ANN, $\mathcal{R}$-ANN, $\mathcal{H}$-ANN, $\mathcal{P}^{LR}_{\textsc{met}}$-CNN, $\mathcal{P}^{HR}_{\textsc{met}}$-CNN, $\mathcal{P}^{LR}_{J}$-CNN and $\mathcal{P}^{HR}_{J}$-CNN. All networks were executed in Keras(v2.2.4) \cite{chollet2015keras} with TensorFlow(v1.14.1) \cite{tensorflow2015-whitepaper} backend. 

\subsection{Choice of hyperparameters}The CNN is composed of three modules with each module formed by two convolutional layers followed by an average-pooling layer. Each convolutional layer consists of sixty-four filters with a size $4\times4$, with a single stride in each dimension. We pad all inputs to maintain the size of the outputs after each convolution. The pool-size is set to be $2\times2$ for all three modules with $2\times2$ stride size. The third module's output is flattened and fed into a dense network of three layers having three hundred nodes each, which we pass into the final layer with the two nodes and softmax activation. The convolutional layers and the dense layers before the final layer have ReLu activations. In total, the CNNs for the high-resolution (low-resolution) images have approximately 3.7 (1.2) million trainable parameters. The information bottleneck principle \cite{tishby2015deep} inspires the ANN architectures. It has close connections to coarse-graining of the renormalization-group evolution and was, in fact, priorly pointed out in reference \cite{mehta2014exact}. We choose the number of nodes in the first layer to be equal to the number of input-nodes, which is then successively reduced after two layers of the same dimension.\footnote{This provides stability of the representations learned at each dimension} 
These reductions in successive nodes are chosen to be five for the $\mathcal{R}$-ANN and $\mathcal{H}$-ANN, while for $\mathcal{K}$-ANN, we consider four due to the low-dimensionality of the input. We stop this process when there is no further reduction possible, or after four such reductions.
We checked two activation functions: sigmoid and ReLu for the ANNs. We found that sigmoid activation gave the best validation accuracy for $\mathcal{R}$-ANN and $\mathcal{H}$-ANN, while it decreased over ReLu activations for $\mathcal{K}$-ANN. In total, the $\mathcal{K}$-ANN, $\mathcal{R}$-ANN, and the $\mathcal{H}$-ANN have 210, 991, and 2790 trainable parameters, respectively. Since this is a first exploratory study, we do not optimize the hyperparameters and use the values specified here for extracting the results. Simplified architecture flowcharts for each of the different networks are given in figure \ref{fig:NN_archi}.

\begin{figure*}[t]
	\centering
	\subfloat[]{\label{fig:binned_y_0_low_a}\includegraphics[height=0.31\textwidth,
		width=0.49\textwidth]{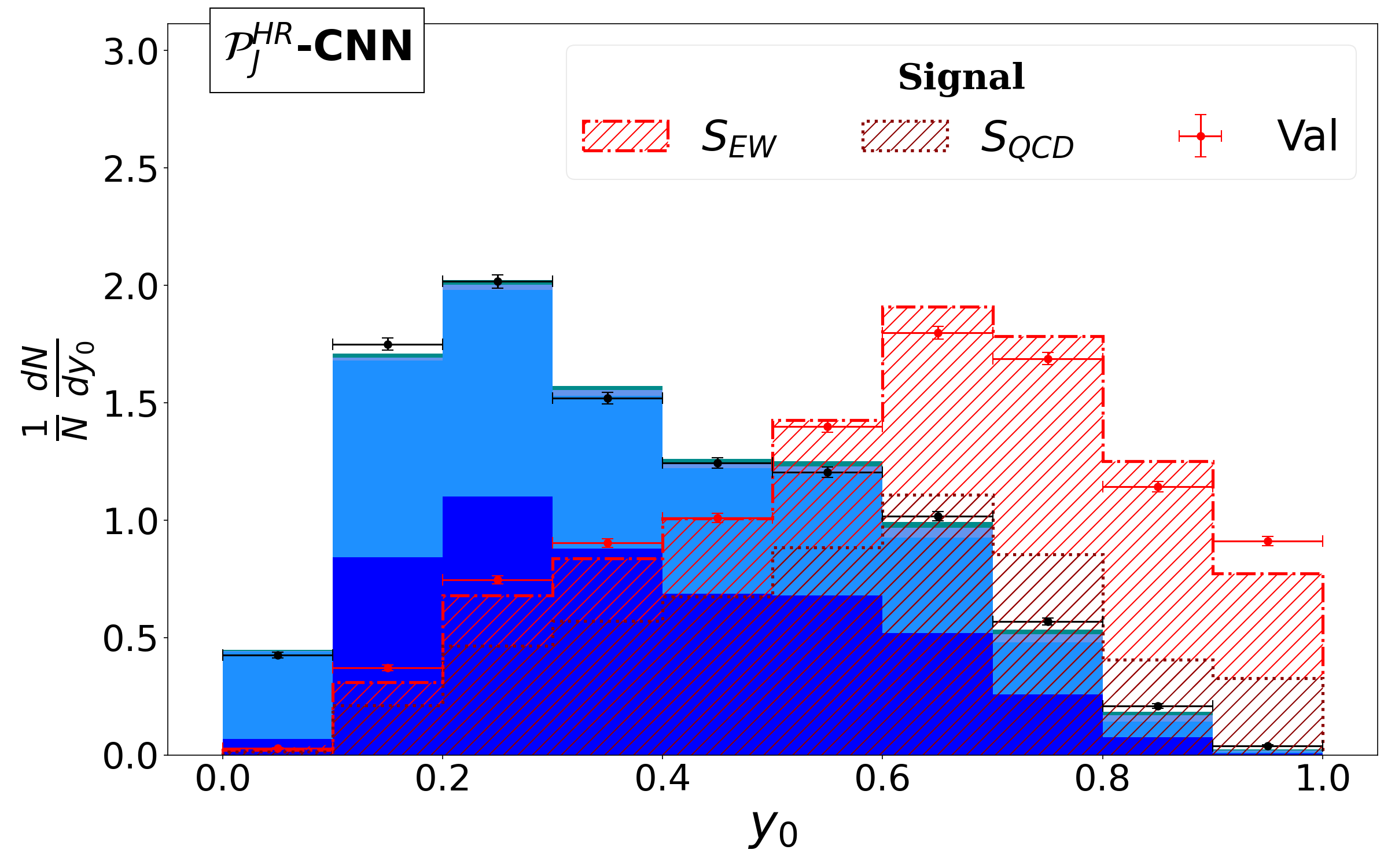}}~
	\hspace{0.4 cm}
	\subfloat[]{\label{fig:binned_y_0_low_b}\includegraphics[height=0.31\textwidth,
		width=0.49\textwidth]{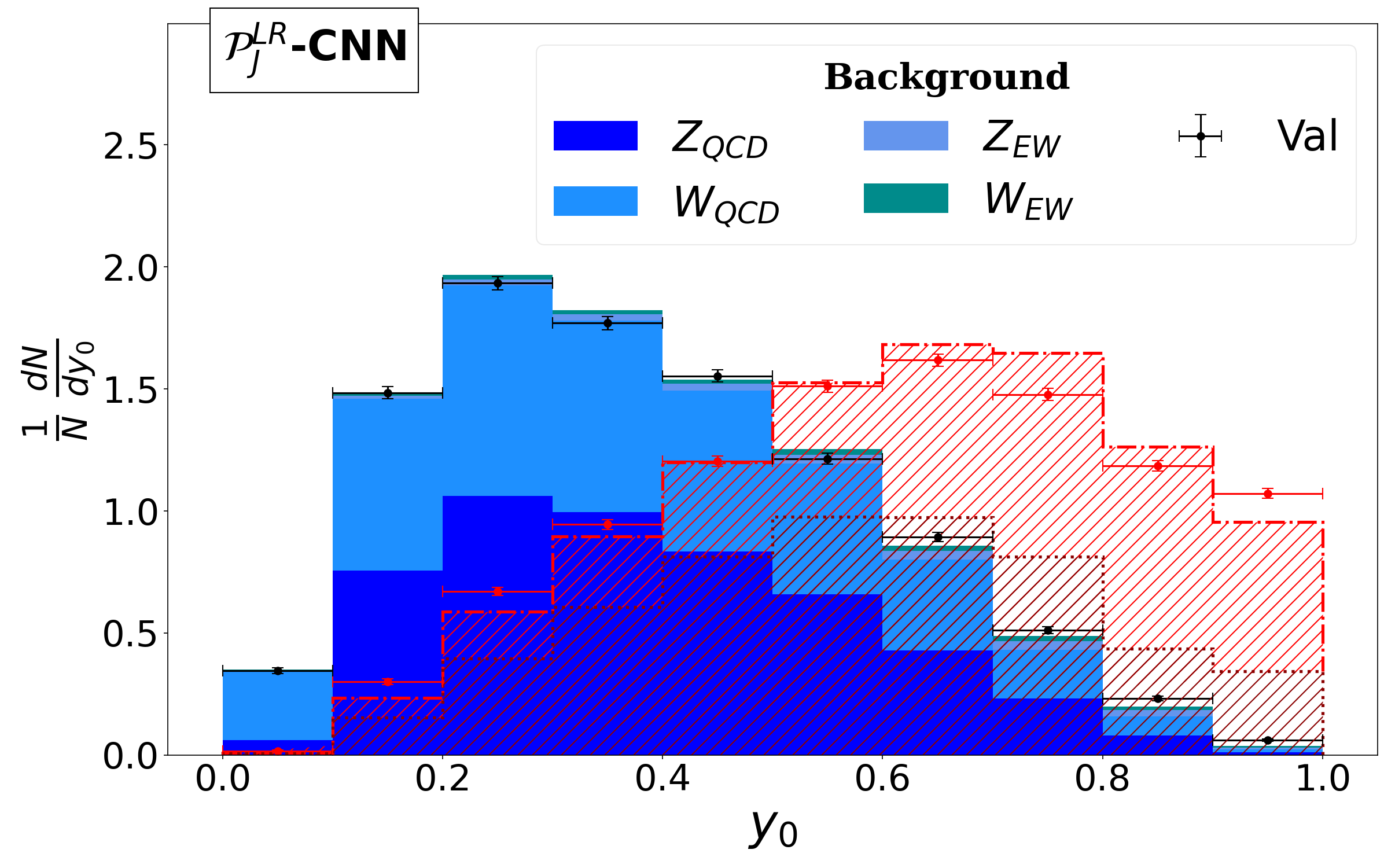}}\\
	\vspace{0.4 cm}
	\subfloat[]{\label{fig:binned_y_0_low_c}\includegraphics[height=0.31\textwidth,
		width=0.49\textwidth]{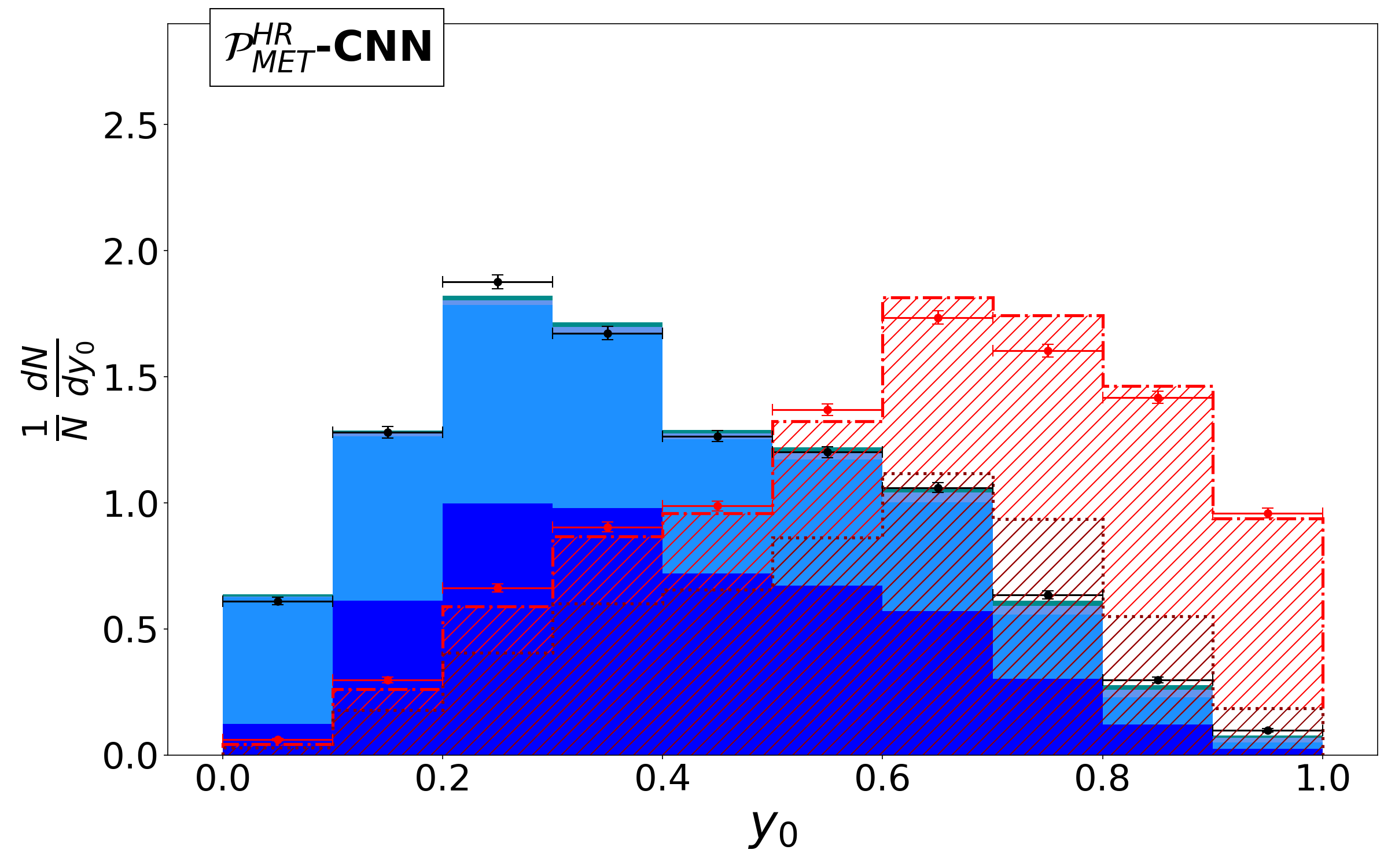}}~
	\hspace{0.4 cm}
	\subfloat[]{\label{fig:binned_y_0_low_d}\includegraphics[height=0.31\textwidth,
		width=0.49\textwidth]{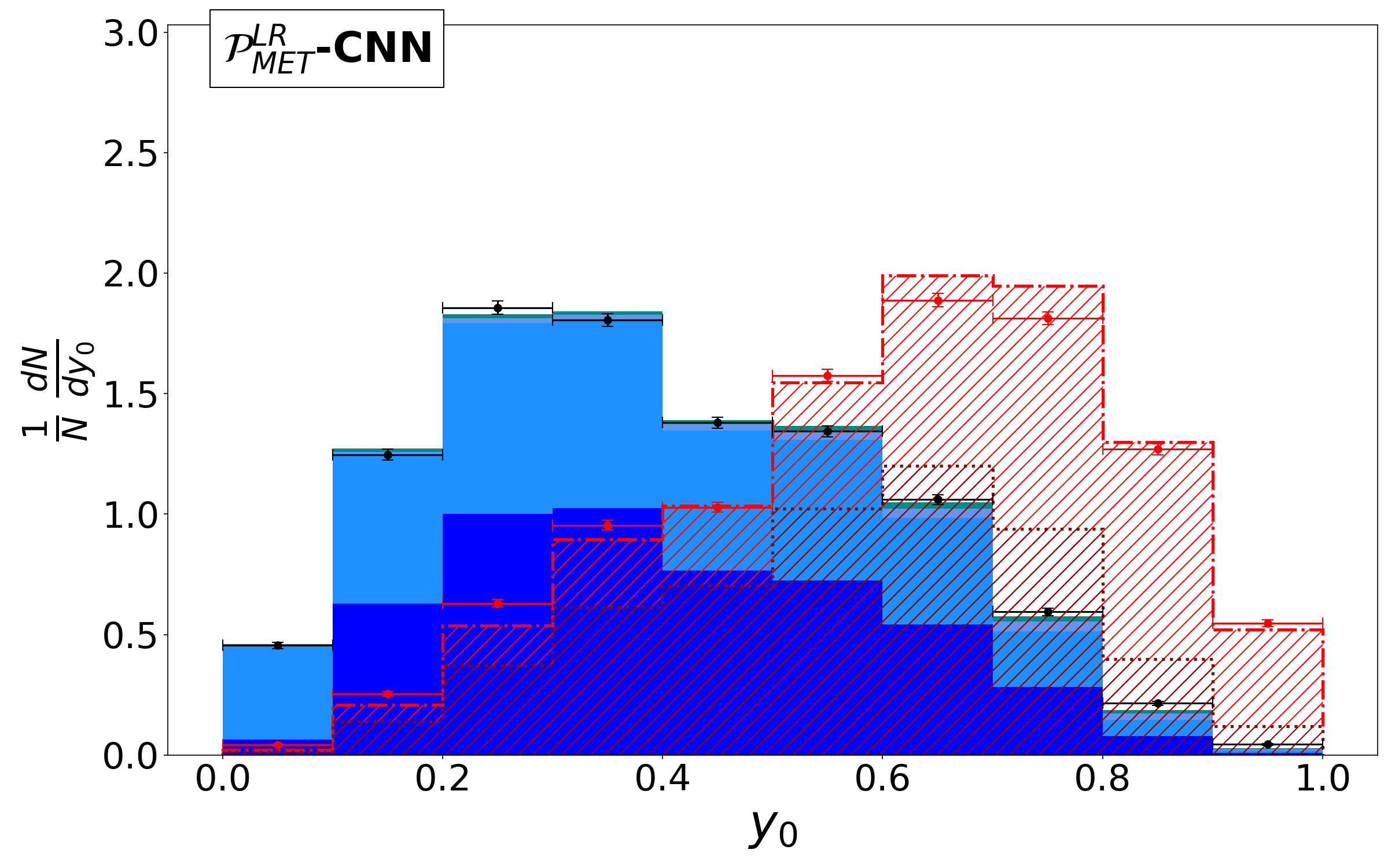}}\\
	\caption{Binned distribution of the network output for (a)
		$\mathcal{P}^{HR}_{J}$-CNN (top-left), (b)
		$\mathcal{P}^{LR}_{J}$-CNN (top-right), (c)
		$\mathcal{P}^{HR}_{\textsc{met}}$-CNN (bottom-left) and (d)
		$\mathcal{P}^{LR}_{\textsc{met}}$-CNN (bottom-right).}
	\label{fig:binned_y_0_low}
\end{figure*} 

We chose categorical-cross entropy as the loss function. The cross-entropy between two probability distributions $y_0$ and $y_t$ is defined as,
\begin{equation}
\label{eq:ce}
L=-\sum_{\vec{x}\in \mathcal{X}} y_t(\vec{x})\;\ln(y_0(\vec{x})) \quad,
\end{equation} where the distributions are functions of the feature-vector $\vec{x}$. It is a measure of how well a modeled distribution $y_0$, corresponding to the network-output, resembles the true distribution of $y_t$, the true values provided during training. For a fixed true-distribution $y_t$, minimizing the cross-entropy essentially minimizes the KL-divergence \cite{kullback1951}, 
\begin{eqnarray}\nonumber
\label{eq:kl_div}
D_{KL}(y_t||y_0)=\sum_{\vec{x}\in \mathcal{X}} y_t(\vec{x})\;\ln(y_t(\vec{x}))-\sum_{\vec{x}\in \mathcal{X}} y_t(\vec{x})\;\ln(y_0(\vec{x})) \quad,
\end{eqnarray} 
which is a measure of the similarity between two distributions, and becomes zero iff they are identical.    
We used Nadam \cite{Dozat2016IncorporatingNM} optimizer with a learning rate of 0.001 to minimize the loss function for all neural-networks.  The optimizer's adaptive nature: smaller updates for frequently occurring features while larger updates for rare features, helps in better convergence for the sparse image-data that we have, with the added benefits of Nesterov accelerated gradient descent \cite{Nesterov1983AMF}. Moreover, the learning-rate is no longer a hyperparameter. For the CNNs, training does not require more than ten epochs to reach optimal validation accuracy. Nevertheless, we train them five times from random initialization for twenty epochs. The ANNs are trained for more epochs since the relatively fewer parameters make the convergence slower. For the ANNs, ReLu activation networks are trained for two hundred epochs. In comparison, sigmoid activation networks are trained for one thousand epochs due to their relative difference in convergence compounded with fewer parameters. A batch-size of three hundred was chosen for training all networks. Each model, including all of its parameters, is stored after every epoch in the Keras-provided ``hdf5" format during training. Out of these, we use the best performing model with the highest validation accuracy for further analysis.

\subsection{Network Outputs} 
\label{sec:net_out}

\begin{figure*}[t]
	\centering	    		
	\includegraphics[height=0.22\textwidth, width=0.32\textwidth] {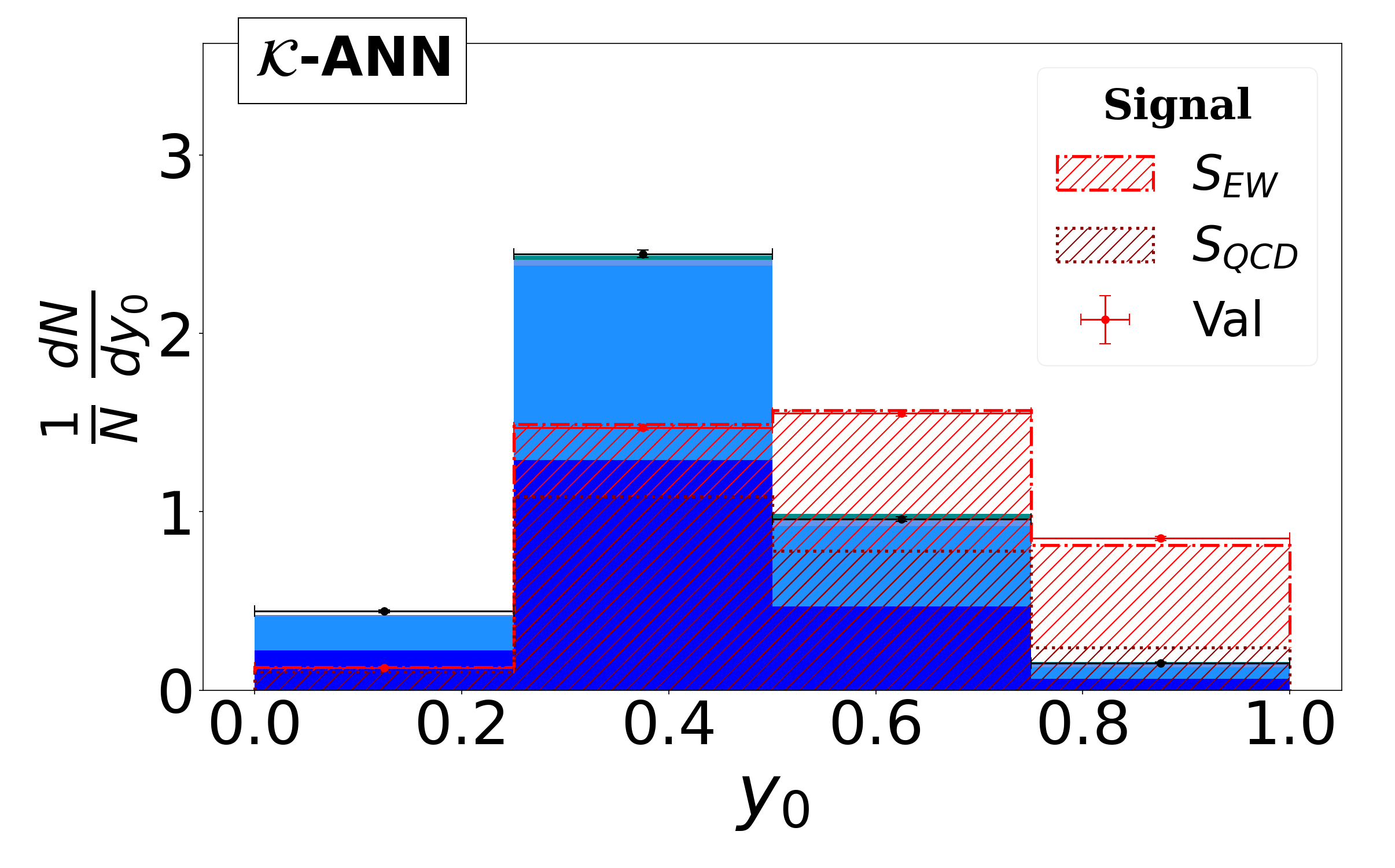}
	\includegraphics[height=0.22\textwidth, width=0.32\textwidth] {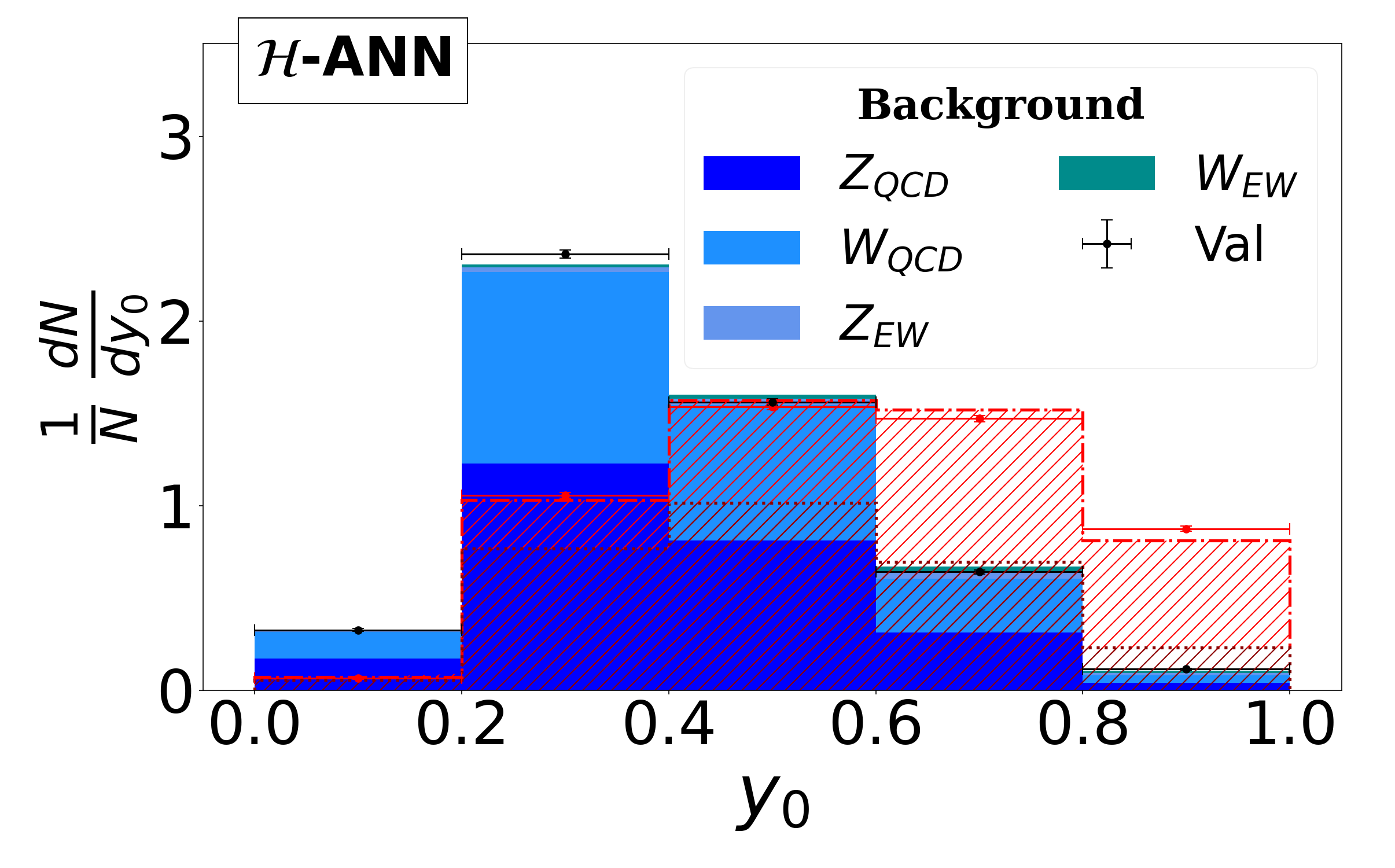}
	\includegraphics[height=0.22\textwidth, width=0.32\textwidth] {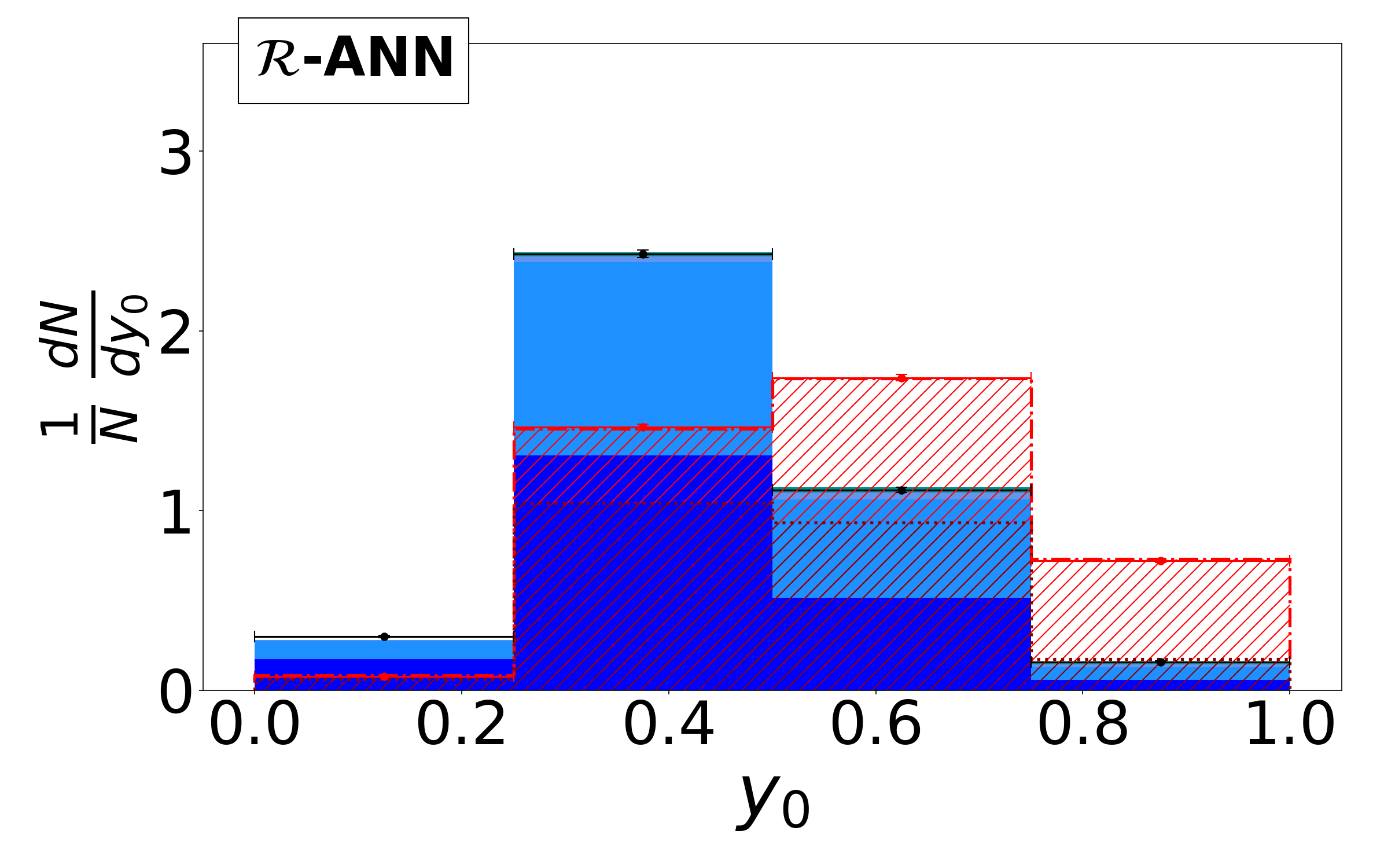}
	\caption{Binned distribution of the network output for (left) $\mathcal{K}$-ANN, (center) $\mathcal{H}$-ANN, and (right) $\mathcal{R}$-ANN.}
	\label{fig:binned_y_0_high}
\end{figure*} 

We extract the network output $y_0$, which is the probability of the event being a signal, from the best performing model from each network class. The class-wise binned distribution of $y_0$, for training and validation datasets of the low-level and high-level feature spaces, are shown in figure \ref{fig:binned_y_0_low} and \ref{fig:binned_y_0_high}, respectively.  These also show the channel wise contribution to their parent class. The choice of binning is set to the same ones used in extracting the bounds on the invisible branching ratio of the Higgs in Sect. \ref{sec:results}. It has been set such that the minimum number of entries of each class for the validation data in the edge bins have enough numbers to reduce the statistical fluctuations to less than 15\%. Contributions of the $S_{EW}$ and $S_{QCD}$ components to the signal class follow a definite pattern. 
As expected, all networks find it difficult to distinguish the $S_{QCD}$ signal from the $QCD$ dominated background. Hence, $S_{QCD}$ contributes more in the bins closer to zero, which is governed by the background class. $S_{EW}$ shows the opposite behavior dominating near one.
This same feature, although a little inconspicuous, is present for the background class's $EW$ subset as well. It may be pointed out that even for traditional analysis methods, there is significant contamination from $S_{QCD}$. A relevant machine-learning paradigm \cite{Metodiev:2017vrx} where mixed samples are used in place of pure ones, could have an interesting application in reducing this $S_{QCD}$ contamination of the signal for precision studies. Another notable feature prominent in the CNN outputs is the relative contribution of the $Z_{QCD}$ and $W_{QCD}$ channels to the background in the first bin, which is dominated by $W_{QCD}$. This can be apprehended from the fact that some of the leptons from $W^\pm$ decay, although not reconstructed, can still make calorimeter deposits on top of the QCD radiation to make itself visible to CNNs.

\begin{figure*}[t]
	\centering
	\includegraphics[width=0.49\textwidth] {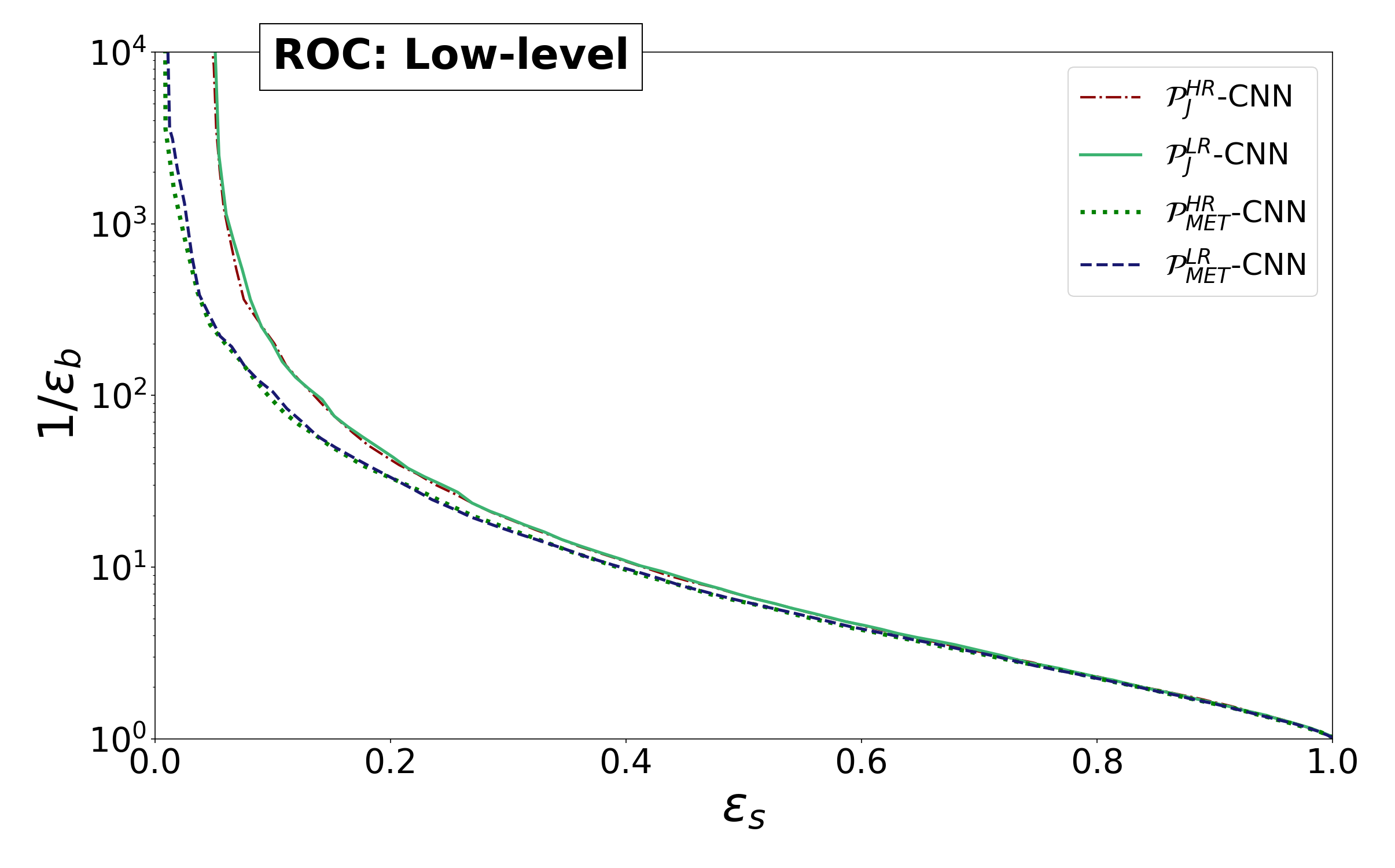}
	\includegraphics[width=0.49\textwidth] {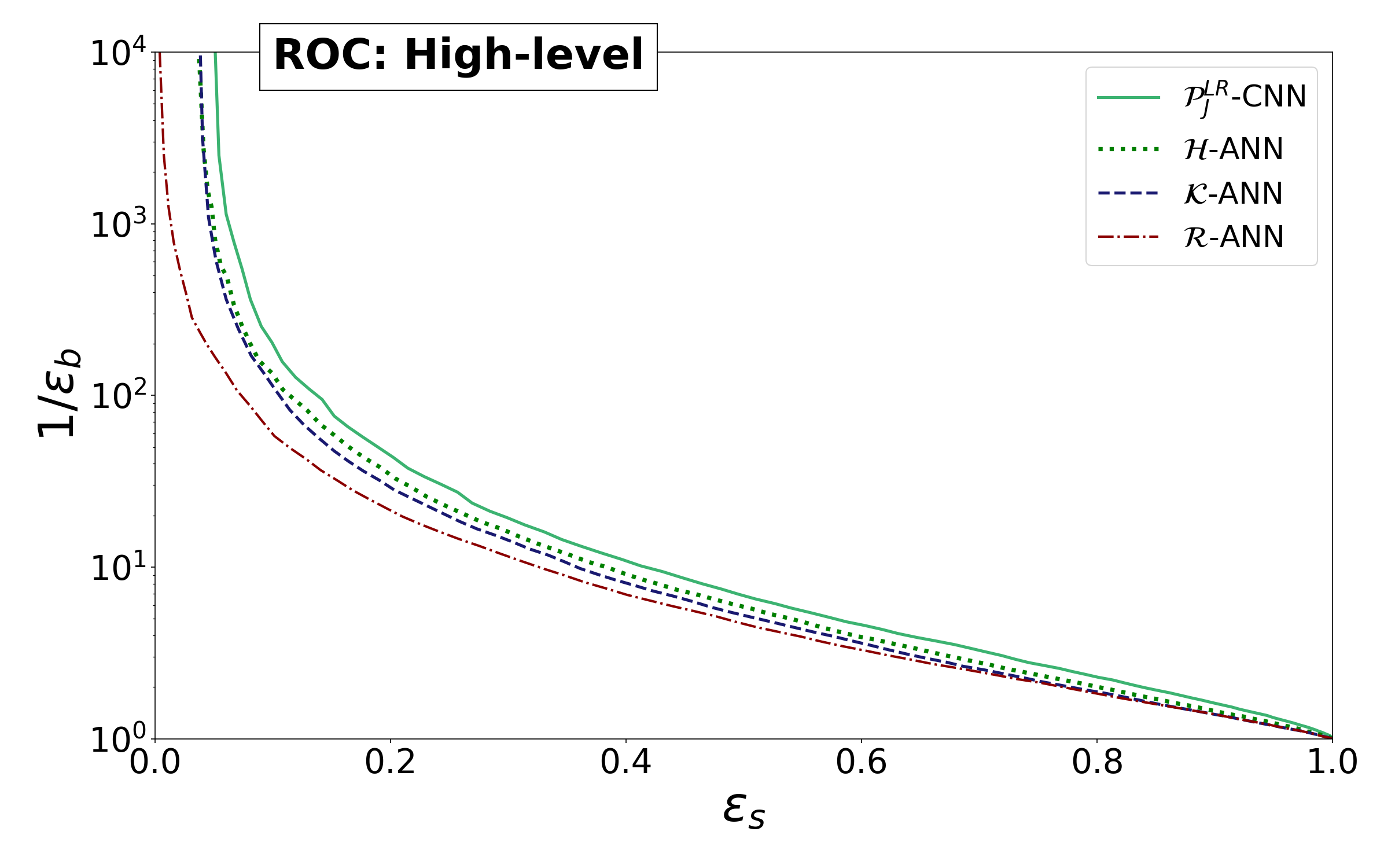}\\  \vspace{0.1cm}
	\includegraphics[width=0.49\textwidth] {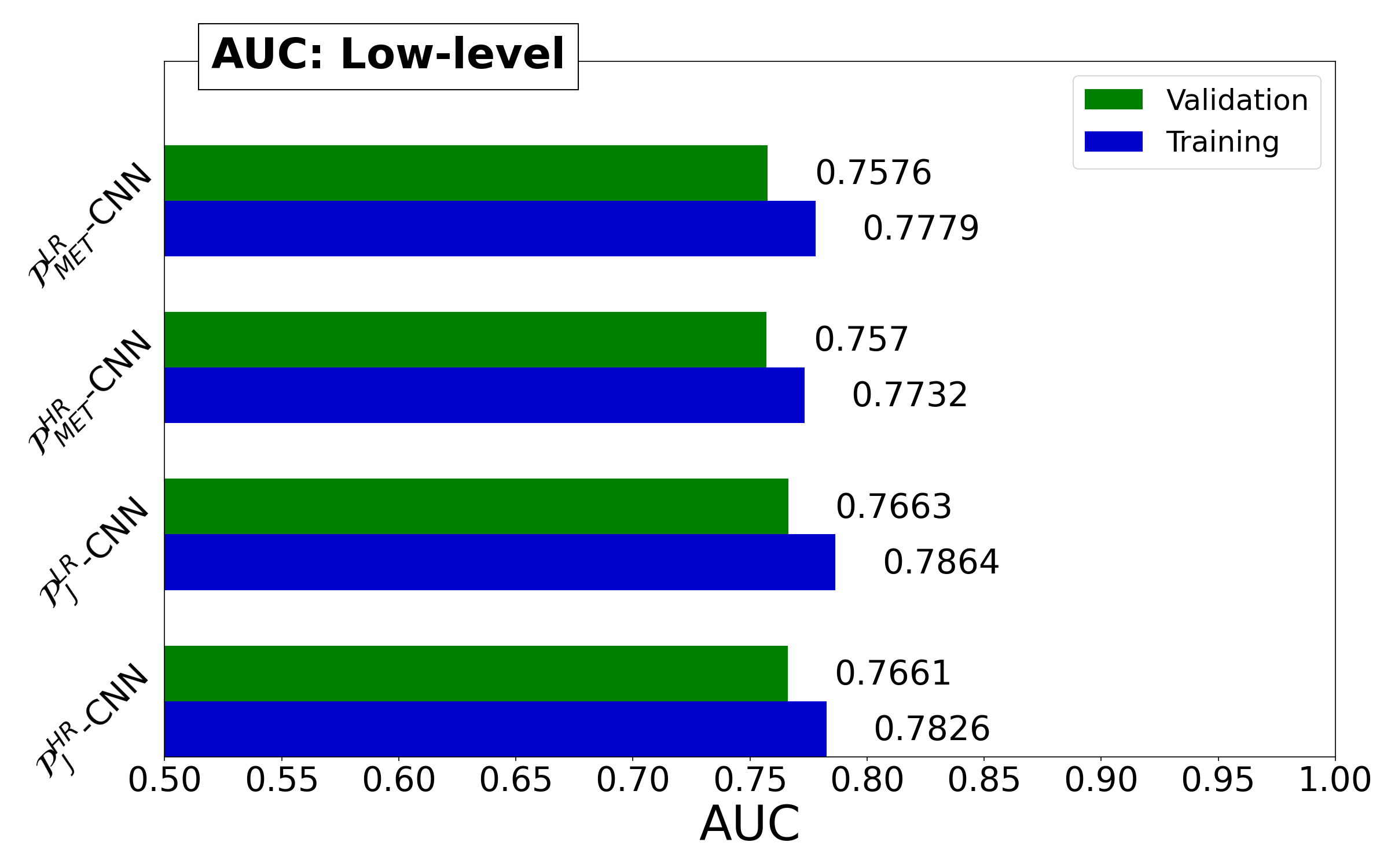}
	\includegraphics[width=0.49\textwidth] {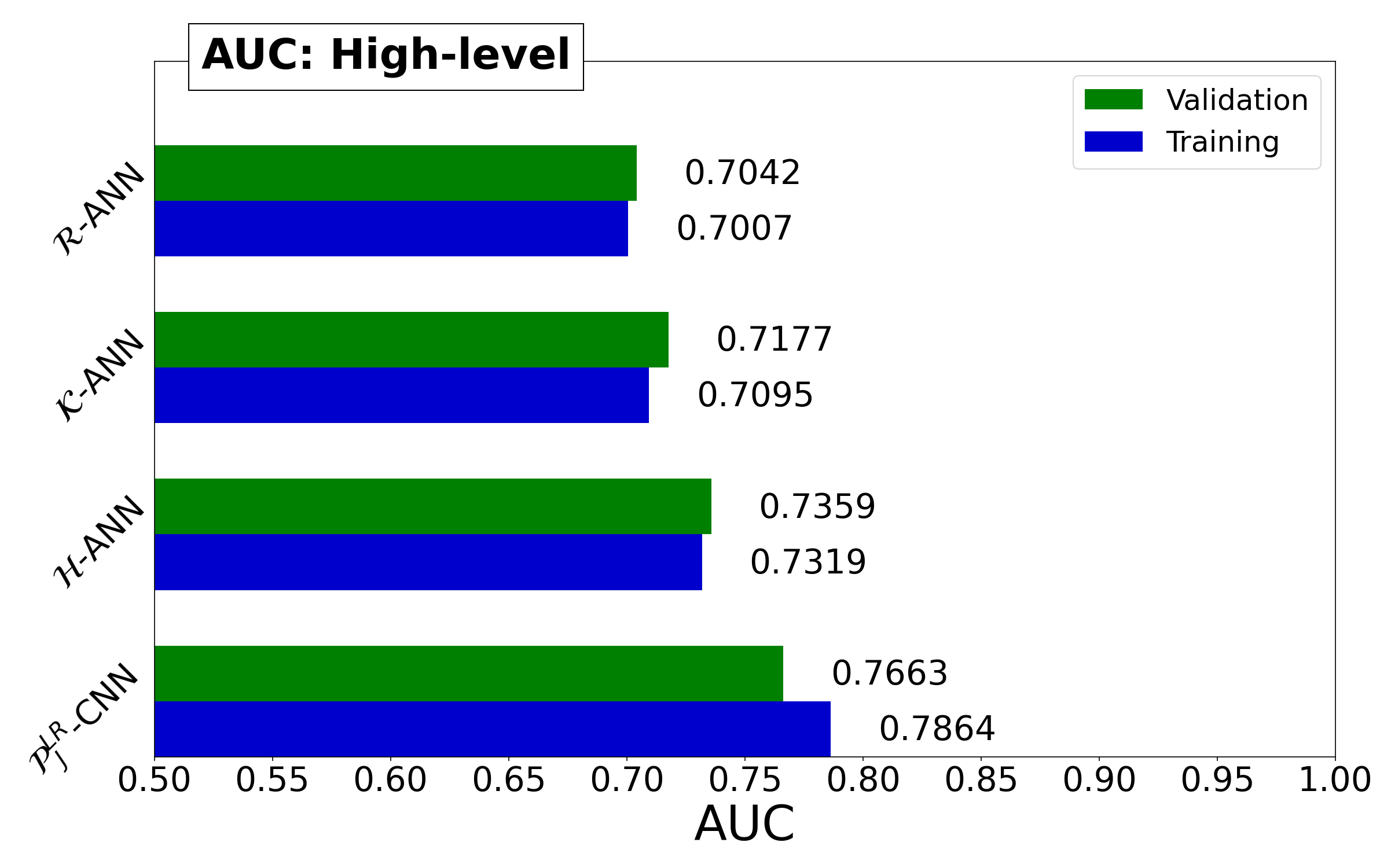}
	\caption{The validation (top panel) ROC-curves and (bottom panel) training/validation AUC for (left) low-level and (right) high-level feature spaces. In order to compare the feature spaces, the highest performing CNN is added to the plots on the right. The x-axis of the ROC-curve is the signal acceptance $\epsilon_S$, while the y-axis is the inverse of background acceptance $\epsilon_B$.}
	\label{fig:roc_auc}
\end{figure*} 

\begin{figure*}[t]
	\centering	
	\includegraphics[width=0.49\textwidth] {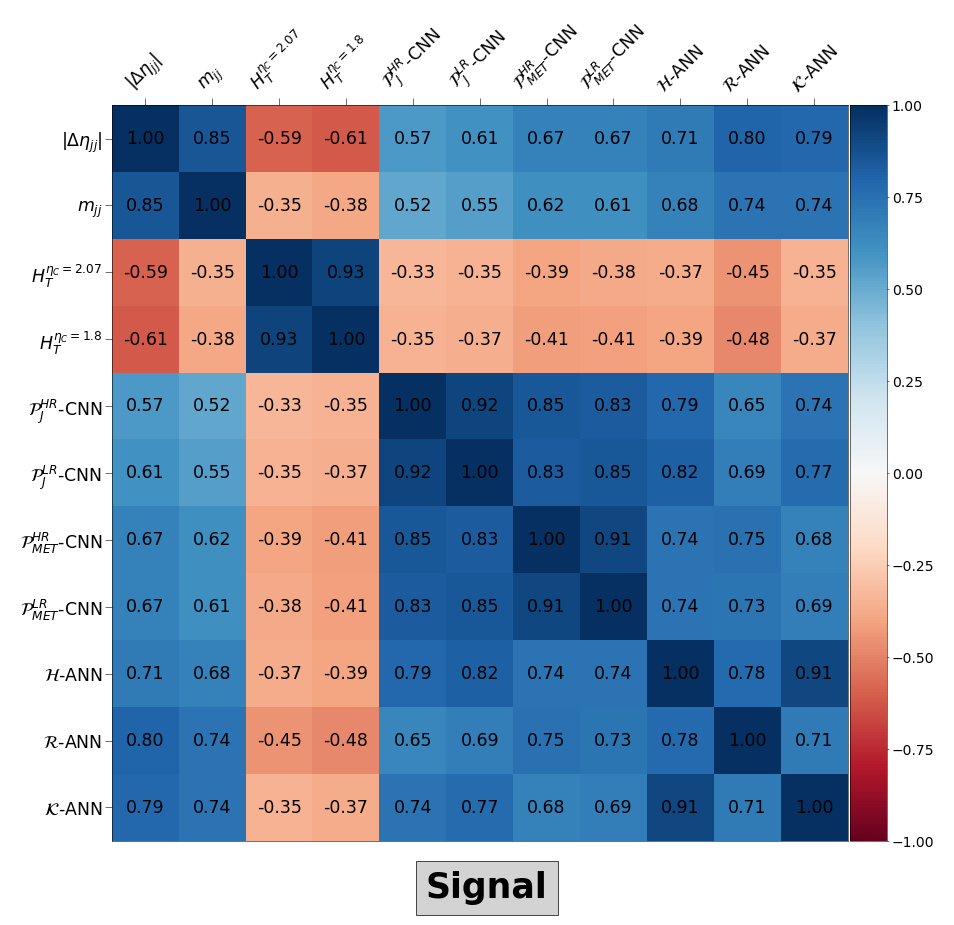}~
	\includegraphics[width=0.49\textwidth] {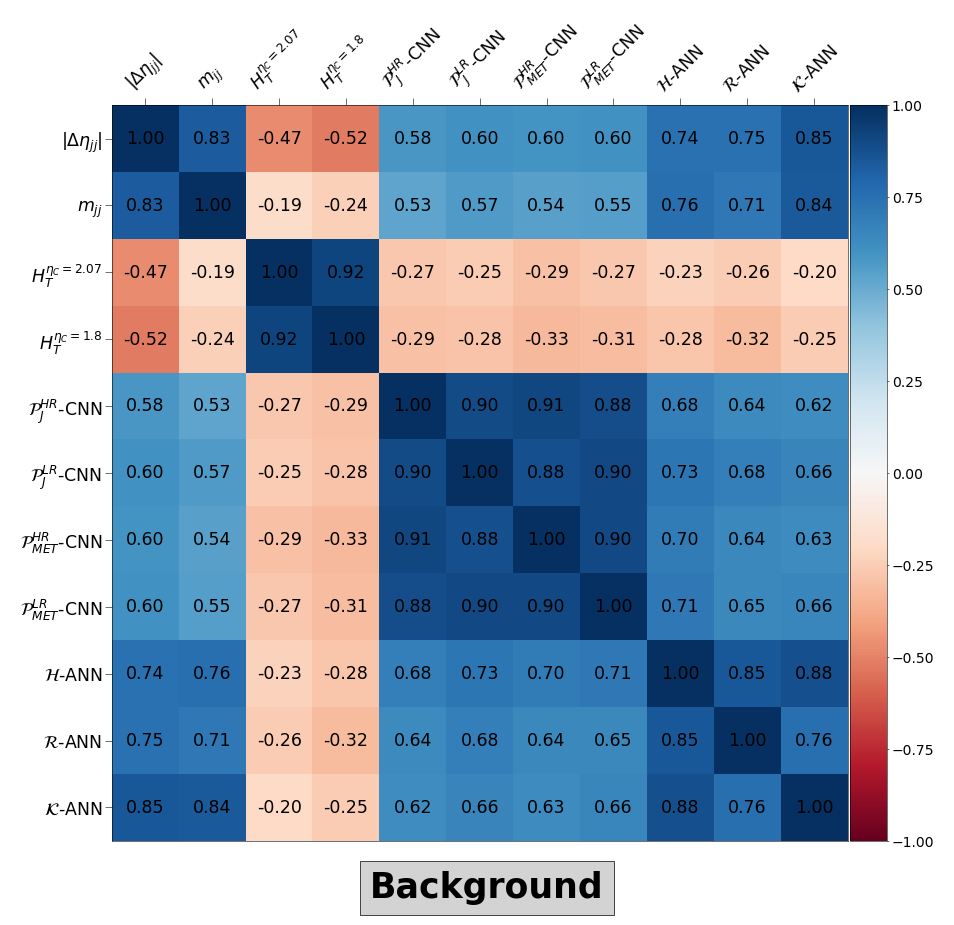}
	\caption{Pearson's correlation coefficients amongst the first four high-level variables with highest separation and the network-outputs for (left) signal and (right) background. These have been calculated using the validation dataset.}
	\label{fig:corr}
\end{figure*}

Receiver operating characteristic (ROC) curves between the signal acceptance $\epsilon_S$, and the background rejection $1/\epsilon_B$; and also the area under the curve (AUC) for all networks are shown in figure \ref{fig:roc_auc}.  The AUCs were calculated using $y_0$ and the true class labels $y_t$ with the scikit-learn(v0.22) \cite{scikit-learn} package. It is interesting to see that the so-called QCD-radiative variables ($\mathcal{R}$) perform almost as good as the kinematic-variables ($\mathcal{K}$) with only less than a percent difference in the validation AUCs. It can be understood by recalling that the radiative variables' definition includes the radiation pattern of the event, including the radiation inside the jet in cumulative $\eta$ bins. This, in principle, has similar information to  $|\Delta\eta_{jj}|$, which is one of the kinematic-variables with high separation. We confirm this by observing the correlations (shown in figure \ref{fig:corr}) between the variables $H_T^{\eta_C =2.07}$ and $H_T^{\eta_C =1.8}$ with $|\Delta\eta_{jj}|$ and $m_{jj}$. They are relatively more correlated with $|\Delta\eta_{jj}|$ than with $m_{jj}$. 
The AUC for our combined variable $\mathcal{H}$-ANN shows that the $\mathcal{R}$ variables may contain some extra information on top of what is extracted from the kinematic variables.
As emphasized earlier, we get less than 0.1 percent difference in the validation AUCs of the low and high-resolution networks. The difference in AUC between $\mathcal{P}_{J}$ and $\mathcal{P}_{\textsc{met}}$, although small, is still significant. It can be understood by looking at figure \ref{fig:avg_img}: there is better feature regularization in $\mathcal{P}_{J}$ due to the choice of $\phi_0$ than in $\mathcal{P}_{\textsc{met}}$. CNNs, in general, are supposed to be robust to these kinds of differences owing to their properties of translational invariance \cite{geom_cnn_rev}. In our case, the presence of fully-connected layers and the relatively small training sample hamper the generalization power of the CNNs. Application of global-pooling instead of using fully-connected layers and an increase in data size coupled with proper hyper-parameter optimization should reduce this difference in AUCs. These can be explored in future studies.

The class-wise linear correlation matrix between the network-outputs, along with the four high-level variables possessing the highest separations, are shown in figure \ref{fig:corr}. As expected, the outputs within the respective subset of networks are highly correlated. The outputs of the ANNs and the CNNs are also correlated significantly. A closer look reflects the addition of information in the high-level feature spaces: the correlations increase as we go from $\mathcal{R}/\mathcal{K}$-ANN to $\mathcal{H}$-ANN. In fact, if we extrapolate this argument in conjunction with the relative increase in AUC, we find that the CNNs have extracted the most information from the low-level data, which is not present in any of the high-level variables. A detailed description of the correlation of high-level variables and the ANN outputs are given in \ref{app:corr}.

\section{Bounds on Higgs invisible Branching Ratio} 
\label{sec:results}

In order to quantify our network performance in terms of  expected improvements in the invisible Higgs search results at LHC,  we obtain expected upper limits on the Higgs to invisible BRs from the distribution of the network output.  We use CL$_s$ method \cite{Junk:1999kv,Read_2002} in the asymptotic approximation \cite{Cowan:2010js}, to calculate the upper limit on the invisible BR at 95\% CL. The method is briefly discussed as follows. In a binned Poisson counting experiment of expected signal $s_i$ and background $b_i$ (which are functions of nuisance parameters jointly denoted by $\boldsymbol{\theta}$) in a bin with observed number $n_i$ of some observable, we can write the likelihood function as:
\begin{equation}
\label{eq:poisson_likelihood}
\mathcal{L}(\mu,\boldsymbol{\theta})=\prod_{i=1}^{N_b}\; \frac{\left(\mu\; s_i(\boldsymbol{\theta})+b_i(\boldsymbol{\theta})\right)^{n_i}}{n_i!}\;e^{-\left(\mu\; s_i(\boldsymbol{\theta})+b_i(\boldsymbol{\theta})\right)}    \quad,
\end{equation}
where $N_b$ is the total number of bins. $N_b$ and the bin-edges for the different variables are chosen as shown in their respective distribution plots (figures \ref{fig:cms_shape}, \ref{fig:kinematic}, \ref{fig:binned_y_0_low} and \ref{fig:binned_y_0_high}).  The profile-likelihood ratio: 
\begin{equation}
\label{eq:ple}
\lambda(\mu)=\frac{\mathcal{L}(\mu,\hat{\hat{\boldsymbol{\theta}}})}{\mathcal{L}(\hat{\mu},\hat{\boldsymbol{\theta}})}\quad,
\end{equation}
where the arguments of the denominator maximizes $\mathcal{L}$, and $\hat{\hat{\boldsymbol{\theta}}}$ conditionally maximizes $\mathcal{L}$ for the particular $\mu$, is used as a test-statistic in the form of log-likelihood,
\begin{equation} 
\label{eq:log_llhood}
t_\mu=-2\ln(\lambda(\mu)) \quad.
\end{equation} 
The distribution of the test statistic for different values of $\mu$, is required to extract frequentist confidence intervals/limits. Since, we have fixed the total weight of the signal events with respect to the background to correspond to the ones expected with the total expected production cross-section from SM for each channel($S_{EW}$ and $S_{QCD}$), $\mu$ corresponds to the invisible branching ratio of the Higgs. In the asymptotic method, for one parameter of interest, approximate analytical expressions for the distribution are derived using a result from Wald \cite{10.2307/1990256}, in the form of a non-central Chi-square distribution. Monte-Carlo simulations required to extract the unknown parameters are by-passed by choosing the best representative data called the Asimov data, by the authors of reference \cite{Cowan:2010js}; which is defined as the data when used to estimate the parameters, produces their true values.

We used HistFactory \cite{Cranmer:2012sba} to create the statistical model, and the RooStats \cite{Moneta:2010pm} package to obtain the expected limits. This provides us with greater ease of handling systematic uncertainties.  As stated before, we also redo the shape-based analysis of reference \cite{Sirunyan:2018owy} with our dataset only considering a few simpler systematics, to consistently gauge the increased sensitivity of the deep-learning approach. We incorporate three overall-systematics: uncertainty of the total cross-section, statistical uncertainty of Monte Carlo simulated events, and approximate luminosity uncertainties. We do not take into account the possible change in the shape of the distributions due to Monte Carlo simulation effects. The per-bin statistical error is taken into consideration by activating each sample's statistical-error while creating the statistical model in HistFactory. This is essentially a shape-systematics that considers the bin-wise change in shape due to the statistical uncertainties. Its inclusion increases the median expected upper-limit by around three percent in the reproduced shape-analysis. The number of events for the analysis with the higher $\textsc{met}$ cut is set to the expected number at 36 fb$^{-1}$ for all background channels. This result is also scaled for the other luminosities. For the ones with the lower $\textsc{met}$ cut, we use the validation data scaled by appropriate weights for the respective luminosities.

\begin{figure*}[t]
	\centering	
	\includegraphics[width=0.48\textwidth] {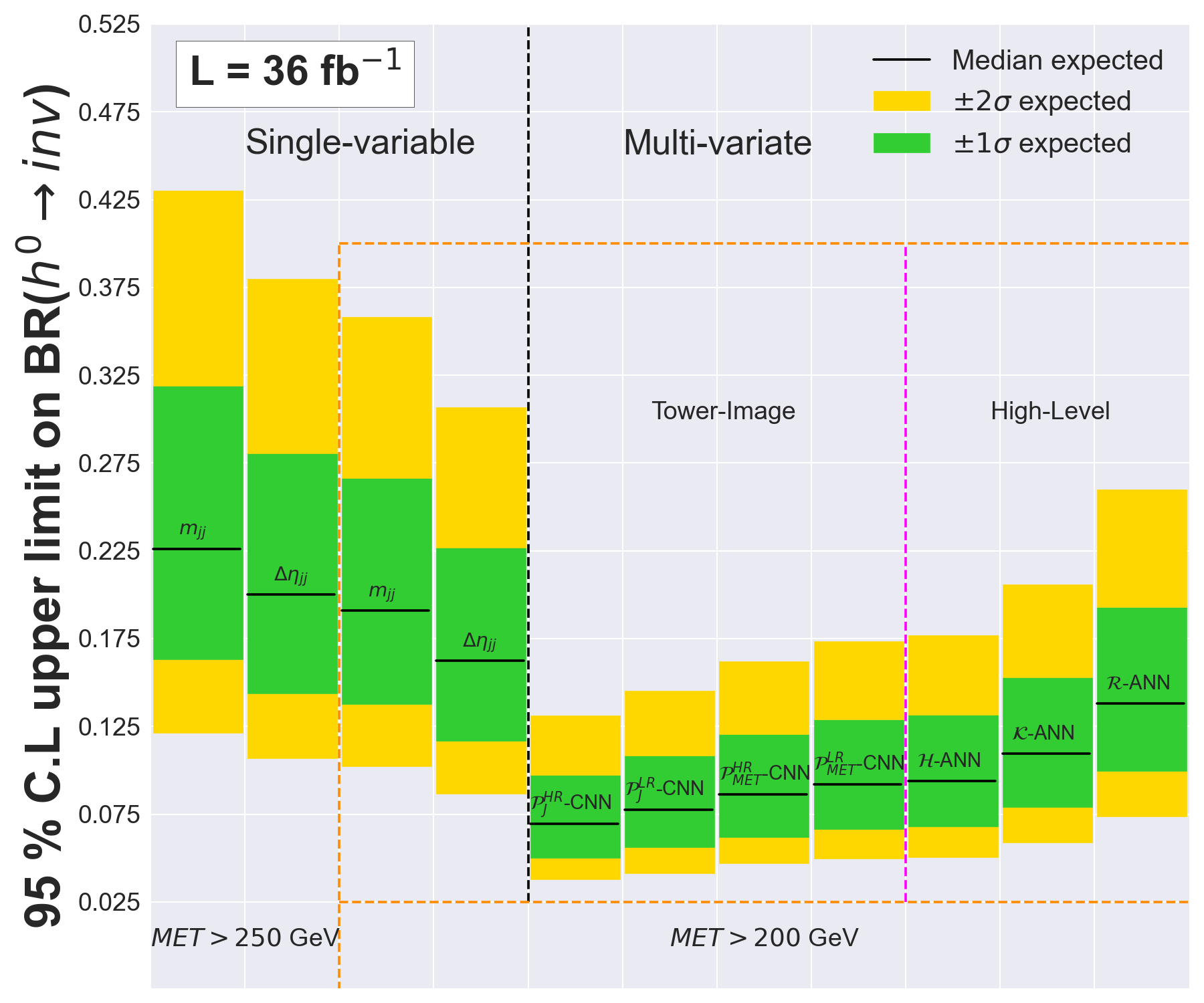} ~
	\includegraphics[width=0.48\textwidth] {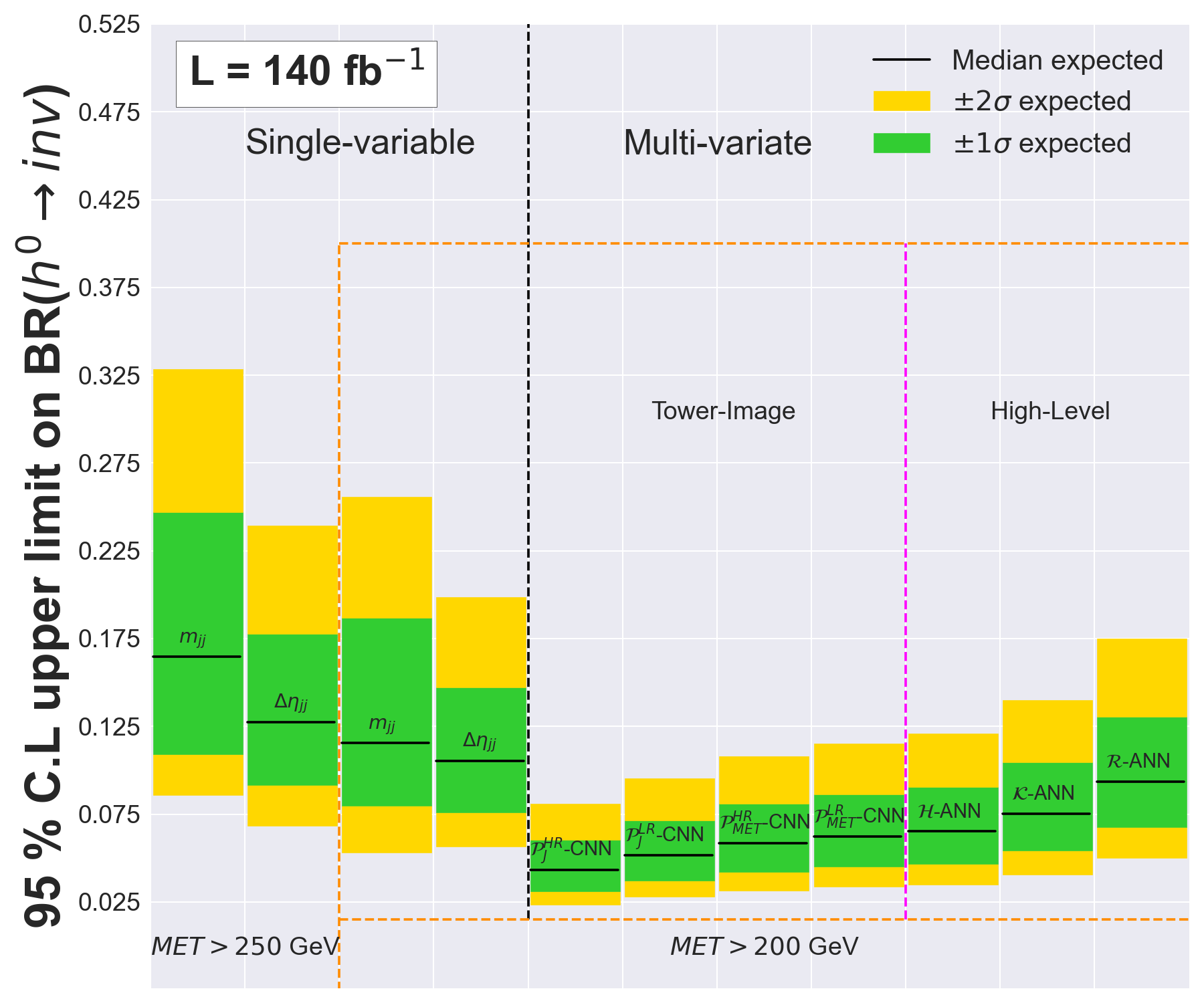} 
	\caption{Expected 95\% C.L median upper limit on the invisible branching ratio of SM Higgs with one and two sigma sidebands for (left) 36 fb$^{-1}$ and (right) 140 fb$^{-1}$ integrated luminosities.}
	\label{fig:result_36}
\end{figure*}

The median expected upper limit on the invisible branching ratio of SM Higgs at 95\% CL along with the one and two sigma error bands are shown in figure \ref{fig:result_36} for integrated luminosities of 36 fb$^{-1}$ and 140 fb$^{-1}$. A short description of the datasets used, and the corresponding median-expected upper limits with 95 \% CL is tabulated in Table \ref{tab:median_up_36}. This also contains the projected limits for 300 fb$^{-1}$, the integrated luminosity expected at the end of LHC Run III. We emphasize that even though we scale to  300 fb$^{-1}$ luminosity, we use the same dataset, and hence, the statistical uncertainties are not reduced. Consequently, our estimation for 300 fb$^{-1}$ is a conservative one. First and foremost, one can notice that the reproduced result of the shape-analysis of reference  \cite{Sirunyan:2018owy} for an integrated luminosity of 36 fb$^{-1}$ is quite consistent, and the difference can be accounted to the excluded background channels and experimental systematics. We repeat this analysis with the weaker selection criteria and see a modest improvement in the median-expected upper-limit. We also perform similar analyses with $|\Delta\eta_{jj}|$ distributions, and get an improvement of 2.9 \% for $\textsc{met}>200$ GeV, and 2.6 \% for $\textsc{met}>250$ GeV cuts. The worst (best) performing neural-network $\mathcal{R}$-ANN ($\mathcal{P}^{HR}_{J}$-CNN) has an improvement of 8.8\% (14.6\%) from the repeated experimental analysis. This, although, is with different cuts, and for the same cut in  $\textsc{met}$, we have an improvement of 5.3\% (12.1\%) for $\mathcal{R}$-ANN ($\mathcal{P}^{HR}_{J}$-CNN). For an integrated luminosity of 140 fb$^{-1}$, we get an improvement of 2.2 \% and 7.3 \% for $\mathcal{R}$-ANN and $\mathcal{P}^{HR}_{J}$-CNN, respectively. The reduced difference for higher luminosities is, of course, expected since the significance does not scale linearly with an increase in data size. An expected median upper-limit of about 3.5\% can be achieved with 300 fb$^{-1}$ of data using the highest performing network, $\mathcal{P}^{HR}_J$-CNN. 

\begin{table*}[t]
	\resizebox{\textwidth}{!}{
		\renewcommand{\arraystretch}{1.5}
		\begin{tabular}{|c|c|c|c|c|c|}\hline
			&&&\multicolumn{3}{c|}{\textbf{Expected median} }\\
			\textbf{Sl.No}&\textbf{Name}&\textbf{Description}&\multicolumn{3}{c|}{\textbf{upper-limit}}  \\
			&&&\multicolumn{3}{c|}{ \textbf{on BR($h^0\to$ inv)}} \\\cline{4-6}
			&&&L = 36 fb$^{-1}$&L = 140 fb$^{-1}$&L = 300 fb$^{-1}$\\\hline 
			1.&$m_{jj}$($\textsc{met}>250$ GeV)&reproduced shape analysis of reference \cite{Sirunyan:2018owy}
			& $0.226^{+0.093}_{-0.063}$ &   $0.165^{+0.082}_{-0.056}$  &    $0.130^{+0.089}_{-0.027}$\\\hline
			2.&$|\Delta\eta_{jj}|$($\textsc{met}>250$ GeV)& $|\Delta\eta_{jj}|$ analysis with shape-cuts of reference \cite{Sirunyan:2018owy}
			&   $0.200^{+0.080}_{-0.056}$   &   $0.128^{+0.050}_{-0.036}$   &    $0.106^{+0.041}_{-0.025}$ \\\hline 
			3.&$m_{jj}$($\textsc{met}>200$ GeV)&$m_{jj}$ shape analysis with weaker cut
			&   $0.191^{+0.075}_{-0.053}$  &   $0.116^{+0.071}_{-0.036}$ &   $0.101^{+0.037}_{-0.045}$\\\hline  
			4.&$|\Delta\eta_{jj}|$($\textsc{met}>200$ GeV)& $|\Delta\eta_{jj}|$ analysis with weaker cut
			&   $0.162^{+0.065}_{-0.045}$   &   $0.105^{+0.042}_{-0.029}$   &   $0.087^{+0.034}_{-0.025}$\\\hline 
			5.&$\mathcal{P}^{LR}_{J}$-CNN&Low-Resolution, $\phi_0=\phi_{j_1}$ 
			&   $0.078^{+0.030}_{-0.022}$   &   $0.051^{+0.020}_{-0.014}$   &    $0.045^{+0.017}_{-0.013}$\\\hline
			6.&$\mathcal{P}^{HR}_{J}$-CNN&High-Resolution, $\phi_0=\phi_{j_1}$
			&    $0.070^{+0.027}_{-0.020}$  &$0.043^{+0.017}_{-0.012}$  &   $0.035^{+0.013}_{-0.010}$\\\hline 
			7.&$\mathcal{P}^{LR}_{\textsc{met}}$-CNN&Low-Resolution, $\phi_0=\phi_{\textsc{met}}$
			&    $0.092^{+0.037}_{-0.025}$   &   $0.062^{+0.024}_{-0.017}$    &   $0.053^{+0.023}_{-0.014}$\\\hline 
			8.&$\mathcal{P}^{HR}_{\textsc{met}}$-CNN&High-Resolution, $\phi_0=\phi_{\textsc{met}}$
			&    $0.086^{+0.035}_{-0.024}$   &    $0.058^{+0.023}_{-0.016}$   &   $0.051^{+0.020}_{-0.014}$\\\hline 
			9.&$\mathcal{K}$-ANN&8 kinematic-variables
			&   $0.101^{+0.052}_{-0.022}$    &    $0.075^{+0.029}_{-0.021}$   &   $0.063^{+0.027}_{-0.017}$\\\hline 
			10.&$\mathcal{R}$-ANN&16 radiative $H_T^{\eta_C}$ variables
			&   $0.138^{+0.055}_{-0.039}$    &    $0.094^{+0.036}_{-0.027}$   &   $0.079^{+0.032}_{-0.022}$\\\hline 
			11.&$\mathcal{H}$-ANN&Combination of $\mathcal{K}$ and $\mathcal{R}$ variables
			&   $0.094^{+0.038}_{-0.026}$    &    $0.065^{+0.026}_{-0.018}$   &   $0.057^{+0.022}_{-0.015}$\\\hline     			
		\end{tabular}
	}
	\caption{Short description of the different analyses shown in figure \ref{fig:result_36} and the expected median upper-limit on BR($h^0\to$ inv) at 95\% CL for each integrated luminosities which also include projections for L = 300fb$^{-1}$.}
	\label{tab:median_up_36}
\end{table*}

The results of the different feature spaces follow the expected trend. For this discussion, we quote the numbers for an integrated luminosity of 36 fb$^{-1}$. Comparing the performance of high-level feature spaces, we see that $\mathcal{R}$ performs the worst while the combined space $\mathcal{H}$ puts the most stringent bounds. The difference is minimal (0.7 \%) with  $\mathcal{K}$-ANN, and appreciable (4.4\%) with $\mathcal{R}$-ANN. Amongst the image-networks, the difference between the low and high-resolution networks is less than a percent (0.8 \% for $\mathcal{P}_J$, and 0.6\% for $\mathcal{P}_{\textsc{met}}$). Differences in performances of the different preprocessing instances are reflected in this analysis: $\mathcal{P}_J$ puts nominally stricter bounds on the branching ratio (1.4 \% for $LR$, and 1.6 \% for $HR$).

Up to now, we demonstrated the capability of our CNN based low-level networks and also ANN-based networks considering particle level data, including detector effects as well as underlying events during our simulations as discussed in section \ref{sec:vbf}.
However, we neglected the effect of simultaneous occurrences of multiple proton-proton interactions (pileup) in our analysis. The amount of pileup is relatively moderate in low luminosity data, but increasingly significant once we move towards high luminosity. We believe that its presence would not alter our primary results substantially from the calorimeter image data. CNN architectures look into the global features of an input image. Calorimeter deposits due to pileup are expected to be similar for different classes since they are independent of the hard scattering processes. The same can be identified as redundant information,  as a consequence of the optimization algorithm effectively searching for dissimilarities between the two classes. 
Optimal pdfs acquired by CNNs remain very similar, whether it is with or without pileup. This issue was analyzed before, where it was shown that unlike high-level methods, deep-learning from calorimeter deposits shows robustness to pileup effects in the classification of jet-image \cite{Baldi:2016fql}. Although, in these studies, the jets have large transverse boosts and mostly reside in the central regions where its effect is reduced. However, various other studies \cite{Bhimji:2017qvb,Andrews:2019wng} have also shown that deep-learning on the full calorimeter information is less prone to pileup effects. These existing results further elucidate our presumption that CNNs would be less affected by higher pileup expected at future runs of LHC. In contrast, the other analyses, including the ANNs trained on high-level feature spaces, can be relatively more affected. 

To present our arguments in perspective, we combined each event (tower-image) with an additional $N$ randomly chosen minimum bias event with CMS switch through Pythia8 and Delphes without any pileup subtraction. At the same time, $N$ follows a Poisson distribution with $<N>=20, 50, 50$ for integrated luminosity 36, 140 and 300 fb$^{-1}$, respectively. Merged tower-image with pileup is then trained and tested for our high-resolution CNN scenario ($\mathcal{P}^{HR}_{J}$-CNN, which can be noted from Sl.No (6) in Table \ref{tab:median_up_36}). We found a very mild depreciation over our estimated median upper-limit at 0.076, 0.059, and 0.045, which all lie within the $1 \sigma$ error band in the branching ratio constraints. Note that no effort was made to mitigate the effects of the pileup during these estimates, which will not be the case in experimental analyses. 
In fact, there are extensive studies \cite{Komiske:2017ubm,Martinez:2018fwc} of using powerful machine-learning algorithms specially designed to reduce pileup contamination of events. A new interpretation of collider events in terms of optimal transport \cite{Komiske:2019fks,Komiske:2020qhg} have also provided promising new techniques for pileup mitigation on top of reinterpretation of existing ones \cite{Cacciari:2007fd,Berta:2014eza}. These developments offer further optimism for better mitigation of pileup effects in the future.

To test the robustness of our proposal, we also consider the effect of an important experimental systematic uncertainty. One of the significant experimental systematic uncertainties affecting the result of this analysis can be the uncertainty on the jet energy scale.  Therefore, we estimate the effect of uncertainties on the jet energy scale for our main results with calorimeter input data in CNN architecture. We vary the pixel-wise input values (which has already gone through the smearing in Delphes) by $10\%$  in upward and downward directions,
	\footnote{Reference \cite{Khachatryan:2016kdb} reports jet energy scale uncertainty for various observables, which lie well within 5\%. However, since such uncertainties are significantly controlled in jets reconstructed with the particle-flow (PF) method, we take a relatively conservative measure for the pixel-wise uncertainty of the measured energies.},
	and record the variation in the shape of the network output without considering any pileup.
	This is added as a coherent shape systematics, and we obtain an increased expected median upper-limit of $0.071^{+0.028}_{-0.019}$ for $\mathcal{P}^{HR}_J$-CNN at 140 fb$^{-1}$ integrated luminosity, which is still better by a factor of almost two when compared to the latest result from ATLAS \cite{Aaboud:2018sfi}.

\section{Summary and Conclusion} 
\label{sec:conc}

The HEP experimental community is one of the frontrunners in utilizing machine learning algorithms for the last several decades in tagging and characterizing different objects and analyzing the massive data samples with the help of neural-network or boosted decision trees. However, recent developments in deep learning approaches have shown immense prospects in a variety of other applications. 

The Large Hadron Collider, after its breakthrough discovery of an SM like Higgs boson, keeps accumulating an enormous amount of data, pinpointing its different properties and also constraining diverse BSM scenarios at the TeV scale. While such high energy data are opening up scope for new analysis techniques filling possible gaps in previous investigations,  it is prudent to review the effectiveness of some of the effective machine learning tools. 

While proposed as an alternative channel for Higgs search, the vector boson fusion (VBF) mechanism has shown tremendous possibility not only in extracting properties of the Higgs boson but also in many other BSM searches. As a whole, this mechanism reckons upon some of the fundamental features of event shape, vastly used to control the backgrounds. 

We choose VBF production of the Higgs boson decaying to invisible particles as a case study for neural networks to learn the entire event topology without any reconstructed objects. We use the compelling capability of Convolutional Neural Networks (CNN) to examine the potential of deep-learning algorithms using low-level variables. Instead of identifying any particular objects, we utilize the entire calorimeter image to study the event topology, which aims to learn the difference in radiation patterns between the two forward jets of the VBF signal. We specifically develop preprocessing steps that preserve the Lorentz symmetry of the events and are essential to maximizing the statistical output of the data.

Apart from low-level variables as calorimeter images for CNN, we also consider two sets of high-level features. One such set is based on the kinematics of the VBF, whereas the other set of variables are designed to portray the radiation pattern $H_T$ calculated in different $\eta$ ranges of the calorimeter. For a comprehensive analysis, we constructed several neural network architectures and demonstrated the comparative performance of CNN and ANN using different feature spaces. All these networks achieved excellent separation between signal and background. However, we found that CNN based low-level $\mathcal{P}^{HR}_{J}$-CNN performs the best among all the networks, which is based on the high-resolution images, although the dependence on image resolution is relatively insignificant. We also note that deep-learning on the full calorimeter information is less prone to pileup effects as well. 
Without relying on any exclusive event reconstruction, this novel technique can provide the most stringent bounds on the invisible branching ratio of the SM-like Higgs boson, which can be expected to be constrained up to 4.3\% (3.5\%) using a dataset corresponding to an integrated luminosity of 140 fb$^{-1}$ (300 fb$^{-1}$). These limits can severely constrain many BSM scenarios, especially in the context of (Higgs-portal) dark matter models. The techniques presented in this work can easily be extended to a more complex event topology.

\section*{Acknowledgement}
\label{sec:ack}
The work of AB, PK, and VSN is supported by the Physical Research Laboratory (PRL), Department of Space, Government of India. Computational work for this work was performed using the HPC resources ( Vikram-100 HPC) and TDP project at PRL. Authors gratefully acknowledge WHEPP'19, where part of this work was initiated. We also thank Regional Centre for Accelerator-based Particle Physics (RECAPP), Harish-Chandra Research Institute, for offering forums for such workshops and stimulating discussions with S. Bhattacharya, R. K. Singh, S. Rai, and B. Mukhopadhyaya.  

\newpage
\appendix
\section{Incorporating finite mass effect of top quark in gluon-fusion events}
\label{app:gf_plots}
\begin{figure}[b]
	\centering	
	\includegraphics[height=0.34\textwidth,width=0.45\textwidth]{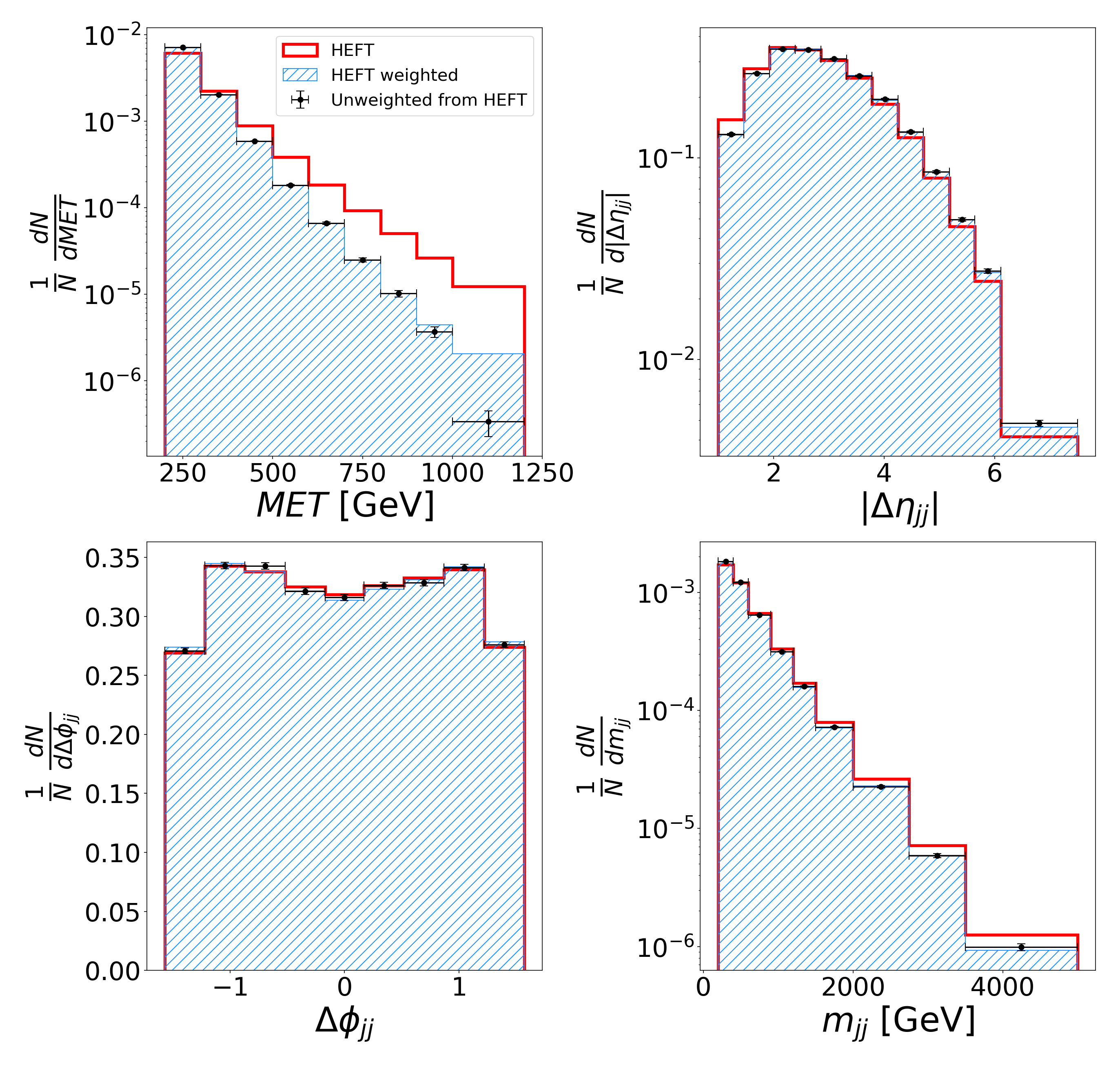}. 
	\caption{Comparative distribution of kinematic variables for HEFT, weighted with finite-top mass effects and unweighted distributions for passed events used in deep-learning training and validation.}
	\label{fig:gghweight}
\end{figure} 

We generate the gluon-fusion production of the Higgs boson by using the Higgs Effective Field Theory (HEFT) model, where the interaction of the Higgs boson with gluons is approximated by an effective vertex calculated by taking the top-quark mass to infinity.
This is a reasonable approximation only when all relevant scales in the physical process are less than $2\;m_t$. The distribution of $p_T$ of the Higgs boson (equivalently $\textsc{met}$ with detector effects introduced via Delphes) has a significant portion of events in regions where the approximation is not valid. We remove this inconsistency by reweighting the $\textsc{met}$ distribution of the events obtained after Delphes. We extract weights (ratio of the full SM results to HEFT) and bins in $p_T$ of the Higgs for the present final state topology from figure 30 on reference \cite{deFlorian:2016spz}. Each event is then assigned the corresponding weight of the bin of its $\textsc{met}$. After reweighting the events, we apply the preselection-cuts and extract the cut efficiency using the weights. 

Since we need unweighted events for the neural network training, the passed events are again unweighted.
This is done in the following steps. We divide all events into sets with unique weights. This is nothing but grouping the events into the extracted bins in $\textsc{met}$. We get mutually exclusive subsets of events $\mathcal{S}_i$, with $i$ being the bin-index. The per-bin weights are divided by their maximum value. We get a weight $w_i\in \left(0,1\right]$ for each $\mathcal{S}_i$. From each set $\mathcal{S}_i$, we randomly choose $w_i$ proportion of events rounded to the closest integer. We show in figure \ref{fig:gghweight}, the distribution of some kinematic-variables of the three datasets: unweighted events generated with HEFT model, weighted events with finite-top mass effects, and unweighted events used in neural network training. The effect of rounding to the nearest integer is seen in the later bins in $\textsc{met}$, where the statistics are weaker due to fewer events.

\section{Characteristics of High-level variables}
\label{app:hl_plots}

\begin{figure*}[t]
	\centering 		
	\includegraphics[height=0.9\textwidth, width=0.9\textwidth]{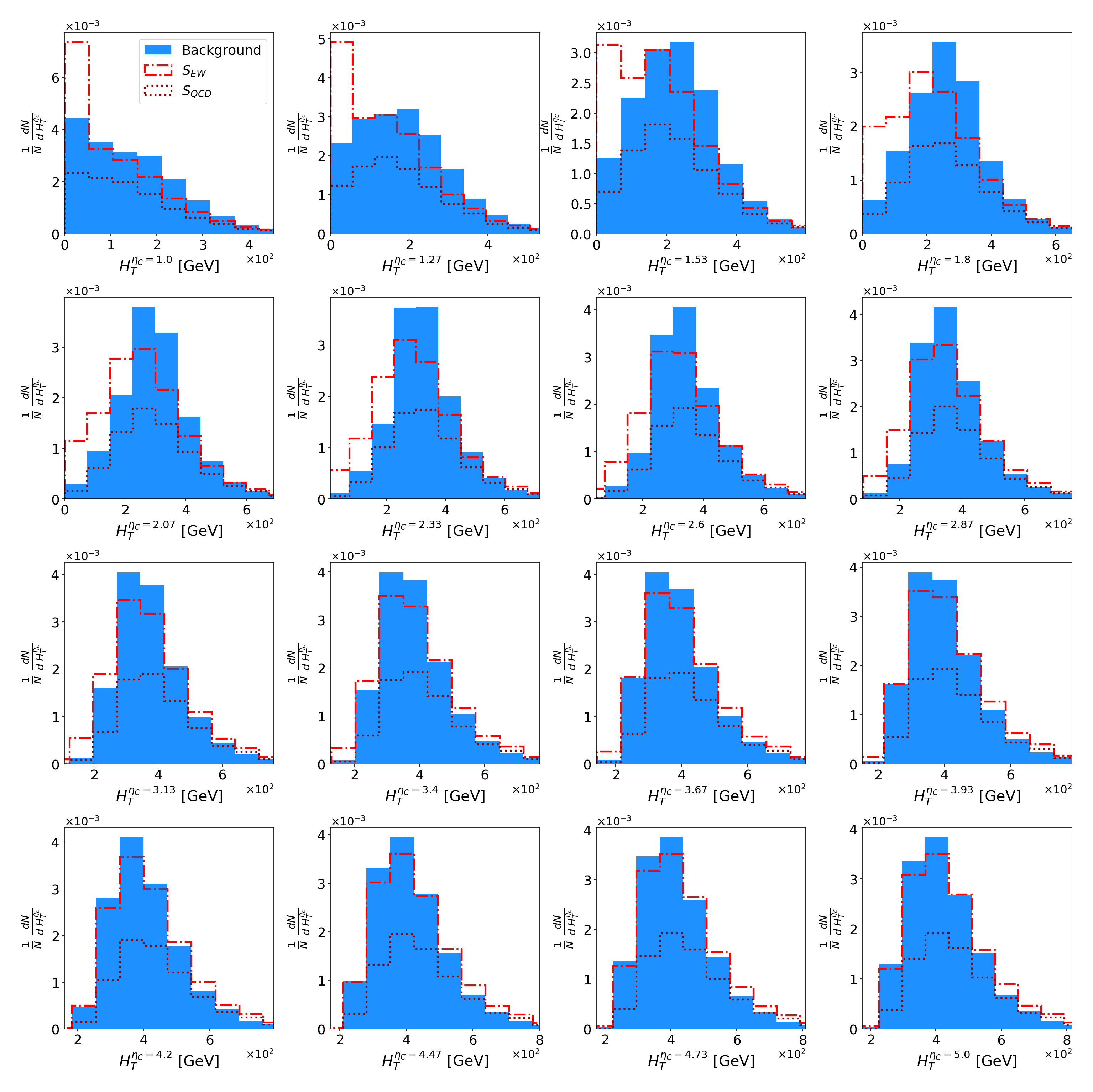}
	\caption{Signal vs Background distribution for all $H_T^{\eta_C}$ variables. We can see that for higher values of $\eta_C$ the signal and background are not that different and the difference grows as we approach the cut value of $\eta$ cut.}
	\label{fig:ht}
\end{figure*}

\begin{figure*}[t]	
	\centering 	
	\includegraphics[height=0.35\textwidth, width=0.49\textwidth]{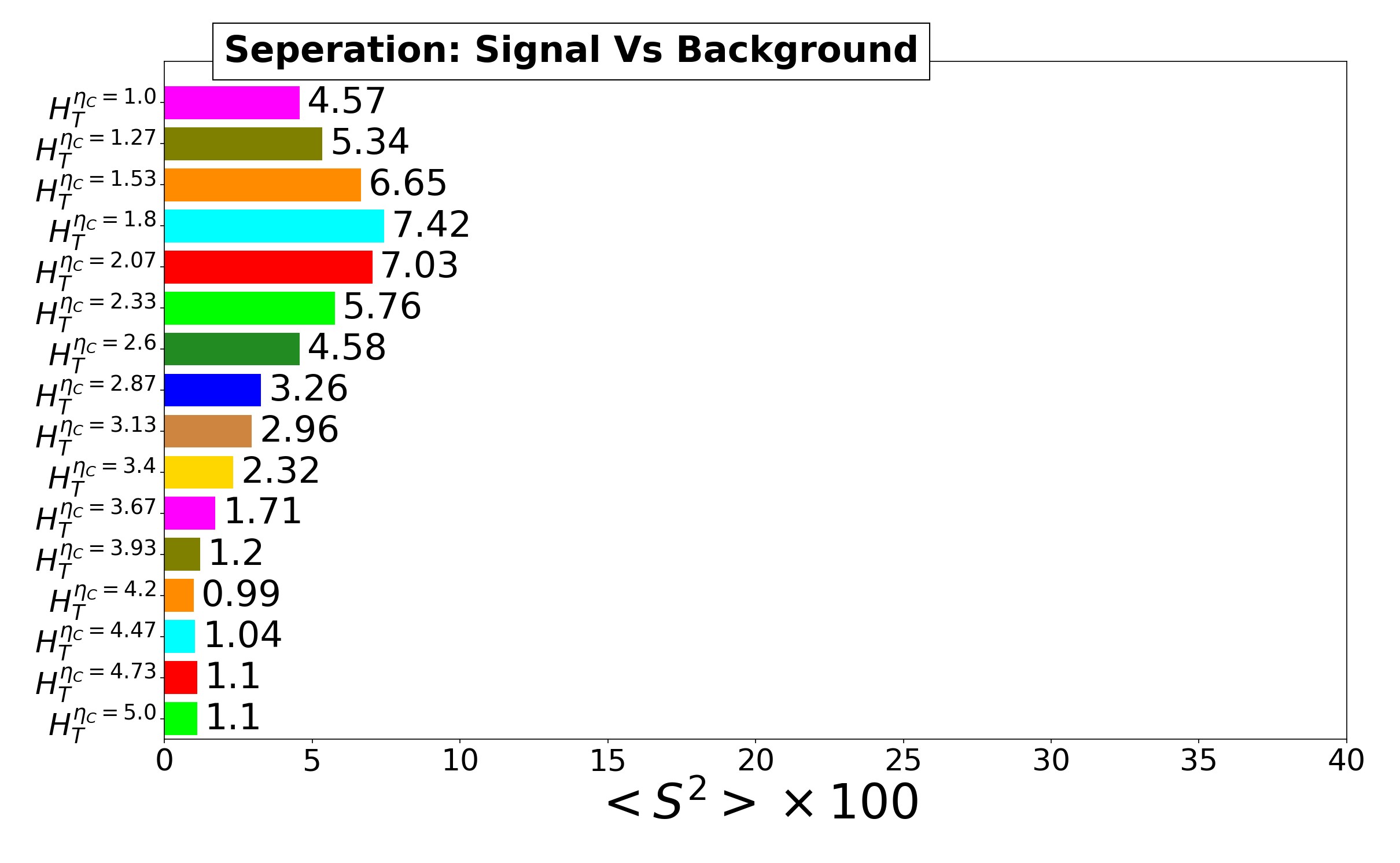}
	\includegraphics[height=0.35\textwidth, width=0.49\textwidth]{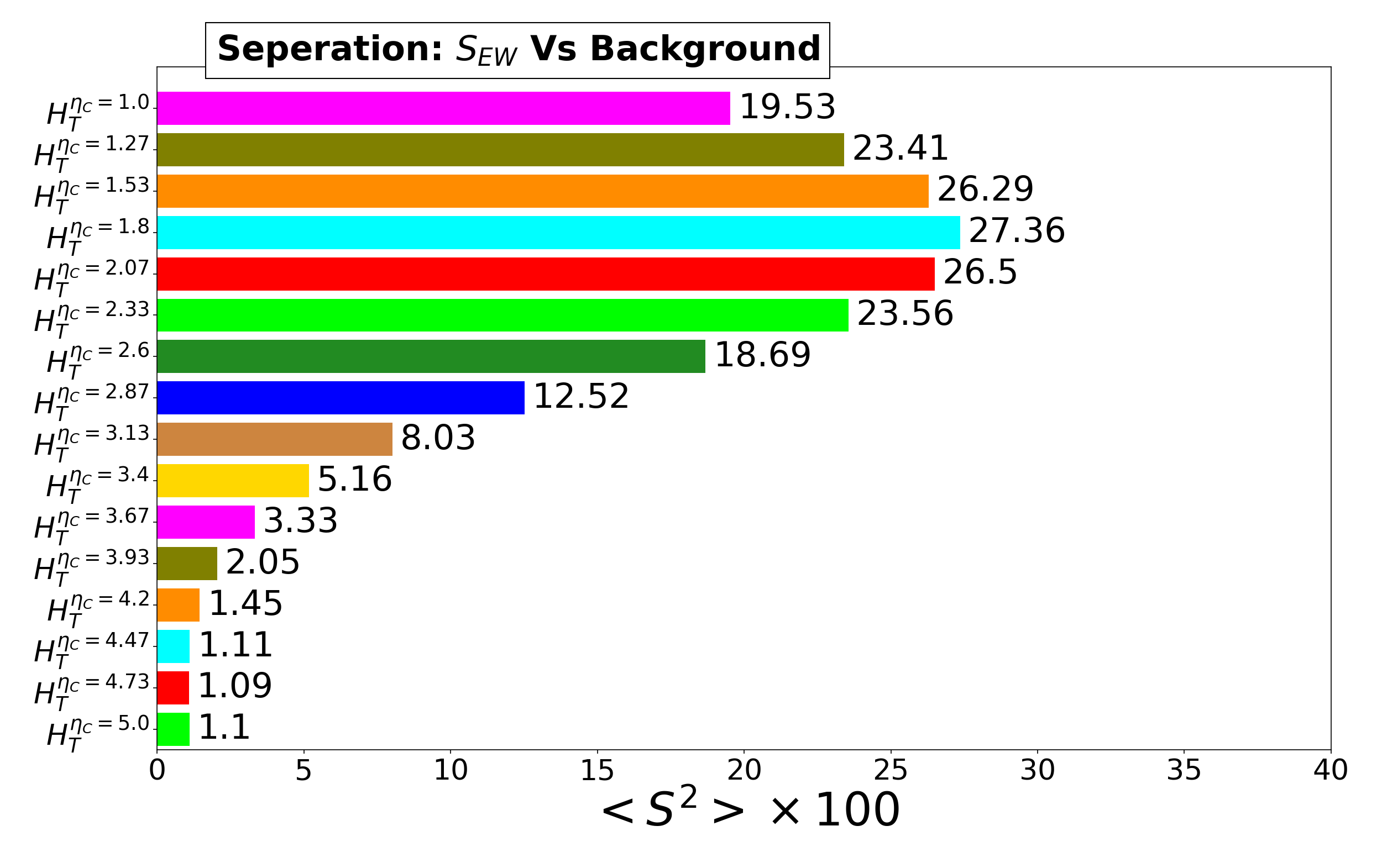}
	\caption{Seperation of all $H_T^{\eta_C}$ variables for (left) signal vs background and (right) $S_{EW}$ vs background. These have been calculated with 25000 events for each of the three datasets with the same binning. We can see that the presence of $S_{QCD}$ significantly reduces the discriminating power of $H_T^{\eta_C}$ variables on the left. }
	\label{fig:ht_var_imp}				
\end{figure*} 

In this section, we take a closer look at the high-level variables, especially the $\mathcal{R}$ variables defined in eq. \ref{eq:ht}. A key element in the extraction of variables belonging to the two spaces $\mathcal{K}$ and $\mathcal{R}$ is that the $\mathcal{K}$ variables are functions of four-momenta of reconstructed objects while the $\mathcal{R}$ variables are functions of four-momenta of tower-constituents (in our case from the Tower class of Delphes). The $\mathcal{R}$ variables do not take into account the tower-resolutions in the strict sense. This may point to a further reduction in the performance of ANNs compared to CNNs, where the tower-resolutions are better modeled.

\begin{figure*}[t]
	\centering		
	\includegraphics[height=0.31\textwidth, width=0.49\textwidth]{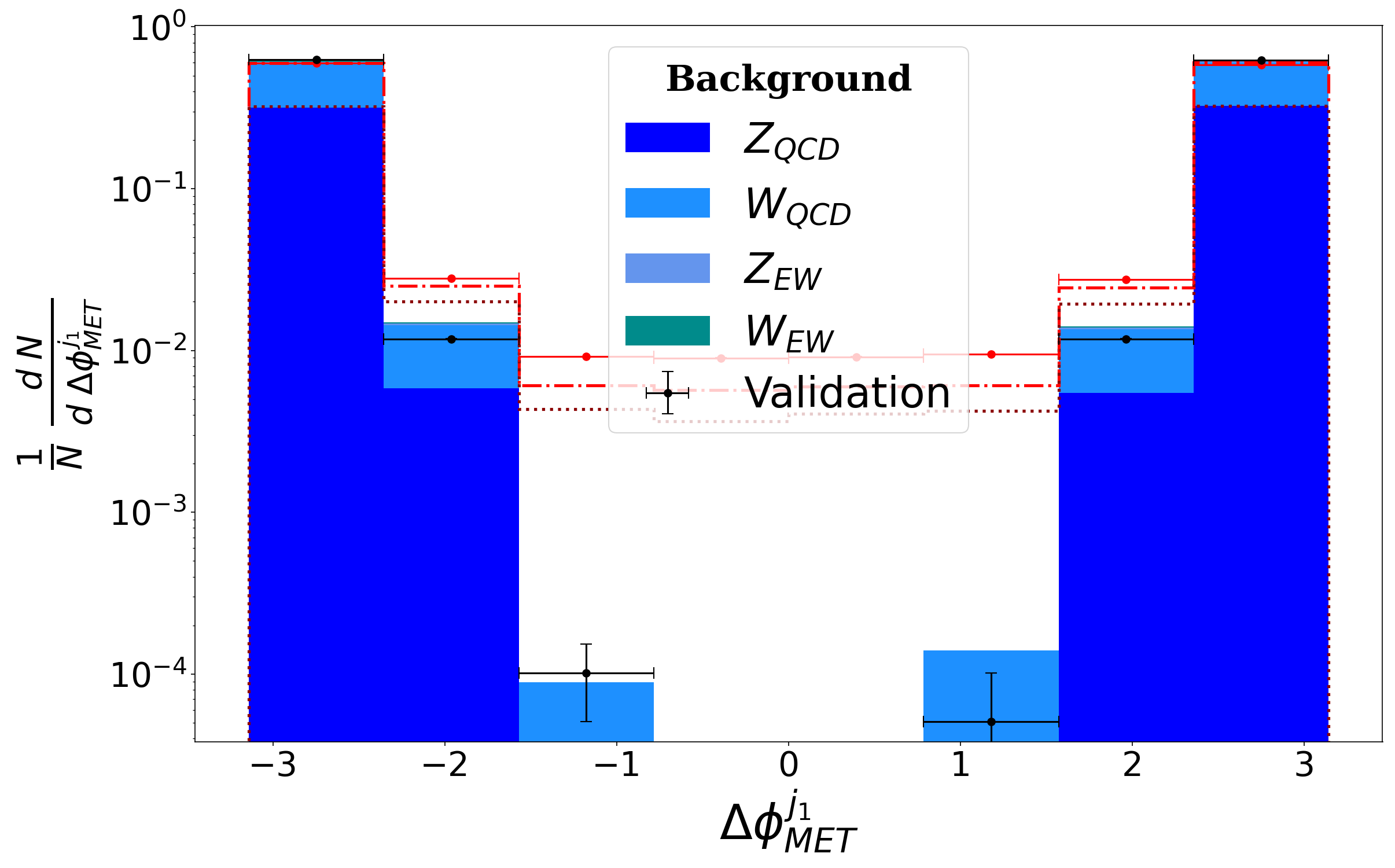}
	\includegraphics[height=0.31\textwidth, width=0.49\textwidth]{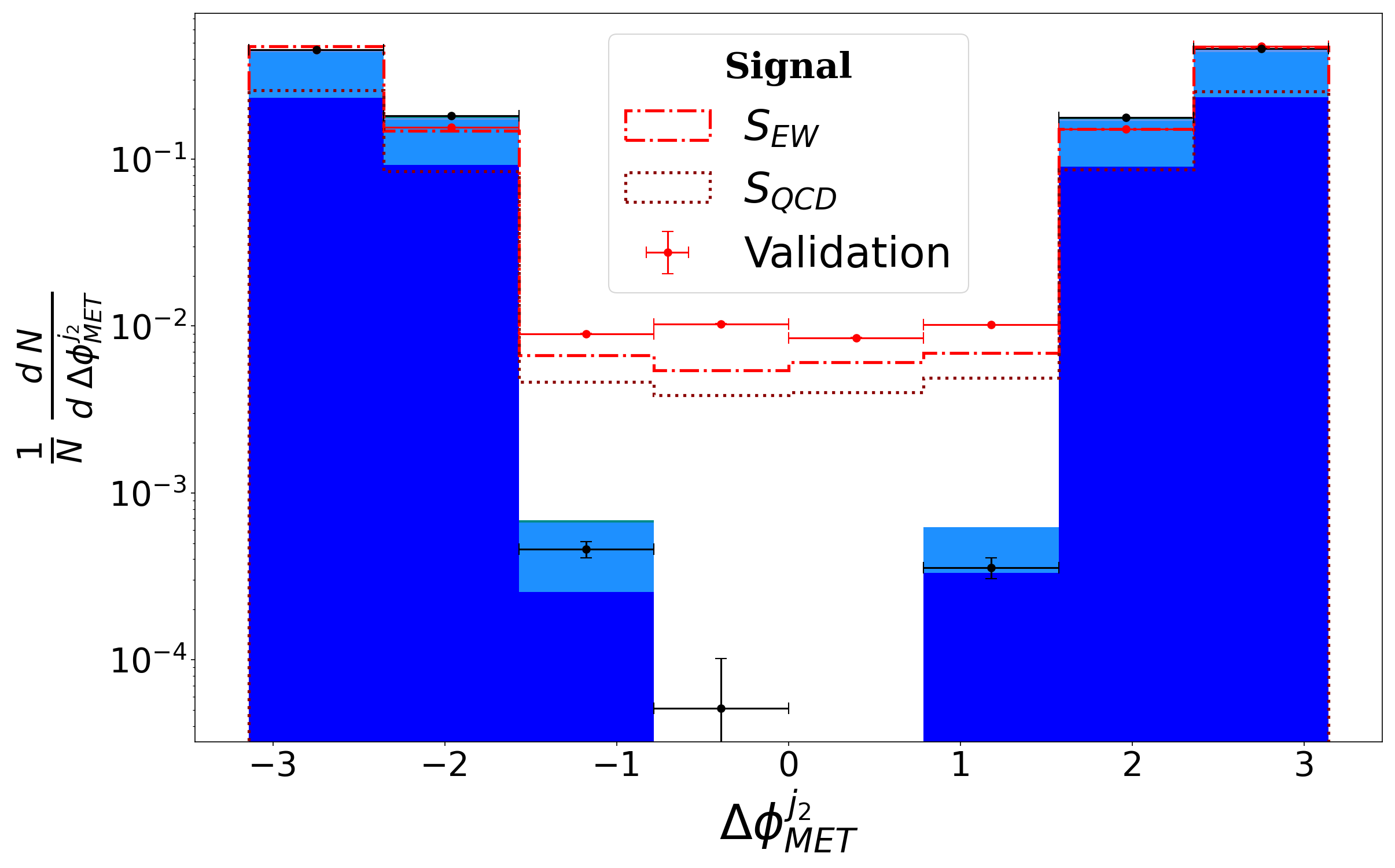}\\
	\includegraphics[height=0.31\textwidth,width=0.49\textwidth]{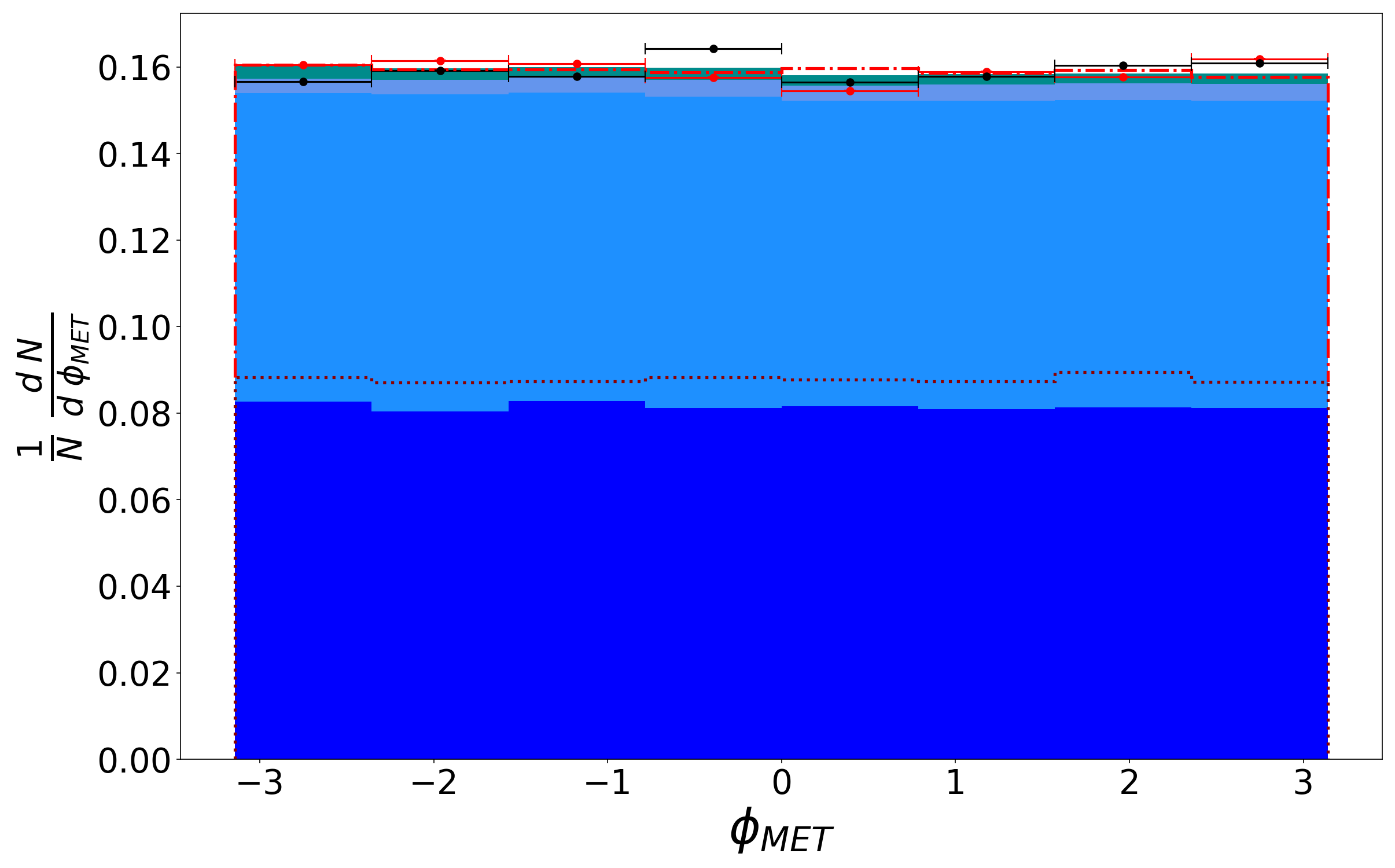}
	\includegraphics[height=0.31\textwidth, width=0.49\textwidth]{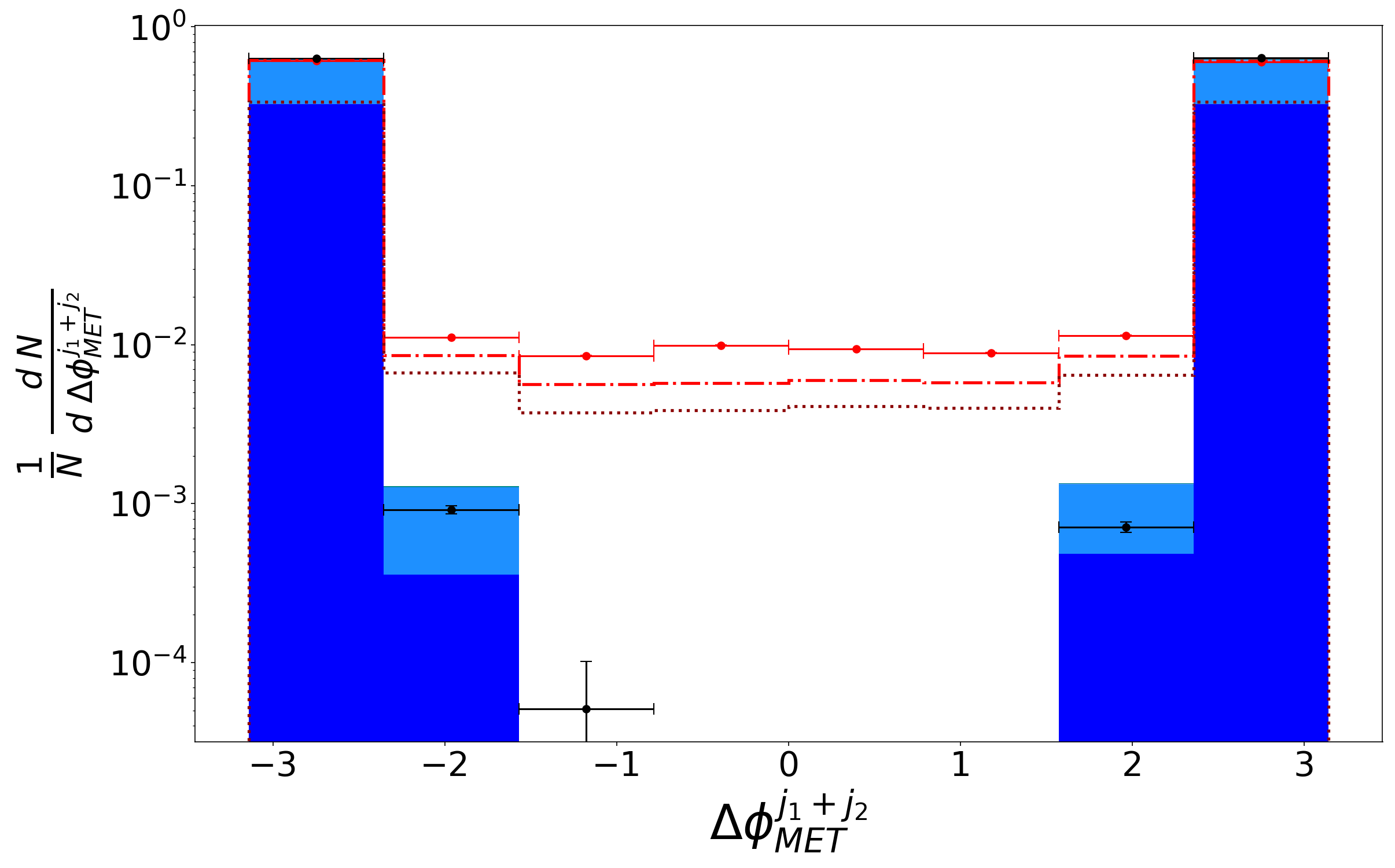}
	\caption{Signal vs Background distribution of the high-level kinematic variables excluded in figure \ref{fig:kinematic}}
	\label{fig:rest_kin}				
\end{figure*}

We show the signal vs background distribution of all $\mathcal{R}$ variables in figure \ref{fig:ht}. The contribution of $S_{EW}$ and $S_{QCD}$ to the total signal is stacked.  The separation, as defined in eq. \ref{eq:separation}, are shown for these variables for the total signal (also, $S_{EW}$) and  background in figure \ref{fig:ht_var_imp}. We can see that the trends in the distribution are in accordance with their respective values of separation. The shape of $S_{QCD}$ and the background distributions are similar for all values of $\eta_C$, and the overall differences, if any, comes from the contribution of $S_{EW}$. The separation is minimal and remains constant for  $\eta_C > 4$. This can be attributed to the fact that above these values, almost all of the calorimeter hits contribute to $H_T^{\eta_C}$. It increases continuously up to $\eta_C=1.8$ and then decreases till $\eta_C=1.0$. The increase is expected from the VBF topology, while the decrease can be attributed to the smallness of the region $[-\eta_C,\eta_C]$.

In figure \ref{fig:rest_kin}, we show the remaining kinematic variables not shown in figure \ref{fig:kinematic}. As can be seen, there is not much discriminatory information in any of these variables: $\phi_{\textsc{met}}$ is uniform for all channels since the beams are unpolarized, while $\Delta \phi^{J}_{\textsc{met}}$ ($J \in\{j_1,j_2,j_1+j_2\}$) has most contributions around $\pm\pi$, due to the imposed separation of two jets $\Delta \phi_{jj}$ and momentum conservation in the recoil of quarks/gluons against heavy bosons ($W^\pm,\;Z^0$ and $h^0$).

\section{Correlation between High-level variables and network-outputs}  
\label{app:corr} 
\begin{figure*}[t]
	\centering		
	\includegraphics[height=0.49\textwidth, width=0.49\textwidth]{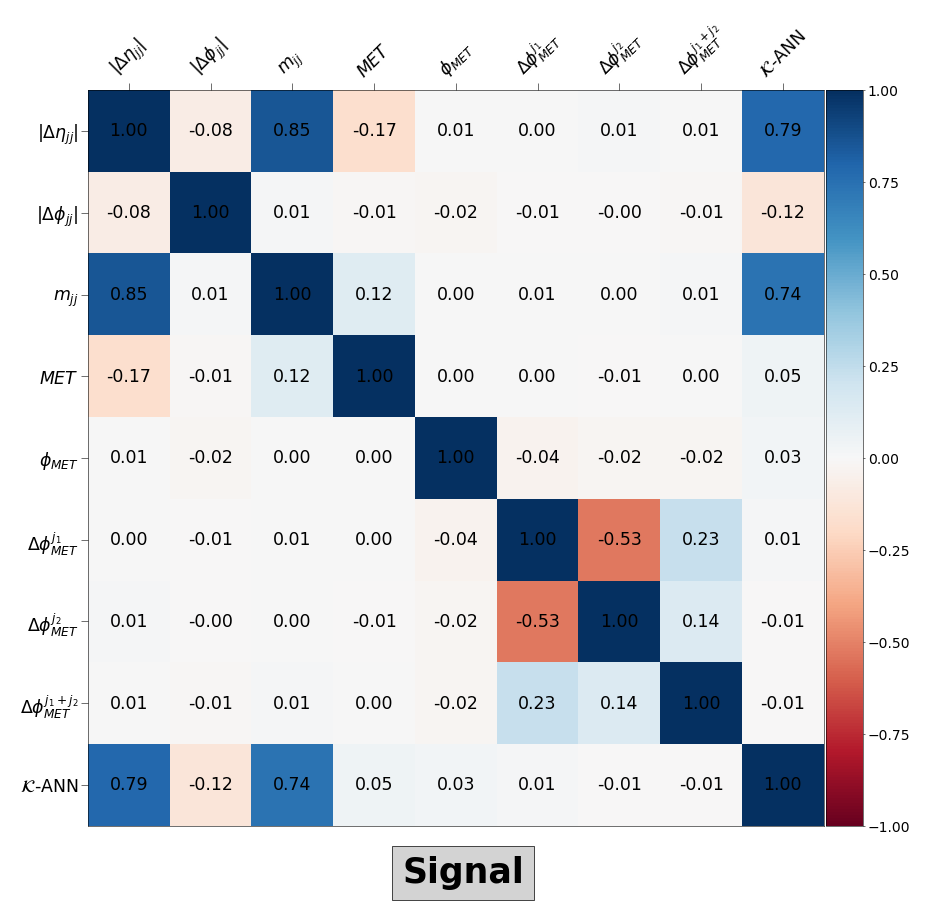}
	\includegraphics[height=0.49\textwidth, width=0.49\textwidth]{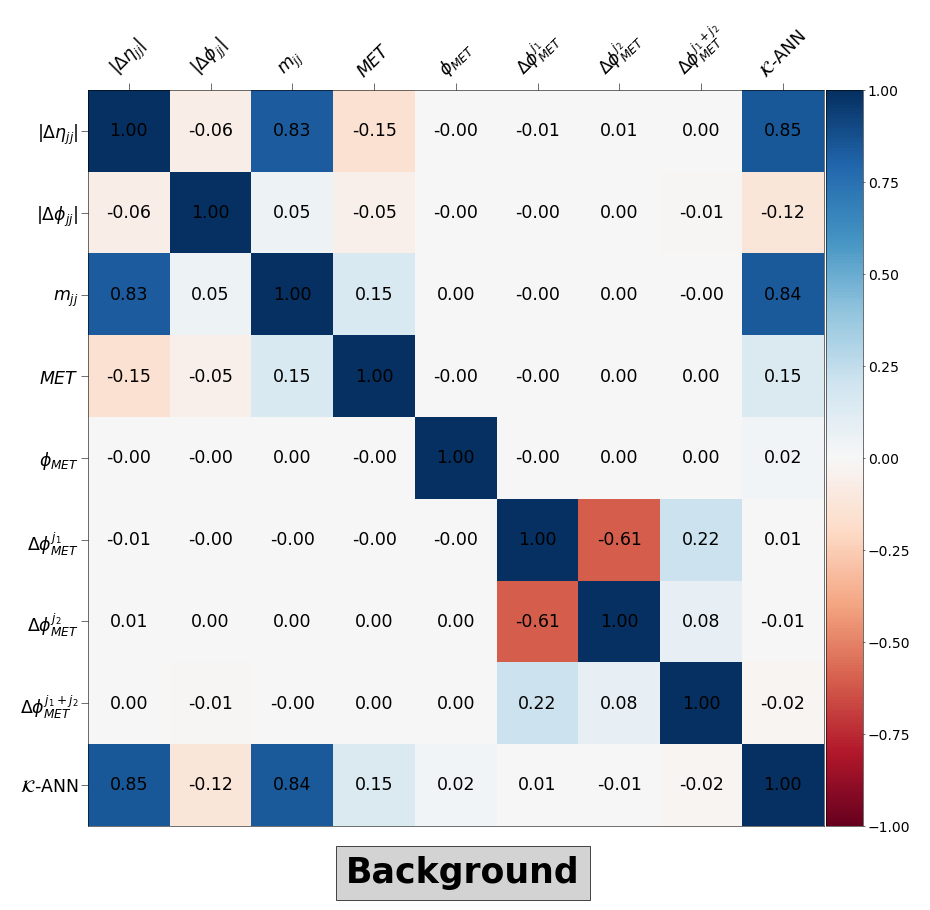}
	\caption{Correlation between the high-level kinematic variables $\mathcal{K}$ and the network-output of $\mathcal{K}$-ANN for (left) signal and (right) background.}\label{fig:corr_high_level}				
\end{figure*}

Salient features of the correlation of important variables with all neural network outputs have been given in the main text (figure \ref{fig:corr}). We examine the correlation of the ANNs with their inputs in this section. All correlations have been calculated using the inbuilt function in NumPy(v1.17.2)\cite{5725236}. 

In figure \ref{fig:corr_high_level} we show the correlations amongst the $\mathcal{K}$ variables including the $\mathcal{K}$-ANN network output for each class. As expected, the $\mathcal{K}$-ANN output is highly correlated with the two most discriminating variables $|\Delta\eta_{jj}|$ and $m_{jj}$. The next highest correlation with $\mathcal{K}$-ANN  is found to be with $\textsc{met}$ for background and $|\Delta\phi_{jj}|$ for signal. Except for $|\Delta\phi_{jj}|$, all other $\phi$ variables are almost uncorrelated with $\mathcal{K}$-ANN for both classes. The uniformity of $\phi_{\textsc{met}}$ results in its negligible correlation with all other variables. In the correlation among $\mathcal{K}$ variables, we can see two distinct sets of variables with comparatively moderate to high correlations formed amongst $\{|\Delta\eta_{jj}|,m_{jj},\textsc{met}\}$ and $\{\Delta \phi^{j_1}_{\textsc{met}},\Delta \phi^{j_2}_{\textsc{met}},\Delta \phi^{j_1+j_2}_{\textsc{met}}\}$. In the first set, $|\Delta\eta_{jj}|$ and $m_{jj}$ are almost completely correlated since, the angular opening between two four vectors $p^\mu_{j_1}$ and $p^\mu_{j_2}$, determine the invariant mass $m_{jj}=(p^\mu_{j_1}+p^\mu_{j_2})^2$. The \textsc{met} shows a moderate correlation with both $|\Delta\eta_{jj}|$ and $m_{jj}$ as momentum conservation forces $|\vec{p}_{j_1}+\vec{p}_{j_2}|$ to be higher for higher $\textsc{met}$. The correlation amongst the second subset can also be explained by transverse momentum conservation in the collision, with contamination from subsidiary QCD radiation and detector effects.

\begin{figure*}[t]
	\centering	
	\includegraphics[height=0.49\textwidth, width=0.49\textwidth]{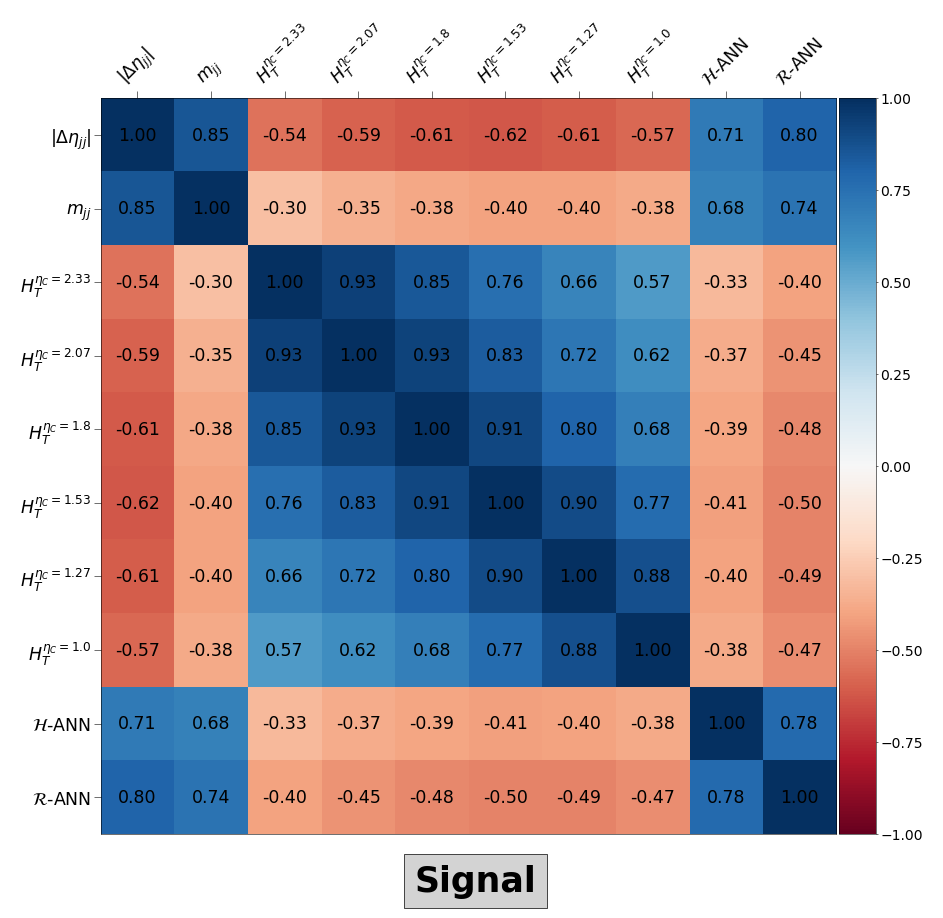}
	\includegraphics[height=0.49\textwidth, width=0.49\textwidth]{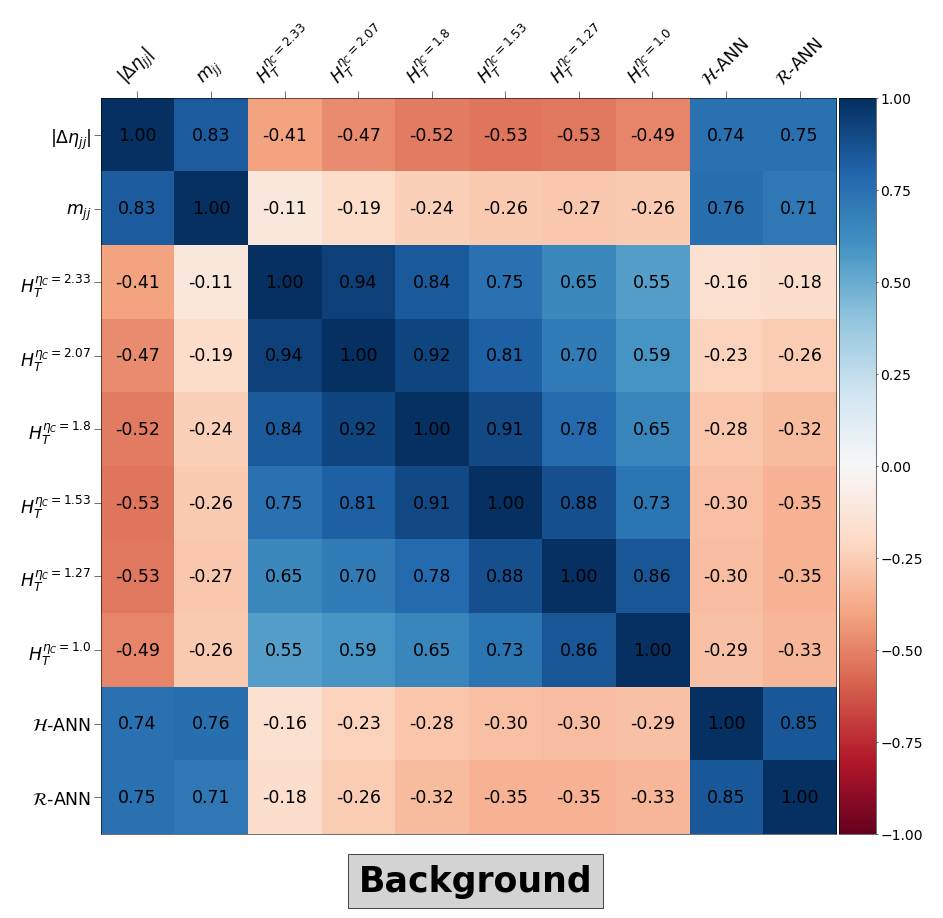}
	\caption{Correlation between the high-level variables $\mathcal{H}$ and the network-outputs of $\mathcal{R}$-ANN and $\mathcal{H}$-ANN for (left) signal and (right) background. For better representation we have chosen variables with non-negligible correlations with the network outputs.}\label{fig:corr_dnn}				
\end{figure*}

The class-wise correlations amongst the outputs of $\mathcal{R}$-ANN and $\mathcal{H}$-ANN along with six variables from $\mathcal{R}$ with high separation, and the two kinematic variables $|\Delta\eta_{jj}|$ and $m_{jj}$ are shown in figure \ref{fig:corr_dnn}. As expected, we see that the $\mathcal{R}$ variables are highly correlated with one another, which decreases with increasing distance in $\eta_C$. Another highlight is the negative correlation between them and the kinematic variables. It can be understood if we recall that the dominant radiation in the tower comes from the two leading jets, and an increase in $|\Delta\eta_{jj}|$ will decrease the calorimeter hits in the central regions.  In the case of correlations between neural-network outputs and their respective inputs, the sign of the correlation is not much relevant for binary classification due to the probabilistic interpretation of the outputs $y_i$: $y_0+y_1=1$ and $y_i>0$. On the contrary, the relative difference in sign and magnitude in correlations between the different input features and the output is relevant. In the case of $\mathcal{H}$-ANN, we can see that in terms of both magnitude (importance as plotted in figure \ref{fig:var_imp}) and sign (as discussed here), the relations amongst $\mathcal{K}$ and $\mathcal{R}$ variables are carried over to their corresponding correlations with the network-output.

\newpage

\bibliographystyle{JHEP}
\bibliography{ref.bib}

\end{document}